\documentclass[12pt]{article}

\usepackage[T1]{fontenc}
\usepackage[utf8]{inputenc}

\usepackage{amsmath,amssymb,slashed,hyperref}
\usepackage{amsfonts}
\usepackage{upgreek}
\usepackage{comment}

\numberwithin{equation}{section}

\usepackage[english]{babel}

\usepackage[mathscr]{eucal}
\usepackage{epsfig}
\usepackage{booktabs}
\usepackage{bbold}
\usepackage{slashed}

\DeclareMathAlphabet{\pxitfont}{OML}{pxmi}{m}{it}
\DeclareMathAlphabet{\pxitfontn}{U}{pxmia}{m}{it}
\def\Qb{\pxitfont Q}

\DeclareFontFamily{U}{euc}{}%
\DeclareFontShape{U}{euc}{m}{n}{<-6>eurm5<6-8>eurm7<8->eurm10}{}%
\DeclareSymbolFont{AMSc}{U}{euc}{m}{n} %
\DeclareMathSymbol{\psitt}{\mathord}{AMSc}{"20}    %%%symbol 20 = psi
\DeclareMathSymbol{\chitt}{\mathord}{AMSc}{"1F}    %%%symbol 1F = chi

\def\U{{\mathrm U}}

\newcommand{\bea}{\begin{array}}
\newcommand{\eea}{\end{array}}
\newcommand{\beq}{\begin{equation}}
\newcommand{\eeq}{\end{equation}}
\newcommand{\beqn}{\begin{eqnarray}}
\newcommand{\eeqn}{\end{eqnarray}}
\newcommand{\pvec}[1]{\vec{#1}\mkern2mu\vphantom{#1}}
\newcommand{\tr}{{\rm tr}}

\newcommand{\Str}{{\rm STr}}
\newcommand{\nnr}{\nonumber\\}
\newcommand{\ex}{{\rm e}}

\newcommand{\eps}{\epsilon}

\newcommand{\CC}{{\mathbb C}}
\newcommand{\RR}{{\mathbb R}}
\newcommand{\ZZ}{{\mathbb Z}}

\newcommand{\fr}{\frac}

\newcommand{\cN}{\mathcal{N}}
\newcommand{\cA}{\mathcal{A}}

\newcommand{\rmd}{{\rm d}}

\newcommand{\da}{{\dot{a}}}

\newcommand{\calK}{\mathcal{K}}

\newcommand{\bos}{{\bar{0}}}
\newcommand{\ferm}{{\bar{1}}}
\def\osp{\mathfrak{osp}}

\def\frak{\mathfrak}
\newcommand{\cupproduct}{\mathbin{\smile}}
\newcommand\myeq{\mathrel{\stackrel{\makebox[0pt]{\mbox{\normalfont ?}}}{=}}}

\font\teneusm=eusm10 
\font\seveneusm=eusm7 
\font\fiveeusm=eusm5
\newfam\eusmfam
\textfont\eusmfam=\teneusm \scriptfont\eusmfam=\seveneusm
\scriptscriptfont\eusmfam=\fiveeusm

\font\tencmmib=cmmib10 \skewchar\tencmmib='177
\font\sevencmmib=cmmib7 \skewchar\sevencmmib='177
\font\fivecmmib=cmmib5 \skewchar\fivecmmib='177
\newfam\cmmibfam
\textfont\cmmibfam=\tencmmib \scriptfont\cmmibfam=\sevencmmib
\scriptscriptfont\cmmibfam=\fivecmmib
\def\cmmib#1{{\fam\cmmibfam\relax#1}}

\font\teneurm=eurm10 
\font\seveneurm=eurm7 
\font\fiveeurm=eurm5
\newfam\eurmfam
\textfont\eurmfam=\teneurm \scriptfont\eurmfam=\seveneurm
\scriptscriptfont\eurmfam=\fiveeurm

\font\teneufm=eufm10 
\font\seveneufm=eufm7 
\font\fiveeufm=eufm5
\newfam\eufmfam
\textfont\eufmfam=\teneufm \scriptfont\eufmfam=\seveneufm
\scriptscriptfont\eufmfam=\fiveeufm

\def\bar{\overline}

\def\hat{\widehat}

\def\p{{\cmmib p}}

\def\N{{\mathcal N}}

\def\Str{{\mathrm{Str}}}
\def\tilde{\widetilde}
\def\t{{\mathbf{t}}}

\def\u{{\frak u}}

\def\2{{\bf 2}}
\def\1{{\bf 1}}
\def\0{{\bf 0}}

\def\bar{\overline}
\def\u{{\frak u}}

\def\psu{\mathfrak{psl}}

\def\g{{\mathfrak g}}

\begin{document}

\thispagestyle{empty}
\begin{flushright}\footnotesize
~%\texttt{preprint no}\\

\vspace{2.1cm}
\end{flushright}

\begin{center}
{\Large\textbf{\mathversion{bold} Analytic Torsion, 3d Mirror Symmetry,\\ And Supergroup Chern-Simons Theories}\par}

\vspace{2.1cm}

\textrm{Victor Mikhaylov}
\vspace{0.7cm}

\textit{Department of Physics, Princeton University, Princeton, NJ 08544} \\
\texttt{victor.mikhaylov@gmail.com}\\

 \vspace{3mm}

\par\vspace{2cm}

\textbf{Abstract}%\vspace{5mm}
\end{center}
\noindent  
We consider topological field theories that compute the Reidemeister-Milnor-Turaev torsion in three dimensions. These are the psl(1|1) and the U(1|1) Chern-Simons theories, coupled to a background complex flat gauge field. We use the 3d mirror symmetry to derive the Meng-Taubes theorem, which relates the torsion and the Seiberg-Witten invariants, for a three-manifold with arbitrary first Betti number. We also present the Hamiltonian quantization of our theories, find the modular transformations of states, and various properties of loop operators. Our results for the U(1|1) theory are in general consistent with the results, found for the GL(1|1) WZW model. We also make some comments on more general supergroup Chern-Simons theories.
\vspace*{\fill}

\begin{center}May 12, 2015\end{center}
\setcounter{page}{1}

\newpage

%\tableofcontents

\section{Introduction}
In this paper, we study the topological quantum field theory that computes the Reidemeister-Milnor-Turaev torsion \cite{Nicolaescu}, \cite{TuraevReview} in three dimensions. This is a Gaussian theory of a number of bosonic and fermionic fields in a background flat complex GL$(1)$ gauge field. It can be obtained by topological twisting from a free hypermultiplet with $\cN=4$ supersymmetry. This theory is very simple and can be given different names~--~the one-loop Chern-Simons path-integral  \cite{RMM}, or the Rozansky-Witten model \cite{RW} with target space $\CC^2$, or the U$(1|1)$ supergroup Chern-Simons theory \cite{RS3} at level equal to one, but we prefer to call it $\psu(1|1)$ supergroup Chern-Simons theory. 

Let us give a brief summary of the paper. In section \ref{esection}, we describe the definition of the theory. We explain that its functional integral computes a ratio of determinants, which depends holomorphically on a background flat GL$(1)$ bundle $\mathcal{L}$. We also define various line operators, the most important of which lead to the Alexander polynomial for knots and links.

In section~\ref{magnetic}, we use mirror symmetry in three dimensions to represent the $\psu(1|1)$ theory as the endpoint of an RG flow, that starts from the twisted version of the $\cN=4$ QED with one fundamental flavor. The computation of the partition function of the QED can be localized on the set of solutions to the three-dimensional version of the Seiberg-Witten equations \cite{WittenSW}. This provides a physicist's derivation of the relation between the Reidemeister-Turaev torsion and the Seiberg-Witten invariants, which is known as the Meng-Taubes theorem \cite{MT}, \cite{TuraevSW}. We consider, in particular, the subtle case of three-manifolds with first Betti number $b_1\le 1$ and show, how the quantum field theory manages to reproduce the details of the Meng-Taubes theorem in this case. Previously, the same RG flow has been used in \cite{Blau} to derive a special case of the Meng-Taubes theorem for the trivial background bundle, when the torsion degenerates to the Casson-Walker invariant. (We elaborate a little more on this in the end of section~\ref{magnetic}.) In comparison to \cite{Blau}, the new ingredient in our paper is the coupling of the QED to the background flat bundle $\mathcal{L}$, so let us explain, how this works. In flat space and before twisting, the QED has a triplet of FI terms $\phi^a$, which transform  as a vector under the SU$(2)_X$-subgroup of the ${\rm SU}(2)_X\times{\rm SU}(2)_Y$ R-symmetry. (In our notations, the scalars of the vector multiplet of the QED transform in the vector representation of SU$(2)_Y$.) These FI terms can be thought of as a vev of the scalars of a background twisted vector multiplet. The vector field $B_i$ of the same multiplet can be coupled in a supersymmetric way to the current of the topological U$(1)$-symmetry of the QED. Upon twisting the theory by SU$(2)_X$, the scalar and the vector fields of the twisted vector multiplet combine into a complex gauge field $B+i\phi$. Invariance under the topological supercharge $\Qb$ requires this background field to be flat. One can easily see that the partition function depends on it holomorphically. In the $\psu(1|1)$ theory, which emerges in the IR, the field $B+i\phi$ gives rise to the complex flat connection that is used in the definition of the Reidemeister-Turaev torsion.

In section~\ref{u11}, we consider the U$(1|1)$ supergroup Chern-Simons theory. It is obtained from the $\psu(1|1)$ theory by coupling it to U$(1)_k\times{\rm U}(1)_{-k}$  Chern-Simons gauge fields. It has been argued previously \cite{RS1}, \cite{RS3} that this theory computes the torsion that we study. We show that, in fact, the U$(1|1)$ theory for the compact form of the gauge group is a $\ZZ_k$-orbifold of the $\psu(1|1)$ theory, and thus, indeed, computes essentially the same invariant. To be more precise, there exist different versions of the U$(1|1)$ theory, which differ by the global form of the gauge group, but they all are related to the $\psu(1|1)$ theory. Mirror symmetry maps the U$(1|1)$ Chern-Simons theory at level $k$ to an orbifold of the same twisted $\cN=4$ QED, or equivalently, to an $\cN=4$ QED with one electron of charge $k$.

In section~\ref{Ham}, we present the Hamiltonian quantization of the theory. This section does not depend on the results of section~\ref{magnetic}, and can be read separately. By considering braiding transformations of the states on a punctured sphere, we recover the skein relations for the multivariable Alexander polynomial. We consider in some detail the Hilbert space of the $\psu(1|1)$ theory on a torus, and the correspondence between the states and the loop operators. We find the OPEs of line operators and the action of the modular group. In fact, as long as the background bundle $\mathcal{L}$ has non-trivial holonomies along the cycles of the Riemann surface, on which the theory is quantized, the Hilbert space is one-dimensional, and our analysis is very straightforward. We also discuss the canonical quantization of the U$(1|1)$ Chern-Simons theory. We consider modular transformations of the states on the torus, and find results very similar to those obtained from the GL$(1|1)$ WZW model \cite{RS2}. To our knowledge, this is the first example of the canonical quantization of a supergroup Chern-Simons theory, that does not assume an {\it a priori} relation to the WZW model.

In section~\ref{gen}, we discuss possible generalizations to other supergroup Chern-Simons theories. We make a summary of properties of such theories. (Some of these have previously appeared in \cite{MW}.) We also present some brane constructions, and consider possible dualities. 

Besides the papers that we have already mentioned, previous work on the topological field theory interpretation of the Meng-Taubes theorem includes \cite{MarinoMoore}, where the subject was approached from the four-dimensional Donaldson theory, and \cite{Donaldson}, where a mathematically rigorous proof of the Meng-Taubes theorem using TQFT was presented. All the mathematical facts about the Reidemeister-Turaev torsion, the Seiberg-Witten invariants and the Meng-Taubes theory, that we touch upon in this paper, can be found in a comprehensive review \cite{Nicolaescu}.

Finally, let us mention that there exists yet another approach \cite{Hutchings} to the Reidemeister-Turaev torsion, which presumably can be given a physical interpretation,~--~in this case, in terms of the first-quantized theory of Seiberg-Witten monopoles. Unfortunately, this will not be considered in the present paper.

\section{Electric Theory}\label{esection}
In this section, we describe the theory, which computes an analytic analog of the Reidemeister-Turaev torsion. Up to some details, it is simply the theory of the degenerate quadratic functional \cite{Schwarz}. One important difference, however, is that we introduce a coupling to a {\it complex} background flat bundle, and consider the torsion as a holomorphic function of it. Our definition is similar but not quite identical to the definition of the analytic torsion, known in the mathematical literature \cite{BK}. The discussion will be phrased in the language of supergroup Chern-Simons theory. Though this might seem like an unnecessary over-complication, it will make our formulas a little more compact, and will also help, when we discuss generalizations in later sections. Throughout the paper, the theory of this section will be called ``electric'', while its mirror, considered in section~\ref{magnetic}, will be called ``magnetic''.

\subsection{The Simplest Supergroup Chern-Simons Theory}\label{simplest}
In this section we introduce the $\psu(1|1)$ Chern-Simons theory. We work on a closed oriented three-manifold $W$. 

The superalgebra $\g\simeq\psu(1|1)$ is simply the supercommutative Grassmann algebra $\CC^{0|2}$. The Chern-Simons gauge field will be a $\CC^2$-valued fermionic one-form $A=A^I \hat{f}_I$, where $\hat{f}_+$ and $\hat{f}_-$ are the superalgebra generators. To make the theory interesting, we want to couple it to a background flat bundle. It could possibly be a GL$(2)$-bundle, where GL$(2)$ acts on $\g$ in the obvious way. However, the definition of the Chern-Simons action requires a choice of an invariant bilinear form. This reduces the symmetry to SL$(2)$, so we couple the theory to a flat SL$(2)$-bundle $\mathcal{B}$. The Chern-Simons action can be written as\footnote{Throughout the paper we use Euclidean conventions, in which the functional under the path-integral is $\exp(-I)$.} 
\beq
I_{\psu(1|1)}=\fr{i}{4\pi}\int_W\Str\,A\,\rmd_{\mathcal{B}}A\,,\label{CSact1}
\eeq
where the supertrace denotes an invariant two-form, Str$(ab)=\eps_{IJ}a^Ib^J$, and $\rmd_{\mathcal{B}}$ is the covariant differential, acting on the forms valued in $\mathcal{B}$. One could eliminate the flat gauge field from $\rmd_{\mathcal{B}}$ by a suitable choice of trivialization of $\mathcal{B}$, but we prefer not to do so. 

The supergroup gauge transformations act by $A\rightarrow A-\rmd_{\mathcal{B}}\alpha$. To fix the gauge, we introduce a $\g$-valued ghost field $C$. Since our gauge symmetry is fermionic, this field has to be bosonic: its two components are complex scalars $C^+$ and $C^-$. We also introduce a bosonic $\g$-valued antighost field $\bar{C}$ and a $\g$-valued fermionic Lagrange multiplier $\lambda$. The BRST generator $\Qb$ is defined to act as
\beq
\delta A=-\rmd_{\mathcal{B}}C\,,\quad 
\delta C=0\,,\quad \delta \lambda=0\,,\quad \delta\bar{C}=\lambda\,.\label{BRST}
\eeq
Next we have to choose an appropriate gauge-fixing action. It will contain in particular the kinetic term for the bosonic fields $C$ and $\bar{C}$, and we want to make sure that this term is positive-definite. To that end, we pick a hermitian structure on our flat bundle and restrict to unitary gauges. We impose a reality condition $\bar{C}^I=-\eps^{IJ}(C^J)^\dagger$. The complex flat connection in $\mathcal{B}$ can be decomposed as $B+i\phi$, where $B$ is a hermitian connection and $\phi$ is a section of the adjoint bundle. We introduce a covariant derivative $D_{i}=\partial_i+iB_i$, and also introduce notations  $\mathcal{D}_i=D_{i}-\phi_i$ for the covariant derivative in the flat bundle $\mathcal{B}$ and $\bar{\mathcal{D}}_i=D_{i}+\phi_i$ for the covariant derivative with the conjugate gauge field. We pick a metric $\gamma$ on $W$ and take the gauge-fixing action to be
\beq
I_{\mathrm{g.f.}}=\left\{\Qb,\int\rmd^3x\sqrt{\gamma}\gamma^{ij}\,\Str\left(\bar{\mathcal{D}}_i\bar{C}A_j\right)\right\}=-\int \rmd^3x\sqrt{\gamma}\gamma^{ij}\,\Str\left(\bar{\mathcal{D}}_i\bar{C}\mathcal{D}_jC-A_i\bar{\mathcal{D}}_j\lambda\right)\,.\label{gfa}
\eeq
The bosonic part of this action is manifestly positive-definite. The gauge-fixing condition is $\bar{\mathcal{D}}_iA^i=0$. The action has a ghost number symmetry U$(1)_{\rm F}$, under which the ghost and the antighost fields have charges $\pm 1$. If the background field satisfies $[\bar{\mathcal{D}}^i,\mathcal{D}_i]=0$, or equivalently $D^i\phi_i=0$, this symmetry is enhanced to SU$(2)$, which rotates $C$ and $\bar{C}$ as a doublet and which we will call SU$(2)_Y$. If we turn off the background gauge field completely, we also recover the ``flavor'' SU$(2)_{\rm fl}$ symmetry, which is the unitary subgroup of the SL$(2)$ automorphism group of the superalgebra. The groups SU$(2)_Y$ and SU$(2)_{\rm fl}$ commute. Together they generate an action of SO$(4)$ on the real four-dimensional space parameterized by $C$ and $\bar{C}$.

In this paper, we will not consider the general SL$(2)$ analytic torsion\footnote{The reason is that the Meng-Taubes theorem, which will be the subject of section~\ref{magnetic}, does not seem to generalize to SL$(2)$ torsion, since only the abelian part of the symmetry is visible in the UV. However, what could be generalized to the SL$(2)$ torsion (and, in fact, to Sp$(2n,\CC)$ torsion) is the Hamiltonian quantization that we consider in section \ref{Ham}. This generalization will be discussed elsewhere.}. From now on, we restrict our attention to the case that the background flat bundle is abelian, ${\mathcal B}={\mathcal{L}}\oplus\mathcal{L}^{-1}$, where\footnote{Throughout the paper, the coefficients in homology and cohomology are assumed to be $\ZZ$, unless explicitly specified otherwise.} $\mathcal{L}\in{\rm Hom}(H_1(W),\CC^*)$. By abuse of notation, we will denote the connection in $\mathcal{L}$ by the same letters $B+i\phi$, where now $B$ is understood to be a connection in a flat unitary line bundle, and $\phi$ is a closed one-form, whose cohomology class determines the absolute values of the holonomies in $\mathcal{L}$.

The abelian background field preserves a U$(1)_{\mathrm{fl}}$-subgroup of the flavor symmetry group SU$(2)_{\rm fl}$. We will furthermore assume that  $\phi$ is chosen to be the harmonic representative in its cohomology class, so that the ${\rm SU}(2)_Y$-symmetry is present.

\subsection{Relation To A Free Hypermultiplet}\label{free}
Our theory can be obtained by making a topological twist of the theory of a free $\cN=4$ hypermultiplet. This is a trivial special case of the general relation between supergroup Chern-Simons and $\cN=4$ Chern-Simons-matter theories, found in \cite{KapustinSaulina}. For completeness, we provide some details.

The R-symmetry group of $\cN=4$ supersymmetry in three dimensions is SU$(2)_X\times{\rm SU}(2)_Y$. The supercharges transform in the $(\bf{2},\bf{2},\bf{2})$-representation of ${\rm SU}(2)_{\rm Lorentz}\times{\rm SU}(2)_X\times{\rm SU}(2)_Y$. A supersymmetric theory can be twisted by taking the Lorentz spin-connection to act by elements of the diagonal subgroup of SU$(2)_{\rm Lorentz}\times{\rm SU}(2)_X$. This leaves an SU$(2)_Y$ doublet of invariant supercharges. We pick one of them, to be called $\Qb$, and use it to define a cohomological topological theory. The ghost number symmetry U$(1)_{\rm F}$ is the subgroup of SU$(2)_Y$, for which $\Qb$ is an eigenvector.

The scalars of the free hyper give rise to the ghost fields ${C}$ and $\bar{C}$. They parameterize a copy of the quaternionic line $\mathbb{H}$, which has a natural action of two commuting SU$(2)$ groups. One of them is identified with the R-symmetry group SU$(2)_Y$, and the other is the flavor symmetry SU$(2)_{\rm fl}$. The hypermultiplet fermions, which transform in the $(\bf{2},\bf{2},\bf{1})$ representation of the Lorentz and R-symmetry groups, upon twisting give rise to a vector field and a scalar, which we identify with the fermionic gauge field $A_i$ and the Lagrange multiplier field $\lambda$. 

Finally, the imaginary part of the flat connection $\phi_i$ originates from the SU$(2)_X$-triplet of hypermultiplet masses. While they are constant parameters in the untwisted theory, they are promoted to a closed one-form in the topological theory, still preserving the $\Qb$-invariance. Different terms in the action (\ref{CSact1}), (\ref{gfa}) can be easily seen to originate from the kinetic and the mass terms for the hypermultiplet scalars and fermions.

\subsection{A Closer Look At The Analytic Torsion}\label{definition}
Here we would like to take a closer look at the invariant that our theory computes. We discuss its properties and relation to other known definitions of the torsion. For simplicity, the manifold $W$ is assumed to be closed, unless indicated otherwise.

\subsubsection{Definition And Properties}\label{defprop}
The partition function of the theory can be written as a ratio of determinants,
\beq
\tau(\mathcal{L})=\fr{{\rm det}\,L_-}{{\rm det}^2\Delta_0}\,.\label{det}
\eeq
Here the operator $L_-=\star(d_B-\phi)+(d_B+\phi)\star$ is acting in $\Omega^1_{\mathcal L}(W)\oplus\Omega^3_{\mathcal L}(W)$, where $\Omega^p_{\mathcal L}(W)$ is the space of $p$-forms valued in ${\mathcal L}$. The twisted Laplacian $\Delta_0=-D_iD^i+\phi_i\phi^i$ is acting in $\Omega^0_{\mathcal L}(W)$. Note that the operator $\Delta_0$ is hermitian, while $L_-$ is hermitian only when $\phi=0$. 

The ratio $\tau({\mathcal{L}})$, by construction, is a holomorphic function of the flat bundle, even though the determinants in (\ref{det}) are not. We can understand the analytic properties of $\tau(\mathcal{L})$ rather explicitly. The absolute value of the torsion can be written in the usual Ray-Singer form as 
\beq
|\tau(\mathcal{L})|=\fr{(\det\Delta_1)^{1/2}}{(\det\Delta_0)^{3/2}}\,,\label{absdet}
\eeq
where $\Delta_1$ is the twisted Laplacian, acting on one-forms. The numerator in this formula vanishes, whenever the twisted cohomology $H^1(W,\mathcal{L})$ is non-empty. This subspace, possibly with the exception of the trivial flat bundle, is the locus of zeros of $\tau(\mathcal{L})$. The denominator vanishes, when the twisted cohomology $H^0(W,\mathcal{L})$ is non-empty, which is precisely when the flat bundle $\mathcal{L}$ is trivial. At this point the function $\tau(\mathcal{L})$ can potentially have a singularity. In fact, if the first Betti number $b_1$ of $W$ is greater than one, the singularity would be of codimension at least two, which is not possible for a holomorphic function. For $b_1=1$, let the holonomies of $\mathcal{L}$ around the torsion\footnote{A cycle is called ``torsion'' if it lies in the torsion part of $H_1(W)$, that is, if some multiple of it is trivial. This use of the word ``torsion'' is totally unrelated to ``torsion'' as an invariant of the manifold. Hopefully, this will not cause confusion.} one-cycles be trivial, and let $\mathbf{t}$ be the holonomy around the non-torsion one-cycle. At $\mathbf{t}=1$, the operators $\Delta_0$ and $\Delta_1$ have one zero mode each. At small $\mathbf{t}-1$, these eigenfunctions become quasi-zero modes with eigenvalues of order $(\mathbf{t}-1)^2$, according to the non-degenerate perturbation theory. Plugging this into (\ref{absdet}), we see that the ratio $\tau(\mathcal{L})$ near the trivial flat bundle is proportional to $1/(\mathbf{t}-1)^2$, that is, has a second-order pole. Finally, for $b_1=0$ the torsion is a function on the discrete set of flat bundles. For the trivial flat bundle and $b_1=0$, it is natural to set $\tau$ to be equal to infinity\footnote{One could say that for the trivial bundle the path-integral is undefined, since it has both bosonic and fermionic zero modes. But it is natural to set it equal to infinity for $b_1=0$, because, thinking in terms of gauge-fixing, the path-integral has a factor of inverse volume of the gauge supergroup, which is infinity, since this volume is zero. Taking $Z(S^3)=\infty$ also makes the factorization formulas of the ordinary Chern-Simons valid in the supergroup case.}. 

Another important property of the torsion is the relation
\beq
\tau(\mathcal{L})=\tau(\mathcal{L}^{-1})\,,\label{cconj}
\eeq
which follows from the charge conjugation symmetry $\mathcal{C}$ that maps the superalgebra generators as $\hat{f}_\pm\rightarrow\pm\hat{f}_\mp$, and the line bundle $\mathcal{L}$ to its dual $\mathcal{L}^{-1}$.

\subsubsection{Details Of The Definition}\label{defdetails}
We would like to make a more precise statement about what we mean by the formal definition (\ref{det}). Let us assume for now that the flat bundle $\mathcal{L}$ is unitary. If we eliminate the ambiguities in the definition of $\tau(\mathcal{L})$ for such bundles, the definition for complex flat bundles will also be unambiguous, by analyticity.

The absolute value (\ref{absdet}) of $\tau(\mathcal{L})$ is (the inverse of) the Ray-Singer torsion, which is a well-defined and metric-independent object. However, as is well-known in the context of Chern-Simons theory \cite{WittenCS}, the definition of the phase of $\tau(\mathcal{L})$ requires more care. With our assumption that $\mathcal{L}$ is unitary, the operator $L_-$ is hermitian and has real eigenvalues. Since the determinant of $L_-$ comes from a fermionic path-integral, it is natural to choose a regularization, in which it is real. The only possible ambiguity then is in its sign. Note that this is mainly interesting in the case when there is torsion in $H_1(W)$, so that the space of flat bundles is not connected, and signs can potentially be changed for different connected components.

Let us suggest a way to define the sign of $L_-$. What we are about to say might not seem particularly natural at first sight, but, as we show later, matches well with known definitions of the analytic and combinatorial torsion. Let us pick a spin structure $s$ on the three-manifold $W$, and take some oriented spin four-manifold $V$, of which $W$ with a given spin structure is a boundary. The line bundle $\mathcal{L}$ can be extended onto $V$, though the extension might not be flat. On $V$ we consider the Donaldson operator $L_4: \Omega^1_{\mathcal{L}}(V)\rightarrow \Omega^0_{\mathcal{L}}(V)\oplus \Omega^{2,-}_{\mathcal{L}}(V)$ that arises from the linearization of the self-duality equations, twisted by the line bundle $\mathcal{L}$. Here $\Omega^{2,-}$ is the bundle of anti-selfdual two-forms. We define the sign of the determinant of $L_-$, and therefore of the torsion $\tau(\mathcal{L})$, using the index of the elliptic operator $L_4$,
\beq
{\rm sign}\,\tau(\mathcal{L})=(-1)^{{\rm ind}(L_4)-{\rm ind}(L_{4,{\rm triv}})}\,,\label{sign}
\eeq
where $L_{4,{\rm triv}}$ is the Donaldson operator coupled to the trivial line bundle. The motivation behind this definition is that, if we were to compute the change of sign of $\det L_-$ under a continuous change of $\mathcal{L}$, we could naturally do it by using the formula (\ref{sign}) with the four-manifold taken to be the cylinder $W\times I$, since the index of $L_4$ on such a cylinder computes the spectral flow of $L_-$.

We started with a choice of a spin structure, but so far it did not explicitly enter the discussion. Its role is the following. For two different choices of the four-manifold, the change in the sign of $\det L_-$ is governed by the index of $L_4$ on a closed four-manifold $V'$, which, according to the index theorem, is
\beq
{\rm ind}(L_4)-{\rm ind}(L_{4,{\rm triv}})=\int_{V'}c_1(\mathcal{L})^2\,.\label{indVV}
\eeq
However, since the spin structure on $W$ can be extended to $V'$, the four-manifold $V'$ is spin, and therefore its intersection form is even, and so is the right hand side of (\ref{indVV}). We conclude that the sign of $\tau(\mathcal{L})$ depends on a spin structure on $W$, but not on the choice of the four-manifold. (This is equivalent to the well-known fact \cite{DW} that a choice of a spin-structure allows to define a half-integral Chern-Simons term for an abelian gauge field.)

It is not hard to calculate the dependence on the spin structure explicitly. Let $s_1$ and $s_2$ be two spin structures on $W$, which differ by some $x\in H^1(W,\ZZ_2)$. Let $V_1$ and $V_2$ be four-manifolds with boundary $W$, onto which $s_1$ and $s_2$ extend. Now the closed four-manifold $V'$, glued from $V_1$ and $V_2$ along their boundary $W$, need not be spin, and its Stiefel-Whitney class $w_2\in H^2(V',\ZZ_2)$ can be non-zero. The intersection form is not even, but its odd part is governed by the Wu's formula, which tells us that ${\bar{c}}_1^2={\bar{c}}_1\cupproduct w_2$, where $\bar{c}_1$ is the mod 2 reduction of $c_1(\mathcal{L})$. (This is true for any $H^2(V',\ZZ_2)$ class, of course.) The Stiefel-Whitney class of $V'$ is determined by $x$. For a given good covering of $V'$, the two spin structures $s_1$ and $s_2$ define a lift of the transition functions of the tangent bundle of $V'$ from SO$(4)$ to Spin$(4)$, and this lift is consistent everywhere, except for a codimension-two chain, lying in $W$. This chain defines the Stiefel-Whitney class of $V'$, but it is also the Poincar\'e dual of the class $x$ in $W$. These arguments allow us to write
\beq
\int_{V'}c_1(\mathcal{L})^2=\int_{V'} c_1(\mathcal{L})\cupproduct w_2=\int_{{\rm PD}(w_2)}c_1(\mathcal{L})=\int_{{\rm PD}(x)}c_1(\mathcal{L})=\int_W c_1(\mathcal{L})\cupproduct x\quad {\rm mod}\,\,2\,,
\eeq
where ${\rm PD}$ stands for Poincar\'e dual. We conclude that under a change of the spin structure by $x$, the sign of $\tau(\mathcal{L})$ changes by the factor
\beq
(-1)^{\int_W c_1(\mathcal{L})\cupproduct x}\,.\label{sign1}
\eeq

It will be useful to rearrange this formula a little. For that we need to recall a couple of topological facts. The topology of a flat line bundle is completely defined by its holonomies around the torsion one-cycles. This is formalized by the following exact sequence,
\beq
H^1(W)\rightarrow H^1(W,\RR)\xrightarrow{\alpha} H^1(W,{\rm U}(1))\xrightarrow{\beta}{\rm tor}\,H^2(W)\rightarrow 0\,,
\eeq
which is associated to the short exact sequence of coefficients $0\rightarrow \ZZ\rightarrow \RR\rightarrow {\rm U}(1)\rightarrow 0$. By Pontryagin duality, $H^1(W,{\rm U}(1))\simeq{\rm Hom}(H_1(W),{\rm U}(1))$ is the abelian group of (unitary) flat line bundles on $W$. The morphism $\alpha$ gives a flat bundle with trivial holonomy around the torsion cycles and given holonomy around the non-torsion cycles\footnote{What one means by non-torsion cycles is not canonically defined, but this does not matter, when the holonomies around the torsion cycles are trivial.}. The morphism $\beta$ maps a given flat bundle to its first Chern class, which depends only on the holonomies around the torsion cycles, by exactness of the sequence. Pick a pair of classes $y_1$ and $y_2$ from ${\rm tor}\,H^2(W)$. Let $\mathcal{L}_1$ be some flat bundle with Chern class $y_1$. Its holonomies around the torsion cycles are completely defined by $y_1$. We can take a holonomy of $\mathcal{L}_1$ around the one-cycle, Poincar\'e-dual to $y_2$. The logarithm of this number gives a pairing ${\rm tor}\,H^2(W)\times{\rm tor}\,H^2(W)\rightarrow \mathbb{Q}/\mathbb{Z}$, which is known as the linking form. An important fact is that it is bilinear and symmetric. (Actually, this pairing is just the U$(1)\times{\rm U}(1)$ Chern-Simons term for flat bundles.)

Returning to the formula (\ref{sign1}), we note that $x\in H^1(W,\ZZ_2)$ defines a $\ZZ_2$-bundle, and (\ref{sign1}) is the holonomy of this bundle around the one-cycle, Poincar\'e dual to $c_1(\mathcal{L})$. Since the linking form is symmetric, this holonomy is equal to the holonomy of $\mathcal{L}$ around the one-cycle, Poincar\'e dual to $c_1(x)$, where, to construct $c_1(x)$, we think of the $\ZZ_2$-bundle defined by $x$ as of a U$(1)$-bundle. This holonomy will be denoted by $\mathcal{L}(c_1(x))$. We conclude that it defines the change of the sign of $\tau(\mathcal{L})$, when the spin structure on $W$ is changed by $x$. To indicate the dependence on the spin structure explicitly, we will sometimes write the torsion as $\tau_s(\mathcal{L})$, so that
\beq
\tau_{x\cdot s}(\mathcal{L})=\mathcal{L}(c_1(x))\,\tau_s(\mathcal{L})\,.\label{schange}
\eeq
It is noteworthy that if the line bundle $\mathcal{L}$ has trivial holonomies around 2-torsion cycles, the definition of $\tau(\mathcal{L})$ is independent of any choices at all.

In fact, even for a generic flat bundle, $\tau_s(\mathcal{L})$ depends on something less than a spin structure. There is a natural map from the set of spin structures to the set of spin-$\CC$ structures with trivial determinant, which is given by tensoring with a trivial line bundle,. This map is not an isomorphism, because in general two different spin structures can map to the same spin-$\CC$ structure. Since the change of the sign of $\tau_s(\mathcal{L})$ under a change of $s$ by an element $x$ of $H^1(W,\ZZ_2)$ depends only on the first Chern class of the line bundle obtained from $x$, the sign of $\tau_s(\mathcal{L})$ really depends only on a spin-$\CC$ structure with trivial determinant, and not on the spin structure itself.

One could consider some trivial generalizations of our definition of the torsion. For example, $\tau_s$ can be naturally defined for an arbitrary spin-$\CC$ structure $s$, not necessarily with trivial determinant. Let $s_0$ be some arbitrary spin-$\CC$ structure with trivial determinant, $s$ be an arbitrary spin-$\CC$ structure, and let $y\in H^2(W)$ be such that $y\cdot s=s_0$. We can set $\tau_s(\mathcal{L})=\mathcal{L}(y)\tau_{s_0}(\mathcal{L})$. Clearly, (\ref{schange}) implies that $\tau_s$ depends only on $s$, and not on the choice of $s_0$. In quantum field theory terms, this modification amounts to adding to the action a local topologically-invariant functional of the background gauge field~--~the Wilson loop of $\mathcal{L}$ around the cycle, Poincar\'e-dual to $y\in H^2(W)$. Another possible  modification of the definition would be to add a Chern-Simons term for the background field $B$. Note that, if we choose the coefficient of this term to be a half-integer, this would eliminate the dependence of $\tau_s$ on the spin structure. In what follows, we will mostly restrict to our most basic definition of $\tau_s$, unless indicated otherwise.

\subsubsection{Comparison To Known Definitions}\label{comparison}
Let us comment on the relation of our torsion to some known definitions from the mathematical literature. A rigorous definition of the complex analytic torsion was given in \cite{BK}. The authors consider essentially\footnote{There are some differences. The discussion in \cite{BK} is more general: the authors consider a manifold of arbitrary odd dimension, and not necessarily one-dimensional flat vector bundles. Another difference from our approach, if phrased in path-integral language, is that in \cite{BK} the gauge-fixing term in the analog of (\ref{gfa}) is defined using the derivative $\mathcal{D}$, rather than its conjugate. This eliminates the need to pick a hermitian structure on the flat bundle, but makes the functional integral representation of the determinant more formal. Finally, the ratio of determinants in \cite{BK} is actually the inverse of ours.} the same ratio of determinants (\ref{det}) and use the $\zeta$-function regularization to define it as a holomorphic function of the flat bundle $\mathcal{L}$. An important difference, however, is that for a unitary flat bundle their torsion is not real, but has a phase, proportional to the eta-invariant of $L_-$. In the language of functional integral, such definition is perhaps more natural \cite{WittenCS}, when the determinant of $L_-$ comes from a bosonic, rather than a fermionic functional integral. The relation to our definition is given by the APS index theorem: to transform the eta-invariant into the index, one needs to subtract what might be called a half-integral Chern-Simons term of the flat connection in the line bundle $\mathcal{L}$. This is why the dependence on a spin structure appeares in our story, but not in \cite{BK}.

In fact, there is a combinatorial definition of torsion, which, as we conjecture, is precisely equal to our $\tau_s(\mathcal{L})$. This is the Turaev's refinement of Reidemeister torsion\footnote{Note that sometimes Reidemeister torsion is defined to be the inverse of what we consider here. With the definition that we use, the absolute value of the combinatorial torsion is equal to the {\it inverse} of Ray-Singer torsion, defined in the usual way.}. We briefly summarize some facts about it. For a detailed review, as well as references, the reader can consult \cite{Nicolaescu}. 

Let $W'$ be a compact three-manifold, which is closed or is a complement of a link neighborhood in a closed three-manifold, so that it has a boundary consisting of a number of tori. (In our language, non-empty boundary will correspond to adding line operators, to which we turn in the next section.) In either case, the Euler characteristic of $W'$ is zero. Reidemeister torsion of $W'$ is defined as the determinant of a particular acyclic complex, twisted by a vector representation of the fundamental group of the manifold.  The determinant of this combinatorially defined complex can be viewed as a discretisation of the functional integral, which computes the analytic torsion. We will assume that the representation of the fundamental group is given by the flat line bundle $\mathcal{L}$. Reidemeister torsion is defined only up to a sign and up to multiplication by a holonomy of $\mathcal{L}$ around an arbitrary cycle in $W'$. This happens because the determinant depends on the basis in the complex, of which there is no canonical choice. Turaev has shown \cite{RT} that this ambiguity can be eliminated, once one makes a choice of what he called an Euler structure\footnote{More precisely, the choice of an Euler structure eliminates the freedom to multiply the torsion by a holonomy of $\mathcal{L}$, while the overall sign can be fixed by choosing an orientation in the homology $H_\bullet(W')$. At least for a closed three-manifold, there exists a canonical homology orientation, defined by the Poincar\'e duality, and we assume that our theory automatically picks this orientation.}. In analytical terms, it is a choice of a nowhere vanishing vector field, up to homotopy and up to an arbitrary modification inside a three-ball. Such vector fields always exist on $W'$, since $\chi(W')=0$. In three dimensions, it is not hard to see that Euler structures are in  a canonical one-to-one correspondence with spin-$\CC$ structures. For a spin-$\CC$ structure $s$, let us denote the Reidemeister-Turaev combinatorial torsion by $\tau^{\rm RT}_s(\mathcal{L})$. Under a change of the spin-$\CC$ structure by an element $y\in H^2(W')$, the torsion changes as
\beq
\tau^{\rm RT}_{y\cdot s}(\mathcal{L})=\mathcal{L}(y)\,\tau^{\rm RT}_s(\mathcal{L})\,,\label{schange1}
\eeq
where, as usual in our notations, $\mathcal{L}(y)$ is the holonomy of $\mathcal{L}$ around the cycle Poincar\'e dual to $y$. The combinatorial torsion also has a charge conjugation symmetry $\mathcal{C}$
\beq
\tau^{\rm RT}_s(\mathcal{L}^{-1})=(-1)^\ell\tau^{\rm RT}_{\bar{s}}(\mathcal{L})=(-1)^\ell\, \mathcal{L}^{-1}(c_1({\rm det}\,s))\tau^{\rm RT}_s(\mathcal{L})\,,\label{cconj1}
\eeq
where $\bar{s}$ is the conjugate of the spin-$\CC$ structure $s$, and $\ell$ is the number of connected components of the boundary of $W'$. The second equality here follows from (\ref{schange1}).

If the three-manifold $W'$ is closed and the spin-$\CC$ structure $s$ has trivial determinant, we claim that $\tau^{\rm RT}_s$ coincides with our analytic torsion $\tau_s$. (Modulo signs, that is, ignoring the dependence on the spin structure, this statement would follow from the results of \cite{BK} and \cite{Huang}.) For a spin-$\CC$ structure with trivial determinant, the properties (\ref{schange1}) and (\ref{cconj1}) reduce to our formulas (\ref{schange}) and (\ref{cconj}), respectively. When the three-manifold $W'$ is not closed but is a complement of a link, the relation between $\tau^{\rm RT}_s$ and our $\tau_s$ should still hold, with an appropriate definition of the analytic torsion in presence of line operators. This will be discussed in the next section.

An important special case is when the flat bundle $\mathcal{L}$ has trivial holonomies around the torsion one-cycles. Then $\tau_s^{\rm RT}(\mathcal{L})$ is a holomorphic function of $b_1(W')$ variables $\mathbf{t}_1,\dots,\mathbf{t}_{b_1}$. Let us also ignore the dependence on $s$, so that we consider $\tau$ modulo sign and modulo multiplication by powers of $\mathbf{t}_\bullet$. This variant of the combinatorial torsion is known as the Milnor torsion. A theorem due to Milnor \cite{Milnor} and Turaev \cite{Turaev} describes its relation to the Alexander polynomial $\Delta$ of $W'$, which is a function of the same variables $\mathbf{t}_1,\dots \mathbf{t}_{b_1}$. If $b_1(W')>1$, then $\tau=\Delta$. If $b_1(W')=1$, then $\tau=\mathbf{t}\Delta/(\mathbf{t}-1)^2$, if $W'$ is a closed three-manifold, and $\tau=\Delta/(\mathbf{t}-1)$, if $W'$ is a complement of a knot in a closed three-manifold. For a closed $W'$, these statements are in agreement with the analytical properties of our $\tau$, described in section \ref{defprop}.

\subsection{Line Operators}\label{eline}
We would like to define some line operators in our theory, in order to study knot invariants. First thing that comes to mind is to use Wilson lines. For these to be invariant under the transformations (\ref{BRST}), they should be labeled by representations of $\mathfrak{pl}(1|1)$. This superalgebra contains $\psu(1|1)$ as well as one bosonic generator, which acts on the fermionic generators with charges $\pm1$. The Wilson lines should be defined with the $\mathfrak{pl}(1|1)$ connection $A+B+i\phi$. In fact, the only irreducible representations of $\mathfrak{pl}(1|1)$ are one-dimensional representations, to be denoted $(m)$, in which the bosonic generator acts with some charge $m$, and the fermionic generators act trivially. Inserting a Wilson loop in representation $(m)$ along a knot $K$ is equivalent to multiplying the path-integral by the $m$-th power of the holonomy of the background bundle $\mathcal{L}$ around the cycle $K$. Though this operator is of a rather trivial sort, it will be convenient to consider it as a line operator. It will be denoted by $L_m$, $m\in \ZZ$. According to the remarks at the end of section \ref{defdetails}, inserting operators $L_m$ around various cycles is equivalent to changing the spin-$\CC$ structure, with which the torsion is defined.

\begin{figure}
 \begin{center}
   \includegraphics[width=11cm]{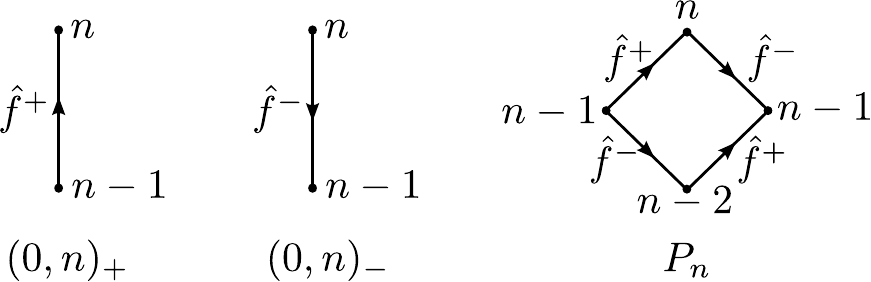}
 \end{center}
 \caption{\small Examples of reducible indecomposable representations of $\mathfrak{pl}(1|1)$. The dots are the basis vectors, and the arrows show the action of the fermionic generators $\hat{f}^\pm$. The numbers $n, n-1, \dots$ are the eigenvalues of the bosonic generator of $\mathfrak{pl}(1|1)$, that is, the U$(1)_{\rm fl}$-charges. The representations $(0,n)_-$ and $(0,n)_+$ are known as the (anti-)Kac modules.}
  \label{reps}
\end{figure}
All the other representations of $\mathfrak{pl}(1|1)$ are reducible, but, in general, can be indecomposable\footnote{That is, they have invariant subspaces, but need not split into direct sums.}. Some examples are shown on fig.~\ref{reps}. (There are more such representations~--~they are listed {\it e.g.} in \cite{sl21},~--~ but we will not need them.) In this paper, we are mostly interested in closed loop operators. Naively, due to the presence of the supertrace, a closed Wilson loop labeled by a reducible indecomposable representation splits into a sum of Wilson loops for the invariant subspaces and quotients by them. If this were true, the indecomposable representations would not need to be considered separately. We will later find that, due to regularization issues, at least for some indecomposable representations the Wilson loops do not actually reduce to sums of Wilson loops $L_m$. This will be discussed in section~\ref{Ham}, but till then we will not consider indecomposable representations.

In the case that the holonomy of the background field along some loop $K$ is trivial, one can construct a line operator by inserting an integral $\oint_K A^\pm$ into the path-integral. Note that these operators transform as a doublet under the SU$(2)_{\rm fl}$ flavor symmetry. These will play the role of creation/annihilation operators in the Hamiltonian picture in section \ref{Ham}, but, again, will not be important till there. 

The most useful line operator can be obtained by cutting a knot (or a link) $K$ out of $W$, and requiring the background gauge field to have a singularity near $K$ with some prescribed holonomy $\mathbf{t}$ around the meridian of the knot complement\footnote{The meridian is the cycle that can be represented by a small circle, linking around the knot. A longitude is a cycle that goes parallel to the knot. The longitude, unlike the meridian, is not canonically defined. Its choice is equivalent to choosing a framing of the knot.}. It is this type of line operators that will give rise to the Alexander knot polynomial.

One has to be careful in defining the determinants (\ref{det}) in presence of such a singularity. In this paper, our understanding of the determinants in this case will be much less complete than in the case of closed three-manifolds. 
We will not attempt to give a rigorous definition, but will simply state some results that are consistent with other approaches to line operators, which are discussed later in the paper, and with known properties of the Alexander  polynomial. Let $\mathbf{t}$ be the holonomy around the meridian of the knot $K$, and $\mathbf{t}_\parallel$ be the holonomy around the longitude. While $\mathbf{t}$ is a part of the definition of the line operator along $K$, $\mathbf{t}_\parallel$ depends on the flat connection and, in particular, on other line operators, linked with $K$. The problem with the determinants (\ref{det}) in presence of line operators is that in general they can be anomalous, that is, they can change sign under large gauge transformations of the background gauge field. Equivalently, one will in general encounter half-integral powers of $\mathbf{t}$ and $\mathbf{t}_{\parallel}$ in the expectation values. One possible resolution is to choose a square root of the holonomies, or, equivalently, to take $\mathcal{L}\simeq\mathcal{L}'^2$, and to consider the knot polynomial as a function of the holonomies of $\mathcal{L}'$. One expects this to produce a version of the Alexander polynomial known as the Conway function. (See section~4 of \cite{Turaev} for a review.) Alternative approach, which we will assume in most of the paper, is to add along the longitude of the knot a Wilson line for the background gauge field. So, we will in general consider combined line operators, labeled by two parameters $\mathbf{t}$ and $m$, with $m$ being the charge for the Wilson line for the background field $B+i\phi$. It will be clear from the discussion of the U$(1|1)$ theory in section~\ref{u11} that for gauge invariance, the charge $m$ should be taken valued in $1/2+\ZZ$. It is more convenient to work with an integer parameter $n=m+1/2$, and we will accordingly label our line operators as $L_{\mathbf{t},\,n}$. Note that, since the longitude cycle is not canonically defined, the definition of these line operators depends on the knot framing. Under a unit change of framing, the Wilson line for the background gauge field will produce a factor of $\mathbf{t}^{n-1/2}$. With suitable choices of framing, half-integral powers of $\mathbf{t}$ will not appear in the expectation values.

The operators $L_{\t,\,n}$ will sometimes be called typical, while $L_n$ and Wilson lines for the indecomposable representations will be called atypical. This terminology originates from the classification of superalgebra representations, as we briefly recall in section~\ref{u11review}.

\section{Magnetic Theory And The Meng-Taubes Theorem}\label{magnetic}
As was explained in section \ref{free}, our Chern-Simons theory can be obtained from the theory of a free $\cN=4$ hypermultiplet by twisting. An alternative description of the same topological theory can be obtained, if we recall that the free hypermultiplet describes the infrared limit of the $\cN=4$ QED with one electron. This is the basic example of mirror symmetry \cite{3dMirror} in three-dimensional abelian theories, which was understood in \cite{KapustinStrassler} as a functional Fourier transform. By metric independence of the topological observables, they can be equally well computed in the UV or in the IR description. We now consider the topologically-twisted version of the UV gauge theory, which we will call the ``magnetic'' description.

(On a compact manifold, the claim that the RG flow from the UV theory leads to a free hypermultiplet depends on the presence of the non-trivial background flat bundle, which forces the theory to sit near its conformally-invariant vacuum. When the background gauge field is turned off, {\it e.g.} as is necessarily the case for a manifold with trivial $H_1$, the situation is more subtle. This and some other details will be discussed in part \ref{s33} of the present section.)

\subsection{The $\cN=4$ QED With One Electron}
We now describe the bosonic fields of the theory. The fermionic fields, as well as the details on the action, are discussed in the Appendix~\ref{magdetails}. Bosonic fields of the vector multiplet are a gauge field $A_i$ and an SU$(2)_Y$-triplet of scalars $Y^{\da}$. (Bosonic gauge field $A_i$ here is completely unrelated to the fermionic gauge field of the electric gauge theory. In fact, the fields of the electric description emerge from the monopole operators of the UV theory.) In the twisting construction we use the SU$(2)_X$-subgroup of the R-symmetry, so the scalars of the vector multiplet will remain scalars. It is convenient to introduce a combination $\sigma=(Y_2-iY_3)/\sqrt{2}$, which has charge $2$ under the ghost number symmetry U$(1)_F$. The remaining component $Y_1$ has ghost number zero. The hypermultiplet contains an SU$(2)_X$-doublet of complex scalars, which upon twisting become a spinor $Z^\alpha$. They have charge one under the gauge group. The imaginary part $\phi$ of the background flat connection originated from the masses in the electric description. Under the mirror symmetry, it is mapped to a Fayet-Iliopoulos parameter. 

The flavor symmetry SU$(2)_{\rm fl}$ is emergent in the infrared limit. In the UV, only its Cartan part is visible~--~it is identified with the shift symmetry of the dual photon. The current for this symmetry is $\fr{-i}{2\pi}\star F$, where $F={\rm d}A$. The real part of the background gauge field couples to this symmetry, so, it should enter the action in the interaction $-\fr{i}{2\pi}\int B\wedge F$. In fact, the whole action of the topological theory has the form
\beq
I_{\rm QED}=\{\Qb,\dots\}+I_{\rm top}\,,\label{imprecise0}
\eeq
where
\beq
I_{\rm top}=-\fr{i}{2\pi}\int (B+i\phi)\wedge F\,.\label{imprecise}
\eeq
(More details are given in Appendix~\ref{magdetails}.) This can be more accurately written as
\beq
\exp(-I_{\rm top})=\mathcal{L}^{-1}\left(c_1(\mathcal{A})\right)\,,\label{precise}
\eeq
where $\mathcal{A}$ is a line bundle, in which $A$ is the connection. The fields $Z^\alpha$ take values in a spin-$\CC$ bundle, and correspondingly, the path-integral includes a sum over spin-$\CC$ structures $s'$. We view this spin-$\CC$ bundle as a spin bundle $S$ for some fixed spin structure $s$, tensored with the line bundle $\mathcal{A}$. We identify the reference spin structure $s$ with the spin structure, which was used in the definition of torsion on the electric side. A change of the spin structure by an element $x\in H^1(W,\ZZ_2)$ is equivalent to twisting the bundle $\mathcal{A}$ by the $\ZZ_2$-bundle, corresponding to $x$. The formula (\ref{precise}) then changes in the same way (\ref{schange}) as the torsion $\tau_s(\mathcal{L})$, in agreement with the mirror symmetry\footnote{Again, $s$ should be more appropriately viewed as a spin-$\CC$ structure with trivial determinant. Of course, we could equally well take an arbitrary reference spin-$\CC$ structure. That would give the trivial generalization of $\tau_s$ to arbitrary spin-$\CC$ structures, as described at the end of section \ref{defdetails}.}. The theory also has a charge conjugation symmetry, which, as on the electric side, implies that the invariants for $\mathcal{L}$ and $\mathcal{L}^{-1}$ are the same.

Note that, instead of (\ref{imprecise}), we could try to use
\beq
\exp(-I_{\rm top})\myeq \mathcal{L}^{-1}\left(\fr{1}{2}c_1({\rm det}\,s')\right)\,.
\eeq
Here ${\rm det}\,s'$ is the determinant line bundle of the spin-$\CC$ bundle, in which the fields $Z^\alpha$ live. However, the factor of $1/2$ inside the brackets means that one has to take a square root of the holonomy of $\mathcal{L}$, and therefore the sign of this quantity is not well defined. This is the same ambiguity that we encountered in section \ref{defdetails}, and it is resolved, again, by picking a reference spin structure $s$.

The functional integral of the magnetic theory can be localized on the solutions of BPS equations $\{\Qb,\psi\}=0$, where $\psi$ is any fermion of the theory. One group of these equations actually tells us that the solution should be invariant under the gauge transformation with parameter, equal to the field $\sigma$. We will only consider irreducible solutions, and therefore $\sigma$ must be zero. We also only consider the case that the background field satisfies ${\rm d}\star\phi=0$, so that the twisted theory has the full SU$(2)_Y$-symmetry. (We have seen on the electric side that ${\rm d}\star\phi=0$ is the condition for this symmetry to be present. On the magnetic side, one can also explicitly check this, as shown in the Appendix~\ref{magdetails}.) This symmetry, together with vanishing of $\sigma$, implies that $Y_1$ is also zero. With this vanishing assumed, the remaining BPS equations take the form of the three-dimensional Seiberg-Witten equations,
\begin{align}
F+\fr{1}{2}\star \left(\mu-e^2\phi\right)&=0\,,\nnr
\slashed{D}Z&=0\,,\label{SW}
\end{align}
where $\mu=i\sigma^{\beta}_{j\,\alpha}\,Z^\alpha\bar{Z}_\beta\,{\rm d}x^j$ is the moment map, with $\sigma^{\beta}_{i\,\alpha}$ being the Pauli matrices contracted with the vielbein, $e^2$ is the gauge coupling, and $\slashed{D}$ is the Dirac operator, acting on the sections of $S\otimes\mathcal{A}$. Generically, the localization equations have a discrete set of solutions, and the partition function of the theory can be written as
\beq
\tau_s(\mathcal{L})=\sum_{\mathfrak{S}} (-1)^{f}\,\mathcal{L}^{-1}(c_1(\mathcal{A}))\,,\label{SWequalT}
\eeq
where the sum goes over the set $\mathfrak{S}$ of solutions of the Seiberg-Witten equations, $\mathcal{A}$ is a line bundle, corresponding to the given solution,  and $(-1)^{f}$ is the sign of the fermionic determinant.

The relation between the Reidemeister-Turaev torsion and the Seiberg-Witten invariant in three-dimensions is the content of the Meng-Taubes theorem \cite{MT} and its refinement due to Turaev \cite{TuraevSW}. We have presented a physicist's derivation of this theorem. Some subtleties that arise for three-manifolds with $b_1\le 1$ are discussed later in this section.

\subsection{Adding Line Operators}
Let us describe the magnetic duals of line operators, which were introduced in section \ref{eline}. The first type of line operators were the integrals of the fermionic gauge field $\int_K A^\pm$. On the magnetic side, their duals will be the integrals of monopole operators, which we will not discuss. The second type were the Wilson lines for the background gauge field. Obviously, their definition will be the same on the magnetic side. 

Non-trivial and interesting line operators were defined by singularities of the background flat connection. We denoted them by $L_{\mathbf{t}\,,n}$ in section \ref{eline}. Since the one-form $\phi$ enters the BPS equations (\ref{SW}) on the magnetic side, the singularity of $\phi$ implies that those equations will have solutions with a singularity along the knot $K$. The line operator is then defined by requiring the fields to diverge near $K$ as in a particular singular model solution. We use notation $W$ for the closed three-manifold, and $W'$ for the manifold, obtained from $W$ by cutting out a small toric neighborhood of the singular line operator. Let $r$ and $\theta$ be the polar coordinates in the plane, perpendicular to $K$. Near the knot, the singularity of the imaginary part of the background gauge field has the form
\beq
\phi=-\gamma\,{\rm d}\theta+\beta\,\fr{{\rm d}r}{r}\,.\label{phi}
\eeq
(We follow the notations of \cite{Surface}.) Note that whenever the parameter $\beta$ is non-zero, the closed two-form $\star\phi$ has a non-vanishing integral around the boundary of the toric neighborhood of the link. This might be forbidden for topological reasons~--~{\it e.g.}, if $K$ is a one-component link in $S^3$. In such cases, $\beta$ cannot be turned on. Even when the parameter $\beta$ can be non-zero, we expect the invariants to be independent of it.

To find the model solution, consider the case that $W$ is the flat space, and $K$ is a straight line. Let $Z^1$ and $Z^2$ be the two components of $Z^\alpha$, and $z=r\exp(i\theta)$ be the complex coordinate in the plane, perpendicular to $K$. We are looking for a time-independent, scale-invariant solution of the Seiberg-Witten equations. The gauge field in such a solution can be set to zero, so that the remaining equations give
\beq
Z^1(Z^2)^\dagger=\fr{e^2(\beta+i\gamma)}{2z}\,,\quad Z^1(Z^1)^\dagger-Z^2(Z^2)^\dagger=0\,, \quad \partial_{\bar{z}}Z^1=\partial_zZ^2=0\,,
\eeq
and the scale-invariant solution is simply $Z^1=a/\sqrt{z}$, $Z^2=b/\sqrt{\bar{z}}$,
where $ab^\dagger= e^2(\beta+i\gamma)/2$ and $|a|^2=|b|^2$. Note that the field $Z^\alpha$ here is antiperiodic around $K$. Since we view $Z^\alpha$ as a spinor field on the closed three-manifold $W$, it should rather be periodic, so, we make a gauge transformation to bring the model solution to the form
\beq
Z^1=\fr{a}{\sqrt{r}}\,,\quad Z^2=\fr{b}{\sqrt{r}}\exp(i\theta)\,,\quad A=-\fr{1}{2}{\rm d}\theta\,.\label{model}
\eeq

To complete the definition of the line operator, we also need to explain, how the topological action (\ref{precise}) is defined in presence of the singularity. The flat bundle $\mathcal{L}$ is naturally an element of ${\rm Hom}(H_1(W'),\CC^*)$. By Poincar\'e duality, it can be paired with an element of the relative cohomology $H^2(W',\partial W')$, and this pairing will define the action. If we forget for a moment about possible torsion, the relative cohomology class that we need is naturally the cohomology class $[F/2\pi]$ of the gauge field strength for a given solution. However, here we encounter the mirror of the problem that we had on the electric side: this class in general is not integral. The reason, roughly speaking, is the antiperiodicity of the field $Z^\alpha$ around $K$, or equivalently, the half-integral term $-\fr{1}{2}{\rm d}\theta$ in the gauge field (\ref{model}) near the line operator. (Depending on the topology, there can also appear a similar term with $\theta$ replaced by the angle along $K$.) This will in general cause half-integral powers of the holonomies $\mathbf{t}$ and $\mathbf{t}_\parallel$ to appear in the torsion invariant. To remove them, just as in section \ref{eline}, one introduces along $K$ a Wilson line for the background gauge field with a half integral charge $n-1/2$. With a suitable choice of framing, this is enough to remove the half-integral powers of holonomies. 

Here we viewed the field $Z^\alpha$ as a section of the spin bundle on $W$, tensored with a line bundle $\mathcal{A}$ with connection $A$. A more systematic way to define these line operators is to allow spin (or spin-$\CC$) structures on $W'$ that do not necessarily extend to $W$. The antiperiodicity of $Z$ in the model solution (\ref{model}) can then be absorbed into the definition of this spin structure. The field $Z^\alpha$ then provides an honest cross-section of the line bundle $\mathcal{A}$ in the neighborhood of the link, and this allows to canonically define an integer-valued relative Chern class $c_1(\mathcal{A})\in H^2(W',\partial W')$. The charge $n$ of the background field Wilson line and the choice of the framing are then absorbed into the choice of the spin-$\CC$ structure. This is the approach taken in the mathematical literature\footnote{There is also another difference of our treatment of line operators from mathematical literature. There, the analogs of  line operators are typically introduced by gluing in an infinite cylindrical end to the manifold $W'$, rather than by considering solutions on $W$ with singularities.}, see {\it e.g.} \cite{Nicolaescu}. This point of view is consistent with the picture that inserting line operators of type $L_m$, or changing the parameter $n$ for operators $L_{\mathbf{t},\,n}$, is equivalent to changing the spin-$\CC$ structure.

We only considered the case that the holonomy of the background flat connection around the meridian of the knot is not unimodular. In the opposite case, we have $\gamma=0$ in eq. (\ref{phi}), and, assuming that the parameter $\beta$ is also zero, the singular model solution seems to disappear. 
This makes it unclear, how to define the magnetic duals of line operators with unimodular holonomy, except by the analytic continuation from $\gamma\ne 0$. This problem looks analogous to the one described in the end of section 4.4.5 of \cite{MW}.

\subsection{More Details On The Invariant}\label{s33}
In our derivation of the relation between the Seiberg-Witten invariant and the Reidemeister-Turaev torsion we ignored some subtleties \cite{MT}, \cite{Nic2}, which occur for three-manifolds with $b_1\le 1$. Here we would like to close this gap. First we look at the UV theory, and then we describe the RG flow to the IR theory in more detail. We will see that the claim that the IR theory is the $\psu(1|1)$ Chern-Simons model sometimes has to be corrected.

\subsubsection{Seiberg-Witten Equations For $b_1\le 1$}
Let us look closer at the Seiberg-Witten counting problem. Our goal here is not to derive something new, but merely to understand, how gauge theory takes care of some subtleties in the formulation of the Meng-Taubes theorem.

Note that in the analogous problem in the context of Donaldson theory in four dimensions, the gauge theory gives the first of the Seiberg-Witten equations roughly in the form $F^++\bar{Z}Z=0$. To avoid dealing with reducible solutions with $F^+=0$, one introduces by hand a deformation two-form in the equation \cite{WittenSW}. In our case, the situation is different: the deformation two-form $e^2\star\phi/2$ is already there. In nice situations,  the counting of solutions does not depend on the choice of this deformation, so any two-form could be taken. But sometimes it is not true, and then it will be important, what deformation two-form is chosen for us by the gauge theory.

The properties of the counting problem depend on the first Betti number $b_1(W)$, whose role here is analogous to $b_2^+$ in four dimensions. For $b_1>0$, a reducible solution has $Z=0$ and $F=e^2\star\phi/2$. For such a solution to occur, the cohomology class of $e^2\star\phi/2$ has to be integral. When in the parameter space we pass through such a point, so that reducible solutions are possible, the counting of solutions can in principle jump. This makes the Seiberg-Witten invariant dependent on the deformation two-form, or the metric and $e^2$, if we prefer to keep the deformation two-form equal to $e^2\star\phi/2$ with fixed $\phi$. Actually, for $b_1>1$ no jumping is possible, since in the space of deformation two-forms we can always bypass the point, where reducible solutions can occur. But for $b_1=1$, non-trivial wall-crossing phenomena do happen. As we change the two-form $e^2\star\phi/2$, and its cohomology class passes through an integer point, the number of solutions with first Chern class $[F/2\pi]$ equal to this integer does change in a known way \cite{Lim}. (For the particular case of $S^1\times S^2$, the Seiberg-Witten counting problem is worked out in detail in the Example 4.1 in \cite{Nicolaescu}.)

There is another related issue. As we explained in section \ref{defprop}, the torsion, to which the Seiberg-Witten invariant is supposed to be equal to, for $b_1=1$ has a second order pole. Just for concreteness, consider the manifold $S^1\times S^2$, for which the torsion is\footnote{The function $\tau(\mathbf{t})$ should have a second order pole at $\mathbf{t}=1$. Also, it cannot have any zeros for $\mathbf{t}\in\CC^*\setminus\{1\}$, since the twisted cohomology $H^1(S^1\times S^2,\mathcal{L})$ for such $\mathbf{t}$ is empty. Imposing also invariance under the charge conjugation $\mathcal{C}$, we recover the stated result, up to a constant numerical factor.} $\tau(\mathbf{t})={\mathbf{t}}/{(\mathbf{t}-1)^2}$, where $\mathbf{t}$ is the holonomy around the non-trivial cycle. If we expand this, say, near $\mathbf{t}=0$, we get a semi-infinite Laurent  series ${\mathbf t}+2{\mathbf t}^2+\dots$. However, it is known that for any given deformation two-form the Seiberg-Witten equations have only a finite number of solutions. 

The resolution of these puzzles is that we need to take the infrared limit of the theory, {\it e.g.} by taking $e^2$ to infinity. This means that we have to take the deformation two-form to be $+\infty\cdot\star\phi$. That is, it should be proportional to $\star\phi$ with a positive coefficient, and, to count solutions with a given Chern class $[F/2\pi]$, we should use a deformation two-form with Chern class much larger than $[F/2\pi]$ in absolute value. This is equivalent to the prescription of Meng and Taubes. Depending on the sign of $\phi$, the two expansions that we get in this way for $S^1\times S^2$ would be ${\mathbf t}+2{\mathbf t}^2+\dots$ and ${\mathbf t}^{-1}+2{\mathbf t}^{-2}+\dots$. One can check that the sign of $\phi$ is such that $|\mathbf{t}|<1$ in the first case and $|\mathbf{t}|>1$ in the second, so that in either case the expansion is absolutely convergent.

Just like for closed three-manifolds, for manifolds with links in them, the Seiberg-Witten counting problem for $b_1(W')=1$ is special. (This case arises {\it e.g.} when one cuts out a one-component knot from a simply-connected manifold.) As we reviewed in the end of section \ref{comparison}, the Reidemeister torsion for such a manifold has a first order pole. Therefore, it has two different Laurent expansions near $\mathbf{t}=0$ and $\mathbf{t}=\infty$. The Seiberg-Witten equations in this case have an infinite number of solutions, with Chern class unbounded from above or from below, depending on the sign of the deformation two-form $e^2\star\phi$. The sign of $e^2\star\phi$ is such that these expansions are absolutely convergent. Unlike the case of a compact three-manifold, here we do not need to explicitly take $e^2$ to infinity, since the deformation two-form $e^2\star\phi$ already diverges near the knot.

When $W$ is a rational homology sphere, that is $b_1=0$, there is no way to avoid reducible solutions in working with the Seiberg-Witten equations. Because of this, a simple signed count of solutions is no longer a topological invariant. Still, one can define a topological invariant by adding an appropriate correction term \cite{Lim}. We will not attempt to derive it from the quantum field theory, but will in what follows use the fact that the definition of the invariant for $b_1=0$ does exist.

\subsubsection{Massive RW Model And The Casson-Walker Invariant}\label{triv}
Let us now turn to the IR theory, which is a valid description, when the size of the three-manifold $W$ is scaled to be large. The topological theory reduces in this case to the Rozansky-Witten (RW) sigma-model \cite{RW} with the target space being the Coulomb branch manifold, which for the $\cN=4$ QED is \cite{SW3d} the smooth Taub-NUT space $X_{\rm TN}$. The U$(1)$ graviphoton translation symmetry is generated on $X_{\rm TN}$ by a Killing vector field $V$. The coupling of the UV theory to the flat gauge field $B+i\phi$ translates into a coupling of the RW model to the same flat gauge field via the isometry $V$. In the untwisted language, the imaginary part $\phi$ of the gauge field would be a hyperk\"aler triplet of mass terms. For this reason, we call our IR theory the massive Rozansky-Witten model. An explicit Lagrangian and more detailed treatment of this theory will appear elsewhere. The coupling of the Rozansky-Witten model to a {\it dynamical} Chern-Simons gauge field has been previously considered in \cite{KapustinSaulina}. We will now see, how and when the massive RW model reduces to the Gaussian $\psu(1|1)$ theory.

First, let us turn off the background flat gauge field. What we get then is the ordinary RW model for $X_{\rm TN}$. The path-integral of that theory has the following structure \cite{RW}. The kinetic terms have both bosonic and fermionic zero modes. The bosonic ones correspond to constant maps to the target space. The integral over the bosonic zero modes thus is an integral over $X_{\rm TN}$. The one-loop path-integral produces a simple measure factor, while most of the higher-loop diagrams vanish. The reason is that all the interactions (which involve the curvature of $X_{\rm TN}$) are irrelevant in the RG sense, and can be dropped, when the worldsheet metric is scaled to infinity. However, some diagrams do survive due to the presence of the zero modes. Overall, the path-integral for each $b_1$ is given by a simple Feynman diagram, which captures the topological information about $W$, times the integral of the Euler density of $X_{\rm TN}$. It is important that the path-integral depends on the target space only by this curvature integral. The Euler numbers happen to be the same for $X_{\rm TN}$ and for the Atiyah-Hitchin manifold $X_{\rm AH}$. This was used in \cite{Blau} to derive a special case of the Meng-Taubes theorem by the following argument. The RW model for $X_{\rm AH}$ can be obtained from the IR limit of the topologically-twisted $\cN=4$ SU$(2)$ Yang-Mills theory \cite{SW3d}, which computes the Casson-Walker invariant \cite{TaubesCasson}, \cite{AtiyahJeffrey}, \cite{BlauCasson}. Since the Rozansky-Witten invariants computed using $X_{\rm TN}$ and $X_{\rm AH}$ are the same, the Casson-Walker invariant is equal to the Seiberg-Witten invariant, when the background bundle is trivial. 

Now let us turn on the background bundle back again. In its presence, the kinetic terms of the RW model have no zero modes. The classical solution, around which one expands, is the map to the fixed point of the vector field $V$, that is, to the conformally-invariant vacuum. In the absence of the zero modes, all the irrelevant curvature couplings can be thrown away. In this way, the RW path-integral reduces to the Gaussian integral of the $\psu(1|1)$ model. It is natural to expect the path-integral to be continuous in $\mathcal{L}$. To the extent that this is true, the torsion $\tau(\mathcal{L})$ evaluated for trivial $\mathcal{L}$ should thus coincide with the Casson-Walker invariant. Note that on the level of Feynman diagrams this is not completely trivial, since for $B+i\phi=0$ the interaction vertices come from the curvature terms, while for $B+i \phi\ne 0$ they come from expanding the Gaussian path-integral in powers of the background gauge field. Still, the actual Feynman integrals should coincide.
 We will not explicitly analyze the diagrams here (most of them were analyzed in \cite{RW}), but will just use the known relation between $\tau$ and the Casson-Walker invariant to check the continuity of the massive RW path-integral in $\mathcal{L}$.

Let $\tau(1)$ denote the torsion evaluated for the trivial background flat bundle\footnote{Note that for $\tau(1)$ the dependence on the spin-$\CC$ structure drops out.}, and CW be the Casson-Walker invariant. For $b_1\ge 2$, it is indeed true that $\tau(1)={\rm CW}$. For $b_1=1$, the torsion has the form
\beq
\tau(\mathbf{t})=\fr{\mathbf{t}\Delta(\t)}{(\t-1)^2}\,,\label{b11}
\eeq
where $\Delta(\t)$ is the Alexander polynomial. Setting $\t=\exp(m)$ and expanding this in $m$, we get
\beq
\tau(\mathbf{t})=\fr{\Delta(1)}{m^2}+\fr{1}{2}\Delta''(1)-\fr{1}{12}\Delta(1)+O(m^2)\,.
\eeq
Dropping the $1/m^2$ term, we define the regularized torsion $\tau_{\rm reg}(1)=\fr{1}{2}\Delta''(1)-\fr{1}{12}\Delta(1)$. This combination, again, is equal to the Casson-Walker invariant for $b_1=1$. However, the presence of the extra divergent piece $\Delta(1)/m^2$ means that the path-integral of the massive RW model in this case is not continuous in its dependence on the background gauge field: for $\mathcal{L}$ approaching the trivial flat bundle, the torsion tends to infinity, while for $\mathcal{L}$ taken to be exactly the trivial flat bundle, the invariant is finite. One can trace the origin of this discontinuity to the wall-crossing in the UV theory. Indeed, for non-zero $\phi$, the Seiberg-Witten invariant is evaluated using the deformation two-form $e^2\star\phi/2$, which in the infrared limit $e^2\rightarrow\infty$ lands us in the infinite wall-crossing chamber. The $1/m^2$ singularity of the torsion for $m\rightarrow 0$ arises from the infinite number of solutions of the Seiberg-Witten equations in this chamber. On the other hand, for trivial $\mathcal{L}$ we have $\phi=0$, and the deformation two-form vanishes for all $e^2$. To evaluate the invariant, one should properly deal with reducible solutions. Instead, we will simply assume that the deformation two-form is non-zero, but infinitesimally small. It is known \cite{Nicolaescu} that in such chamber the Seiberg-Witten invariant is equal to $\fr{1}{2}\Delta''(1)$, which, again, is the Casson-Walker invariant, up to a correction $-\fr{1}{12}\Delta(1)=-\fr{1}{12}|{\rm tor}\,H_1(W)|$, which, presumably, would be recovered with an appropriate treatment of the reducible solutions. Thus, one can say that the discontinuity at trivial $\mathcal{L}$ in the massive RW model for $b_1=1$ is a ``squeezed version'' of the wall-crossing in the UV theory\footnote{It is a ``squeezed version'', because the wall-crossing condition is not conformally-invariant, and thus we cannot see all the walls in the IR theory, but only see a discontinuity at $\phi=0$. This can be contrasted with the situation in the Donaldson theory in four dimensions, where the wall-crossing condition is conformally-invariant, and the walls can be seen both in the UV and in the IR descriptions \cite{MooreWitten}.}

Finally, for $b_1=0$, assuming that the torsion subgroup ${\rm tor}~H_1(W)$ is non-empty, the Reidemeister-Turaev torsion is a function on the discrete set of flat bundles. For non-trivial $\mathcal{L}$, the Seiberg-Witten counting problem computes the torsion, while for trivial $\mathcal{L}$ it computes the Casson-Walker invariant, which now is not related to the torsion, since there is no way to continuously interpolate to the trivial $\mathcal{L}$, starting from a non-trivial $\mathcal{L}$. In fact, for $b_1=0$ the Casson invariant is computed by a two-loop Feynman integral \cite{RW}, and it is clearly not possible to obtain it from the one-loop torsion.

Let us summarize. The UV topological theory, and thus the Seiberg-Witten counting problem, is equivalent to the massive RW theory. For non-trivial $\mathcal{L}$, this theory reduces to the $\psu(1|1)$ Chern-Simons theory and computes the Reidemeister-Turaev torsion. For trivial $\mathcal{L}$, it computes the Casson-Walker invariant, which for $b_1>0$ can be obtained from a limit of the $\psu(1|1)$ invariant, while for $b_1=0$ is not related to it. Our results agree with the mathematical literature \cite{Nicolaescu}, \cite{Nic2}.

\section{U$(1|1)$ Chern-Simons Theory}\label{u11}
In a series of papers \cite{RS1,RS2,RS3}, it has been shown that the Alexander polynomial and the Milnor torsion can be computed from the U$(1|1)$ Chern-Simons theory. We would like to revisit this subject and to show, how it fits together with our discussion in previous sections. We point out that for the compact form of the bosonic gauge group, the U$(1|1)$ Chern-Simons theory is simply an orbifold of the $\psu(1|1)$ theory. (A direct analog of this statement is well-known in the ABJM context.) In particular, it contains no new information compared to the $\psu(1|1)$ Chern-Simons with a coupling to a general background flat bundle~$\mathcal{L}$, and computes, indeed, essentially the same invariant.

\subsection{Lie Superalgebra $\mathfrak{u}(1|1)$ }\label{u11review}
We start with a brief review of the superalgebra $\mathfrak{u}(1|1)$. A more complete discussion can be found {\it e.g.} in \cite{sl21}. Let $\hat{f}_+$ and $\hat{f}_-$ be the fermionic generators, and $\hat{t}_\ell$ and $\hat{t}_r$ the generators of the left and right bosonic $\u(1)$ factors. It will also be convenient to use a different basis in the bosonic subalgebra, which is $\hat{E}=\hat{t}_r+\hat{t}_\ell$ and $\hat{N}=(\hat{t}_r-\hat{t}_\ell)/2$. The element $\hat{N}$ acts on the fermionic subalgebra by the U$(1)_{\rm fl}$ transformations, and the element $\hat{E}$ is central. Explicitly, the non-trivial commutation relations are
\beq
[\hat{N},\hat{f}_\pm]=\pm \hat{f}_\pm\,,\quad \{\hat{f}_+,\hat{f}_-\}=\hat{E}\,.
\eeq
The group of even automorphisms of $\mathfrak{u}(1|1)$ is generated by the charge conjugation $\hat{E}\rightarrow-\hat{E}$, $\hat{N}\rightarrow-\hat{N}$, $\hat{f}_\pm\rightarrow\pm \hat{f}_\pm$, rescalings $\hat{f}_\pm\rightarrow a_\pm \hat{f}_\pm$, $\hat{E}\rightarrow a_+a_- \hat{E}$ with $a_\pm\in\RR\setminus 0$, and shifts $\hat{N}\rightarrow\hat{N}+b\hat{E}$, $b\in\RR$.

As for any Lie superalgebra, the representations of $\u(1|1)$ can be usefully divided into two classes~--~the typical and the atypical ones. (For a very brief review of superalgebra representations, with applications to Chern-Simons theory, see section~3 of \cite{MW}.) The typicals are precisely the ones, in which the central generator $\hat{E}$ acts non-trivially. They are two-dimensional, and the generators, in some basis, act by matrices
\beq
\hat{E}=w\left(\bea{cc}1&0\\0&1\eea\right)\,,\quad \hat{N}=\left(\bea{cc}n&0\\0&n-1\eea\right)\,,\quad
\hat{f}_+=\left(\bea{cc}0&w\\0&0\eea\right)\,,\quad
\hat{f}_-=\left(\bea{cc}0&0\\1&0\eea\right)\,,\label{matrices}
\eeq
with $w\ne 0$. These will be called representations of type $(w,n)$. To be precise, one has to make a choice, whether to assign a bosonic or a fermionic parity to the highest weight vector. This effectively doubles the number of representations. In our applications, the representations will be labeling closed Wilson loops, which come with a supertrace. Therefore, different parity assignments will be just a matter of overall sign, and we will mostly ignore this.

In the atypical representations the generator $\hat{E}$ acts trivially, and therefore they can be equivalently thought of as representations of $\mathfrak{pl}(1|1)$. These have already been described in section \ref{eline}. Note that the indecomposable representation $(0,n)_-$ of fig.~\ref{reps} can be obtained as a degeneration of the typical representation $(w,n)$ for $w\rightarrow 0$. With a suitable rescaling of the generators $\hat{f}^\pm$ before taking the limit, one can similarly obtain the representation $(0,n)_+$ of fig.~\ref{reps}. The representations $(0,n)_-$ and $(0,n)_+$ are known as the atypical Kac module and anti-Kac module.

Let us also write out some tensor products. Tensoring any representation with the one-dimensional atypical $(n)$ simply shifts the $\hat{N}$-charges. The other tensor products are
\begin{align}
(w_1,n_1)\otimes (w_2,n_2)&=(w_1+w_2,n_1+n_2)\oplus (w_1+w_2,n_1+n_2-1)'\,,\quad w_1+w_2\ne 0\,;\label{repope1}\\
(w,n_1)\otimes (-w,n_2)&=P_{n_1+n_2}\label{repope2}\,,
\end{align}
where the indecomposables $P_n$ were defined on fig.~\ref{reps}. The prime on the second representation in the r.h.s. of (\ref{repope1}) means that the highest weight vector in it has reversed parity. The set of representations $(n)$, $(w,n)$, $P_n$ is closed under tensor products.

The superalgebra $\u(1|1)$ possesses a two-dimensional family of non-degenerate invariant bilinear two-forms, which can be obtained by taking a supertrace over a $(w,n)$ representation with $w\ne 0$. Note that all the representations $(w,n)$ for different values of $w\ne 0$ and $n$, and therefore also the corresponding invariant forms, are related by the superalgebra automorphisms.

\subsection{Global Forms}\label{global}
There exist different versions of Chern-Simons theory based on the superalgebra $\u(1|1)$, and here we would like to classify them. To define such a theory, one needs to pick a global form of the gauge group, and also to choose an invariant bilinear form, with which to define the action. These data should be consistent, in the sense that the action should be invariant under the large gauge transformations. Theories related by the superalgebra automorphisms are equivalent. We can use this symmetry to bring either the invariant bilinear form, or the lattice, which defines the global form of the group, to some simple canonical form. To classify the theories, it is convenient to take the first approach.

Let $\g_\bos\simeq \RR^2$ be the bosonic subalgebra of $\u(1|1)$. The $\mathfrak{u}(1|1)$ gauge field, in components, is $A=A^N\hat{N}+A^E\hat{E}+A^+\hat{f}_++A^-\hat{f}_-$. For the bosonic part of the gauge field, we will also use expansion in a different basis, $A^N\hat{N}+A^E\hat{E}\equiv A^\ell\,\hat{t}_\ell+A^r\,\hat{t}_r$. 
The action of the theory can be written as
\beq
I_{\u(1|1)}=I_{\rm bos}+I_{\psu(1|1)}(\mathcal{L}_{A^N}\otimes \mathcal{L})+I_{\rm g.f.}\,,\label{u11act}
\eeq
where $I_{\rm bos}$ is the Chern-Simons term for the bosonic gauge field, $I_{\psu(1|1)}$ is the action (\ref{CSact1}), coupled to the line bundle $\mathcal{L}_{A^N}$ with connection $A^N$, and to some background flat bundle $\mathcal{L}$. Finally, $I_{\rm g.f.}$ is the gauge-fixing action (\ref{gfa}) for the fermionic part of the gauge symmetry.

By using the superalgebra automorphisms, we bring the bosonic Chern-Simons term to the form
\beq
I_{\rm bos}=\fr{i}{4\pi}\int_W A^r{\rm d} A^r - A^{\ell}{\rm d}A^\ell\,.\label{bosCS}
\eeq
(As usual, this formula is literally true only for topologically-trivial bundles. More generally, it is implicitly understood that the action is defined by integrating Chern classes of a continuation of the bundle to some four-manifold.)

Different versions of the theory will correspond to different choices of the global form of the bosonic subgroup G$_\bos$ of U$(1|1)$. A global form is fixed, once we choose a cocharacter lattice $\Gamma_{\rm coch}\subset \g_\bos$, that is, the lattice by which to factorize the vector space $\g_\bos$ to get the torus G$_\bos$. The first constraint on possible choices of the lattice $\Gamma_{\rm coch}$ comes from the fact that the fermionic generators of $\u(1|1)$ should transform in a well-defined representation of G$_\bos$. In the basis dual to $(\hat{t}^\ell,\hat{t}^r)$, the corresponding weight has coordinates $(-1,1)$, and we require that this vector be contained in the dual lattice $\Gamma_{\rm ch}\simeq\Gamma_{\rm coch}^*$.

We also need to make sure that the action (\ref{bosCS}) is invariant under the large gauge transformations. This will be true, if the number
\beq
\fr{1}{2}\int_V c_1^r\wedge c_1^r - c_1^\ell\wedge c_1^\ell\label{square}
\eeq
is integer on any closed spin four-manifold $V$. (We restrict to spin four-manifolds, because we already have a choice of a spin structure on $W$.) Here $c_1^{r,\ell}=[{\rm d}A^{r,\ell}/2\pi]$ are the $H^2(V,\RR)$-valued Chern classes for some extension of the G$_\bos$-bundle onto $V$. 

The classes $c^r$ and $c^\ell$ for different G$_\bos$-bundles form a lattice in $H^2(V,\RR)\oplus H^2(V,\RR)$, which is naturally isomorphic to $\Gamma_{\rm coch}\otimes H^2(V)$ (modulo torsion). Any element of this lattice can be expanded as $v_1\omega_1+v_2\omega_2$, where $\omega_1$ and $\omega_2$ are arbitrary classes in $H^2(V)$, and $v_1$ and $v_2$ are the generators of the lattice $\Gamma_{\rm coch}$. The quadratic form (\ref{square}) can be explicitly written as
\beq
a_{11}\int_V\fr{1}{2}\omega_1\wedge\omega_1+a_{12}\int_V\omega_1\wedge\omega_2+a_{22}\int_V \fr{1}{2}\omega_2\wedge\omega_2\,,\label{form2}
\eeq
with $a_{11}=(v_1^r)^2-(v_1^\ell)^2$, $a_{12}=v_1^rv_2^r-v_1^\ell v_2^\ell$ and $a_{22}=(v_2^r)^2-(v_2^\ell)^2$. For (\ref{form2}) to be an integer for arbitrary $\omega_1$ and $\omega_2$, the three coefficients $a_{ij}$ should be integers. (We used again the fact that the intersection form on a spin four-manifold is even.) This condition is precisely equivalent to the requirement for $\Gamma_{\rm coch}$ to be an integral lattice in $\RR^{1,1}$.
We conclude that U$(1|1)$ Chern-Simons theories are labeled by integral lattices in $\RR^{1,1}$, whose dual contains the vector $(-1,1)$.

\subsection{The Orbifold}
To show that the theory is an orbifold of $\psu(1|1)$ Chern-Simons, it is convenient to rewrite it in a different way. Let us use the basis $(\hat{E},\hat{N})$ in $\g_{\bos}$, in which the $\RR^{1,1}$ scalar product is $(u,v)=u^Nv^E+u^Ev^N$. Let $k$ and $\nu$ be some positive integers, and $\xi$ be an integer or a half-integer, defined modulo $k$. By taking $v_1=(k/\nu,0)$ and $v_2=(\xi/\nu,\nu)$ as the generators, for any such set we define a lattice, which actually has the right properties to serve as $\Gamma_{\rm coch}$. The opposite is also true: any lattice $\Gamma_{\rm coch}$ has a basis of this form, and it is unique modulo shifting $\xi$ by a multiple of $k$. (The parameter $k$ is  actually the area of the fundamental domain of $\Gamma_{\rm coch}$.) This can be seen as follows. Let $v_1=(a,b)$ and $v_2=(c,d)$ be some generators of $\Gamma_{\rm coch}$. The condition that the weight of the fermionic part of the superalgebra is a well-defined weight of G$_\bos$ means that $b$ and $d$ are integers. Let $\nu$ be their greatest common divisor. Then, by Euclidean algorithm, there exists an SL$(2,\ZZ)$-matrix of the form
\beq
\left(\begin{array}{cc} d/\nu&-b/\nu\\p&q\end{array}\right)\,,
\eeq
with some $p$ and $q$. Transforming the basis of the lattice with this matrix, we find a basis of the form $v_1=(a',0)$, $v_2=(b',\nu)$. (We choose $a'$ to be positive.) The integrality of the lattice means that $a'\nu\in\ZZ$ and $2b'\nu\in\ZZ$, so we can indeed parameterize the basis vectors in terms of $k$, $\xi$ and $\nu$. Residual SL$(2,\ZZ)$-transformations of the basis shift $\xi$ by multiples of $k$.

Now we can make a superalgebra automorphism $\hat{E}'=\fr{k}{\nu}\hat{E}$, $\hat{N}'=\hat{N}+\fr{\xi}{\nu^2}\hat{E}$ to transform this basis into $v_1'=(1,0)$, $v_2'=(0,\nu)$, at the expense of changing the action from its canonical form (\ref{bosCS}) to
\beq
I_{\rm bos}=\fr{i}{2\pi}\int_W \fr{k}{\nu}A^N{\rm d}A^E+\fr{\xi}{\nu^2}A^N{\rm d}A^N\,.\label{newaction}
\eeq

The path-integral involves a sum over topological classes of bundles, which are parameterized by the first Chern classes of the $A^E$ and $A^N$ bundles, which take values in $H^2(W)$ and $\nu H^2(W)$, respectively. For every topological type, let us write the gauge field $A^E$ as a sum of some fixed connection $A^{E}_{(0)}$ and a one-form $a^E$. Integrating over $a^E$ produces a delta-function, which localizes the integral to those connections $A^N$, which are flat. The $\psu(1|1)$ part of the path-integral can then be taken explicitly, and we get for the U$(1|1)$ partition function,
\beq
\tau^{{\rm U}(1|1)}_s(\mathcal{L})=\int{\rm D}A^N\sum_{c_1^E}\delta(k{\rm d}A^N/2\pi)\,\mathcal{L}_{A^N}(kc_1^E)\,\exp(\xi{\rm  CS}(\mathcal{L}_{A^N}))\,\tau_s(\mathcal{L}^\nu_{A^N}\otimes\mathcal{L})\,.\label{pathint}
\eeq
Here for convenience we changed the integration variable $A^N\rightarrow \nu A^N$. The origin of different terms here is as follows. The sum over the (integral) Chern classes $c_1^E$ is what remained from the functional integral over $A^E$. The delta-function came from the integration over $a^E$. The holonomy of the flat bundle $\mathcal{L}_{A^N}$ around the Poincar\'e dual of $kc_1^E$ is just a rewriting of the exponential of the Chern-Simons term $kA^N{\rm d}A^E/2\pi$. The Chern-Simons term for $\mathcal{L}_{A^N}$ with coefficient $\xi$ came from the $A^N{\rm d}A^N/2\pi$ term in the action (\ref{newaction}). Finally, $\tau_s$ is the $\psu(1|1)$ torsion evaluated for a flat bundle, which is the $\nu$-th power of $\mathcal{L}_{A^N}$, tensored with some background flat bundle $\mathcal{L}$.

Essentially the same path-integral as (\ref{pathint}) was considered in section 2.2 of \cite{WittenSL2Z}. It was noted that the sum over $c_1^E$ is proportional to the delta-function, supported on flat bundles with $\ZZ_k$-valued holonomy, since the pairing between $H^2(W)$ and the group of flat bundles is perfect. (That paper actually considered $k=1$.) Using this, we finally get
\beq
\tau^{{\rm U}(1|1)}_s(\mathcal{L})=\fr{1}{k}\sum_{\mathcal{L}_k}\exp(\xi{\rm  CS}(\mathcal{L}_k))\,\tau_s(\mathcal{L}^\nu_k\otimes\mathcal{L})\,,\label{orbifolded}
\eeq
where the sum goes over all $\ZZ_k$-bundles $\mathcal{L}_k$. The factor of $k$ appeared from the delta-function in (\ref{pathint}). To be precise, the explanations that we gave are sufficient to fix this formula only up to a prefactor. For manifolds with $b_1=0$, the normalization (\ref{orbifolded}) can be recovered from the considerations in section 2.2 of \cite{WittenSL2Z}. We expect that it is correct in general. The factor of $1/k$ has a natural interpretation in terms of the orbifold~--~it is the volume of the isotropy subgroup, which is $\ZZ_k$.

An important special case is the U$(1|1)$ Chern-Simons defined with the most natural global form of the group, where we simply set $\exp(2\pi i\hat{t}_\ell)=\exp(2\pi i\hat{t}_r)=1$. The action is (\ref{bosCS}) with an integer factor $k$ in front of it. By making an automorphism transformation, this theory can be mapped to the form (\ref{newaction}) with $\xi=k/2$ and $\nu=1$. Interestingly, it becomes independent of the spin structure, if $k$ is odd. This is because the sign of the fermionic determinant is changing in the same way as the half-integral Chern-Simons term for $A^N$. For the general version of the theory, the dependence on the spin structure drops out when $\nu/2+\xi\in\ZZ$. In what follows, we restrict to the version of the theory with $\xi=0$ and $\nu=1$.

Let us make some terminological comments. We call the theory U$(1|1)$ Chern-Simons, and not $\mathfrak{gl}(1|1)$ or $\u(1|1)$, because we need to choose a reality condition and a global form for the bosonic subgroup~--~and we take it to be U$(1)\times{\rm U}(1)$. One could in principle consider other real and global forms. Those theories, if well-defined, would not need to be related to the $\psu(1|1)$ theory by orbifolding. For the $\psu(1|1)$ theory, we do not use the name PSU$(1|1)$, because there is no bosonic subgroup, and therefore no choice of the real form or the global form. This theory is naturally associated to the complex Lie superalgebra.

In this paper, we will not attempt to derive a relation between the supergroup Chern-Simons theory and the WZW models. However, if such a relation does exist, then what we have explained in this section would imply some correspondence between the U$(1|1)$ and the $\psu(1|1)$ WZW models. A duality of this kind is indeed known \cite{symplectic}, although its derivation does not look similar to ours.

\subsection{Magnetic Dual}\label{mirroru11}
The dual magnetic description of the theory is, of course, simply the orbifold of the QED of section \ref{magnetic}. (This fact can also be independently derived from brane constructions, as we review later in section \ref{branes}.) For the polynomial (\ref{SWequalT}), summing over flat bundles has simply the effect of picking only powers of holonomies, which are multiples of $k$. Equivalently, note that the action of the magnetic theory will have the form analogous to (\ref{u11act}), but with $I_{\psu(1|1)}+I_{\rm g.f.}$ replaced by the QED action. The field $A^N$ couples to the QED topological current $iF/2\pi$. Integrating over $A^N$, we simply get that the Chern class of the QED gauge field is the $k$-th multiple of the Chern class of the $A^E$ bundle. Since this bundle is arbitrary, we conclude that the orbifold of the magnetic theory is just the same QED, but with a constraint that the Chern class of the gauge field takes values in $kH^2(W)$. This can be equivalently viewed as\footnote{I thank N.~Seiberg for pointing this out.} an $\cN=4$ QED with one electron of charge $k$.

The $\u(1|1)$ partition function $\tau^{{\rm U}(1|1)}_{s}(\mathcal{L})$ inherits from the torsion $\tau_s$ the dependence on the spin-$\CC$ structure with trivial determinant. As we noted in the end of section \ref{defdetails}, the definition of $\tau_s(\mathcal{L})$ can be easily extended to construct a torsion, which depends on an arbitrary spin-$\CC$ structure, with no constraint. The same applies to $\tau^{{\rm U}(1|1)}_{s}(\mathcal{L})$. Now, consider the limit $k\rightarrow\infty$. Since now we sum essentially over all flat bundles, the U$(1|1)$ partition function cannot depend on the unitary part of the flat connection in $\mathcal{L}$. Therefore, by holomorphicity, it will not depend on $\mathcal{L}$ at all. We denote this version of the torsion by $\tau_{s,\infty}$. This is a number, which depends only on $W$ and on the choice of a spin-$\CC$ structure. Looking at the magnetic side, it is clear that this number is precisely the signed count of solutions to the Seiberg-Witten equations, with the fields $Z^\alpha$ valued in a given spin-$\CC$ bundle $s$. We conclude that the version of the electric theory with $k=\infty$ has these integers as its partition function. We note that this version of the torsion invariant has been defined and studied in \cite{TuraevInteger} and \cite{TuraevSW}. The fact that it is an integer was demonstrated by purely combinatorial methods.
One pedantic comment that we have to make is that $\tau_{s,\infty}$ is completely independent of $\mathcal{L}$ only for a manifold with $b_1>1$. For $b_1=1$, it does depend on the orientation in $H^1(W,\RR)$, induced by the absolute value of the holonomy of $\mathcal{L}$, since we need to choose the chamber, in which the Seiberg-Witten invariant is computed.

\subsection{Line Operators}
In the U$(1|1)$ theory, we can define some Wilson loops. For the atypical representations, these are essentially the operators that were already defined earlier in section \ref{eline} for the $\psu(1|1)$ theory. These are the operators $L_n$, labeled by one-dimensional atypicals $(n)$, as well as Wilson lines for the indecomposable representations, whose role we still have to clarify.

For the typical representations $(w,n)$, we want to claim that the Wilson lines are actually equivalent to the twist line operators of type $L_{\mathbf{t}\,,n}$ with $\mathbf{t}=\exp(2\pi i w/k)$. This relation is the usual statement of equivalence of Wilson lines and monodromy operators in Chern-Simons theory. (For U$(1|1)$, this relation was first suggested in \cite{RS3}.) The argument adapted to the supergroup case is given\footnote{In fact, for U$(1|1)$ the statement is quite obvious. The two-dimensional representation $(w,n)$ can be obtained by quantizing a pair of fermions, living on the Wilson line. After gauging these fermions away, one is left with a singularity in the gauge field, which is equivalent to the monodromy $\mathbf{t}$. The ubiquitous shift of $n$ by $1/2$ can be understood as a shift of the weight by the Weyl vector of the superalgebra $\u(1|1)$. The combination $m=n-1/2$, which appeared in section \ref{eline}, is the ``quantum-corrected'' weight.} in section 3.2 of \cite{MW}. One consistency check can be made by looking at the transformation of these operators under the charge conjugation symmetry $\mathcal{C}$. As can be seen from (\ref{matrices}), the representation changes as $(w,n)\rightarrow (-w,1-n)$, while the twist operator changes as $L_{ \mathbf{t}\,,n}\rightarrow L_{\mathbf{t}^{-1},1-n}$, as follows from its definition in section \ref{eline}. This is consistent with the identification of the operators. Note also that the boson-fermion parity of the highest weight vector of the representation $(w,n)$ is changed under the charge conjugation. A Wilson loop with a supertrace will consequently change its sign. This can be taken as an explanation of the factor $(-1)^\ell$ in the formula (\ref{cconj1}) for the charge conjugation transformation of torsion in presence of the boundary. For $\mathbf{t}=\exp(2\pi i w/k)$, we will also denote the operators $L_{\mathbf{t}\,,n}$ by $L_{w,\,n}$. Hopefully, this will not cause confusion.

\section{Hamiltonian Quantization}\label{Ham}
It is a well-established fact that the quantization of the Chern-Simons theory with an ordinary compact gauge group leads to conformal blocks of a WZW model \cite{WittenCS, Elitzur, OnHol, Axelrod}. For the supergroup case, it is often assumed that a similar relation holds \cite{RS1, RS3,Horne}, however, to our knowledge, no derivation of this statement is available in the literature, and the properties of the supergroup theories in the Hamiltonian picture are fundamentally unclear. In this section, we take an opportunity to bring some clarity to the subject by explicitly quantizing the Chern-Simons theories, which were considered in previous sections. Since these theories are essentially Gaussian, the quantization is straightforward. In this paper, we do not attempt to derive a relation to the conformal field theory. 

\subsection{Generalities}\label{ghosts}
In the quantization of an ordinary, bosonic Chern-Simons theory on a Riemann surface $\Sigma$, the classical phase space to be quantized is the moduli space of flat connections on $\Sigma$. Dividing by the gauge group typically introduces singularities, which, however, do not play much role~--~the correct thing to do is to throw them away by replacing the moduli space of flat connections by the moduli space of stable holomorphic bundles. In the supergroup case, this approach does not seem to lead to consistent results. Reducible connections here can lead to infinite partition functions (as in the case of the theory on $S^3$), and that should somehow be reflected in the canonical quantization. 
The correct approach, we believe, is to consider the theory with gauge-fixed fermionic part of the gauge symmetry. The Hilbert space of the supergroup Chern-Simons should then be constructed by taking the cohomology of the BRST supercharge in the joint Hilbert space of gauge fields and superghosts. Due to ``non-compactness'' of the fermionic directions, even in the ghost number zero sector this cohomology is not equivalent to throwing the ghosts away. 

First we consider the quantization of the $\psu(1|1)$ Chern-Simons theory. We take the three-manifold to be a product $\RR_t\times\Sigma$, where $\RR_t$ is the time direction, and $\Sigma$ is a connected oriented Riemann surface. Non-zero modes of the fields along $\Sigma$ do not contribute to the cohomology of $\Qb$, and can be dropped. Zero-modes are present, when the cohomology $H^\bullet(\Sigma,\mathcal{L})$ of the de Rham differential on $\Sigma$, twisted by the connection in the flat bundle $\mathcal{L}$, is non-trivial. When $H^1(\Sigma,\mathcal{L})$ is non-empty, there is a moduli space of fermionic flat connections on $\Sigma$. This gives a number of fermionic creation and annihilation operators, and a finite-dimensional factor for the Hilbert space,~--~in complete analogy with the ordinary, bosonic Chern-Simons. This will be illustrated in examples later in this section. The zeroth cohomology $H^0(\Sigma,\mathcal{L})$ is non-empty, if and only if the flat bundle $\mathcal{L}$ is trivial on $\Sigma$. In this case, the cohomology is one-dimensional, since we have assumed $\Sigma$ to be connected. The ghosts and the time component $A_0$ of the fermionic gauge field now have zero modes, which organize themselves into the quantum mechanics of a free superparticle in $\RR^{4|4}$, with the action
\beq
-\int{\rm d}t\,\Str(-A_0\dot{\lambda}+\dot{\bar{C}}\dot{C})\,.
\eeq
(Here for simplicity we did not write the coupling to the external gauge field.) 
The Hilbert space\footnote{Here and in what follows, by ``Hilbert space'' we really mean the space of states. It does not, in general, have an everywhere-defined non-degenerate scalar product.}, before we reduce to the cohomology of $\Qb$, is the space of functions on $\CC^2$ (with holomorphic coordinates given by the components $C^\pm$ of the scalar superghost), tensored with the four-dimensional Hilbert space of the fermions $\lambda^\pm$ and $A_0^\pm$. We can write the states as
\beq 
\psi_0|0\rangle+\psi_+\lambda^+|0\rangle+\psi_-\lambda^-|0\rangle+\psi_{+-}\lambda^+\lambda^-|0\rangle\,,\label{wavefunc}
\eeq
where $|0\rangle$ is annihilated by $A_0^\pm$, and $\psi_{\bullet}$ are functions of $C$ and $\bar{C}$. We recall from eq. (\ref{BRST}) that the BRST differential transforms $\bar{C}$ into $\lambda$. If we treat $\lambda^\pm$ as the differentials ${\rm d}\bar{C}^\pm$ and identify the wavefunctions (\ref{wavefunc}) with differential forms on $\CC^2$ with antiholomorphic indices, then $\Qb$ acts as the Dolbeault operator.  Thus, formally, the Hilbert space of the ghost system $\mathcal{H}_{\rm gh}$ is the Dolbeault cohomology\footnote{In the context of general Rozansky-Witten theories this statement~--~with $\CC^2$ replaced by a compact hyper-K\"ahler manifold~--~appeares already in the original paper \cite{RW}.} of $\CC^2$ with antiholomorphic indices. 

Since $\CC^2$ is non-compact, it is not obvious, how to make precise sense of this statement. Certainly, the path-integral of the theory on some three-manifold with a boundary produces a $\Qb$-closed state on the boundary. But to divide by $\Qb$-exact wavefunctions, we need to specify, what class of states is considered. 
For example, one could consider differential forms with no constraints on the behavior at infinity. This would lead to the ordinary Dolbeault complex. By the $\bar{\partial}$-Poincar\'e lemma, the cohomology is supported in degree zero, and consists simply of holomorphic functions on $\CC^2$. This space will be denoted by $H^{0,0}_{\bar\partial}$, and the states will be called non-compact. In our applications, we can usually restrict to states, which are invariant under the U$(1)_F$ ghost number symmetry. In $H^{0,0}_{\bar\partial}$, such states are multiples of $v_0=|0\rangle$, the constant holomorphic function. Another possibility is to look at the cohomology with compact support\footnote{For our purposes, the cohomology with compact support and the integrable cohomology will be considered as identical.}. By Serre duality, it is the dual of the space of holomorphic functions, and lives in degree $(0,2)$. We will denote this space by $H^{0,2}_{\bar{\partial},{\rm comp}}$, and call the corresponding states compact. The U$(1)_F$-invariant states here are multiples of $v_1=\delta^{(4)}(C,\bar{C})\lambda^+\lambda^-|0\rangle$.

To understand the interpretation of these states in our theory, we need to recall some properties of the torsion. Let $W'$ be a three-manifold with boundary $\Sigma$, together with some choice of the flat bundle $\mathcal{L}$ and, possibly, line operators inside. Let the holonomies of $\mathcal{L}$ be trivial on $\Sigma$, so that $H^0(\Sigma,\mathcal{L})$ is non-empty. If the flat bundle $\mathcal{L}$ is completely trivial even inside $W'$, and, in particular, $W'$ contains no line operators $L_{\mathbf{t}\,,n}$, we call the manifold with this choice of the flat bundle unstable. In the opposite case, we call it stable. Let $W$ be a connected sum of two three-manifolds $W_1$ and $W_2$ along their common boundary $\Sigma$, with no holonomies of $\mathcal{L}$ along the cycles of $\Sigma$. There are three possibilities. If both $W_1$ and $W_2$ are stable, the path-integral on $W$ vanishes, because of the fermionic zero modes,~--~this property of the torsion is known as ``unstability''. If both $W_1$ and $W_2$ are unstable, the path-integral is not well-defined, because of the presence of both fermionic and bosonic zero modes. Finally, if one of $W_1$, $W_2$ is stable, and the other is unstable, the functional integral generically has no zero modes, and the torsion is a finite number.

We claim that our functional integral for an unstable three-manifold $W'$ with boundary $\Sigma$ naturally yields a state for the ghosts in the non-compact cohomology $H^{0,0}_{\bar\partial}$. Indeed, the zero modes of $C$, $\bar{C}$ and $\lambda$ are completely free to fluctuate inside $W'$, and therefore the wavefunction as a function of $C$ is constant and should not contain insertions of $\lambda$,~--~so, it is a multiple of $v_0$. On the other hand, if the manifold $W'$ is stable, we get a state in the compact cohomology $H^{0,2}_{\bar\partial,{\rm comp}}$. The holonomies of the flat bundle inside $W'$ do not allow the zero modes of the ghosts and $\lambda$ to freely go to infinity. Modulo $\Qb$, the wavefunction in this case is a multiple of the state $v_1$. The natural pairing between the compact and the non-compact cohomology yields a finite answer for a closed three-manifold, glued from a stable and an unstable piece. If, on the other hand, we try to pair two stable manifolds, we get zero, since we have too many insertions of the operators $\lambda^\pm$ in the product of the wavefunctions. If we try to pair two non-compact, unstable states, the result is not well-defined, because one encounters both bosonic and fermionic zero modes\footnote{For a manifold $W$ glued from two unstable pieces, depending on the situation, it can be natural to define the torsion to be infinity, or zero, or some finite number, by perturbing $\mathcal{L}$ away from the singular case. However, it does not seem to be possible to give any universal meaning to the pairing of non-compact wavefunctions in the ghost Hilbert space.}. This is consistent with the properties of the torsion, described above.

In the special case that $\Sigma$ is a two-sphere with no punctures, the ghost Hilbert space $\mathcal{H}_{\rm gh}$ is all of the Hilbert space. Since it is not one-dimensional, the topological theory contains non-trivial local operators. They are in correspondence with $\bar\partial$-closed $(0,p)$-forms on $\CC^2$. Again, one might think that all of these, except for the holomorphic functions, are $\Qb$-exact, and therefore decouple, but this is not in general true due to the non-compactness of the field space. Let us introduce a special notation $\mathcal{O}_1$ for the operator $\lambda^+\lambda^-\delta^{(4)}(C,\bar{C})$, which we will need in what follows.

\subsection{The Theory On $S^1\times\Sigma$ }
Let us illustrate in some examples, how this machinery works. First we compute the invariants for the theory on $S^1\times \Sigma$, with $\Sigma$ a closed Riemann surface with no punctures. Then we add punctures and derive the skein relations for the Alexander polynomial. In the whole section~\ref{Ham}, we typically ignore the overall sign of the torsion, and its dependence on the spin structure.

\subsubsection{No Punctures}
Consider a three-manifold $S^1\times\Sigma$, where $\Sigma$ is a Riemann surface of genus $g$. Let the flat bundle $\mathcal{L}$ have a holonomy $\mathbf{t}$ along the $S^1$, and no holonomies along the cycles of $\Sigma$. We would like to compute the  torsion $\tau(\mathbf{t})$ of this manifold. For simplicity, we take $|\mathbf{t}|=1$.

The topological theory on this manifold reduces to the quantum mechanics of zero modes of the fields on $\Sigma$. The components of the gauge field $A^\pm$, tangential to $\Sigma$, will produce $4g$ fermionic zero modes, which can be grouped into $2g$ pairs of fermions, corresponding to some choice of  $a$- and $b$-cycles on $\Sigma$. For each pair of the fermions, the action is defined with the kinetic operator $i\partial_t+B_0$, where $B_0$ is the background gauge field in the time direction. If we denote the determinant of this operator by $d(\mathbf{t})$, the gauge fields contribute a factor of $d^{2g}(\mathbf{t})$ to the torsion. The time component of the gauge field $A_0^\pm$ together with the Lagrange multiplier $\lambda$ give two more pairs of fermions with the same action, and hence a factor of $d^2(\mathbf{t})$. Finally, the zero-modes of the superghosts $C^\pm$ and $\bar{C}^\pm$ give two complex scalars, which contribute a factor of $d^{-4}(\mathbf{t})$. The torsion altogether is $\tau(\mathbf{t})=d^{2g-2}(\mathbf{t})$. Using the zeta-regularization,\footnote{One needs to use the identity $\exp\left(-\zeta'(0,a)-\zeta'(0,1-a)\right)=2\sin(\pi a)$ for the derivative $\partial_s\zeta(s,a)$ of the Hurwitz zeta-function. In the text we ignored the factor of $-i$, which results from this computation, since we are not interested in the overall sign of $\tau(\mathbf{t})$.} one readily computes $d(\mathbf{t})=\mathbf{t}^{1/2}-\mathbf{t}^{-1/2}$. For the torsion of $S^1\times \Sigma$, we get
\beq
\tau(\mathbf{t})=(\mathbf{t}^{1/2}-\mathbf{t}^{-1/2})^{2g-2}\,.\label{tausigma}
\eeq

Let us derive the same result by a Hilbert space computation. The torsion can be computed by taking the supertrace $\Str_{\mathcal H}\mathbf{t}^{\hat{J}}$ over the Hilbert space, where $\hat{J}$ is the generator of the U$(1)_{\rm fl}$-symmetry. In this formalism, it is obvious that the contribution of a single pair of fermions is indeed $d(\mathbf{t})=\mathbf{t}^{1/2}-\mathbf{t}^{-1/2}$. The contribution of the superghosts $C$ and $\bar{C}$ can also be easily computed. We set $\mathbf{t}=\exp(i\alpha)$. The quantum mechanics of the complex field $C^+$ is the theory of a free particle in $\RR^2$, and we need to find the trace of the rotation operator $\exp(i\alpha \hat{J})$ over its Hilbert space,
\beq
\tr\,\exp\left(i\alpha \hat{J}\right) = \int \fr{{\rm d}^2\vec p\,{\rm d}^2\vec x}{(2\pi)^2}\exp(i\vec p\vec x)\exp(-i\pvec{p}'\vec x)=\fr{1}{4\sin^2(\alpha/2)}\,,
\eeq
where $\pvec{p}'$ is the vector obtained from $\vec p$ by a rotation by the angle $\alpha$. This is equal to $-d^{-2}(\mathbf{t})$, and together with a similar contribution from $C^-$ leads to the correct result $d^{-4}(\mathbf{t})$.

In the computation above, the trace was taken over the whole Hilbert space of the ghost system, and not over the cohomology of $\Qb$, since it is not clear in general, what one should mean by this cohomology. However, it is curious to observe that one can obtain the same results by tracing over the non-compact (or over the compact) Dolbeault cohomology. Indeed, $H^{0,0}_{\bar\partial}$ is the space of holomorphic functions on $\CC^2$, which can be expanded in the basis generated by the monomials $1$, $C^+$, $C^-$, $(C^+)^2$, {\it etc.} The trace of $\mathbf{t}^{\hat{J}}$ over this space can be written as 
\beq
\mathbf{t}^0+\ex^{-\eps}(\mathbf{t}^{-1}+\mathbf{t})+\ex^{-2\eps}(\mathbf{t}^{-2}+\mathbf{t}^0+\mathbf{t}^2)+\dots\,,
\eeq
where we introduced a regulator $\eps>0$. The sum of this convergent series for $\eps\rightarrow 0$ is equal to $-d^{-2}(\mathbf{t})$, which is the correct contribution of the ghost system to the torsion. I do not know, if this computation should be taken seriously.

\subsubsection{Surfaces With Punctures}\label{punctures}
Next, let us incorporate some line operators. Consider a Riemann surface $\Sigma$ of genus $g$ with $p\ge 2$ punctures, corresponding to $p$ parallel line operators $L_{\t_1,\,n_1}\,\dots L_{\t_p,\,n_p}$, stretched along the $S^1$. For consistency, we assume $\t_1\t_2\dots\t_p=1$. Let there also be a background holonomy $\t$ around the $S^1$. We introduce the number $N=\sum_i(n_i-1/2)$, which measures the total U$(1)_{\rm fl}$-charge. For $N=0$, the configuration is symmetric under the charge conjugation (up to the substitution $\t\rightarrow\t^{-1}$ for all the holonomies.) 

Due to the presence of line operators, the cohomology $H^0(\Sigma,\mathcal{L})$ is empty, and the Hilbert space does not contain the ghost factor $\mathcal{H}_{\mathrm {gh}}$. However, the cohomology $H^1(\Sigma,\mathcal{L})\equiv H^1$ is in general non-empty, so there will be $h={\rm dim}\,H^1$ zero modes of the fermionic gauge field $A^+$ and $h$ zero modes of the field $A^-$. Our Lie superalgebra is a direct sum, and correspondingly it is convenient to choose a polarization, in which the modes of $A^+$ are the creation operators, and the modes of $A^-$ are the annihilation operators. The Hilbert space is 
\beq
(\det H^1)^{-1/2+N/h}\otimes\wedge^\bullet H^1\,.\label{hilbert4}
\eeq
It contains states with charges ranging from $-h/2+N$ to $h/2+N$, with
\beq
\mathsf{N}(q)={h\choose{q+h/2-N}}\label{dimension}
\eeq
states of charge $q$. (The overall power of $\det H^1$ was chosen so as to ensure that for $N=0$ the spectrum of U$(1)_{\rm fl}$-charges is symmetric.) Taking the supertrace of $\t^{\hat{J}}$ over this Hilbert space, we find the invariant for $S^1\times \Sigma$,
\beq
\tau(\mathbf{t})=\t^N(\t^{1/2}-\t^{-1/2})^h\,.
\eeq
As we will see, $h=-\chi=2g-2+p$. An important special case is that $\Sigma$ is $S^2$ with two marked points. Then $h=0$, the Hilbert space is one-dimensional, and the invariant $\tau$ is equal to one, up to an overall power of $\t$.

Let us give a more explicit description of the twisted cohomology for the simple case of $\Sigma\simeq S^2$. In the presence of a singular background field, corresponding to an insertion of a line operator $L_{\t,\,n}$ along some knot $K$, the behavior of the dynamical fields of the $\psu(1|1)$ theory near $K$ is determined by a boundary condition, which is described in Appendix~\ref{bc}. It says that the superghost fields $C^\pm$ should vanish near $K$, while the components of the fermionic gauge field $A^\pm$, perpendicular to $K$, are allowed to have a singularity, which however has to be better than a pole. This boundary condition is elliptic. The cohomology $H^1$, therefore, can be represented by $\mathcal{L}$-twisted one-forms, which lie in the kernel of the operator $d+d^*$ on $\Sigma$ and which near the marked points are less singular than $1/r$. Just for illustration, we can write an explicit formula for these one-forms. For that, pick a complex structure on $\Sigma$, and let the marked points be $z_1,\dots,z_p$. The cohomology will be represented by holomorphic $(1,0)$- and antiholomorphic $(0,1)$-forms. Let us write $\t_i=\exp(2\pi i\,a_i)$, with $a_i\in (0,1)$, for the holonomies. (We assume that the bundle $\mathcal{L}$ is unitary.) Note that the sum $\sum a_i$ is a positive integer. Any twisted holomorphic one-form can be written as
\beq
\omega=\prod_{i=1}^p (z-z_i)^{a_i} P(z){\rm d}z\,,
\eeq
with some rational function $P(z)$, which is allowed to have simple poles at points $z_i$, according to our boundary condition. Assuming that infinity is not among the marked points, we should have $\omega\sim {\rm d}z/z^2+o(1/z^2)$ at large $z$. Writing $P(z)$ as $\sum P_i/(z-z_i)$, the condition at infinity gives $1+\sum a_i$ linear equations on the coefficients $P_i$, so the space of twisted holomorphic forms is of dimension $p-1-\sum a_i$. Similarly, the space of twisted antiholomorphic forms has dimension $\sum a_i-1$, and the total dimension of $H^1$ is $p-2$, in agreement with the formula $h=-\chi$.

\begin{figure}
 \begin{center}
   \includegraphics[width=17cm]{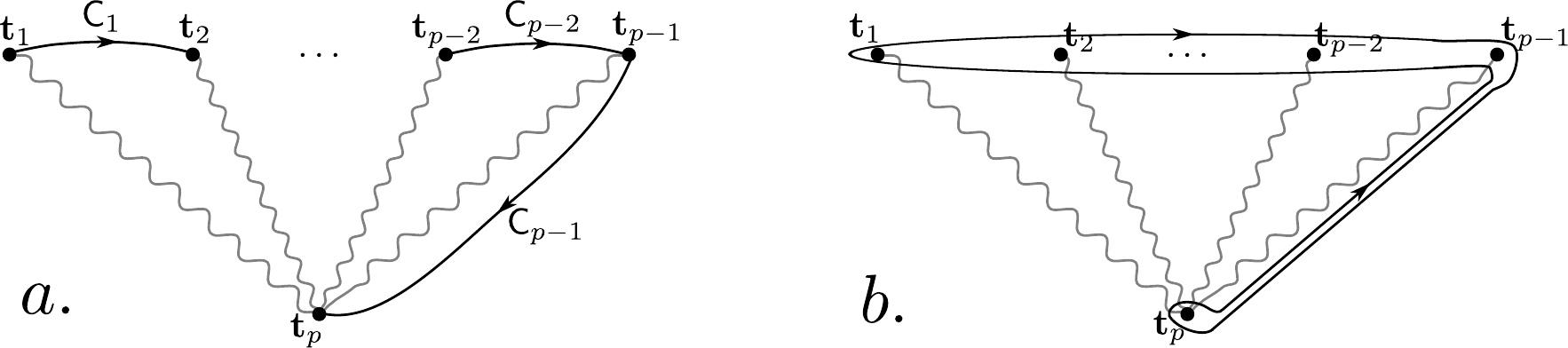}
 \end{center}
 \caption{\small {\large a.}~Marked points, basis contours, and a particular choice of cuts on the $p$-punctured sphere. Locally-constant sections of $\mathcal{L}$ pick a factor of $\t_i$ upon going counter-clockwise around the $i$-th puncture. {\large b.}~This contour is trivial, since it can be pulled off to infinity. This gives a relation $(1-\t_1)\mathsf{C}_1+\dots+(1-\t_1\dots\t_{p-1})\mathsf{C}_{p-1}=0$.}
  \label{contours}
\end{figure}
Instead of working with cohomology, it is more convenient to look at the dual homology, which for $S^2$ with marked points is generated by contours, connecting different punctures\footnote{I am grateful to E.~Witten for the suggestion to look at the homology and for helpful explanations.}. (The differential forms, which behave better than $1/r$ near the punctures, can be integrated over such contours, and the integrals do not change, when the forms are shifted by differentials of functions that vanish at the punctures. Moreover, the pairing between this version of homology and the twisted cohomology is non-degenerate.) The basis in the homology consists of $p-2$ contours $\mathsf{C}_1,\dots, \mathsf{C}_{p-2}$, shown on fig.~\ref{contours}a. One might think that the contour $\mathsf{C}_{p-1}$ should also be included in the basis, but actually it can be expressed in terms of $\mathsf{C}_1,\dots, \mathsf{C}_{p-2}$, using the relation of fig.~\ref{contours}b. On a general Riemann surface, one obtains in the same way that the dimension of the homology is $h=-\chi=p-2+2g$.

\begin{figure}
 \begin{center}
   \includegraphics[width=13cm]{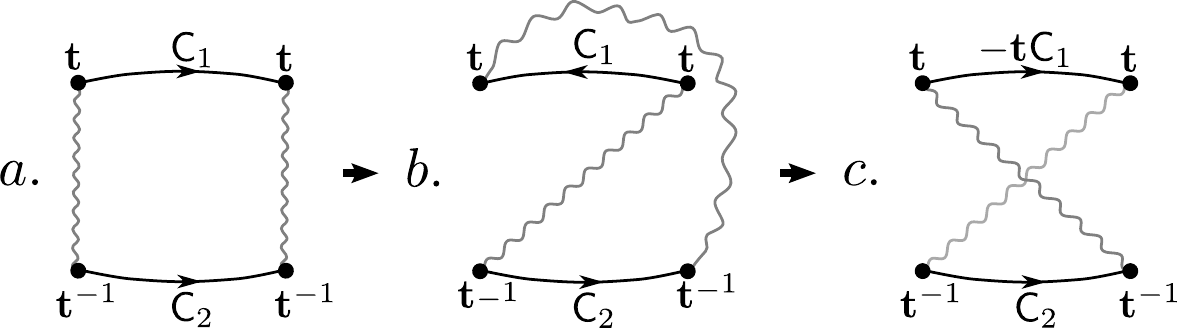}
 \end{center}
 \caption{\small {\large a.}~Two cuts and two basis contours for a four-punctured sphere. {\large b.}~The result of the braiding transformation. {\large c.}~We moved the contour $\mathsf{C}_1$ across the cut and reversed its orientation, which produced a factor of $-\mathbf{t}$.}
  \label{halfcontours}
\end{figure}
It is possible to find modular transformations of states in the Hilbert space. For that, one needs to find the action of large diffeomorphisms on the twisted cohomology $H^1$, or, equivalently, on the basis contours in the dual homology. To give an example of such argument, we derive the skein relations for the Alexander polynomial\footnote{Rather similar contour manipulations are used in \cite{RS1} to  obtain braiding transformations of the states from the CFT free-field representation. The contours in question are then integration contours for the screening fields. In fact, the two computations seem to be directly related, since the screening fields are the CFT currents, which in the Chern-Simons theory correspond to the gauge fields $A^\pm$, whose modes are the cohomology that we are considering. To make the connection more precise, one needs to switch to the holomorphic polarization.}. Consider a Riemann sphere with four punctures, two of which are labeled by holonomies $\mathbf{t}$, and two by $\mathbf{t}^{-1}$. This configuration arises on the boundary of a solid three-ball with two line operators $L_{\t,\,n}$ inside. We set the parameters $n$ equal to one-half, so that the line operators are expected to have trivial framing transformations and to give rise to the Conway function. (We have to mention once again that our understanding of these line operators is incomplete. This will lead to some uncontrollable minus signs in their expectation values.) The twisted cohomology $H^1$ on the four-punctured sphere is two-dimensional. A pair of basis contours $\mathsf{C}_1$ and $\mathsf{C}_2$ for the dual homology is shown on fig.~\ref{halfcontours}a. We make a large diffeomorphism, which exchanges the two punctures labeled by $\mathbf{t}$. This leads to the configuration of fig.~\ref{halfcontours}b. We move the upper cut through the contour $\mathsf{C}_1$. This multiplies $\mathsf{C}_1$ by a factor of $\mathbf{t}$. This brings us to the configuration of fig.~\ref{halfcontours}c, where we have also reversed the orientation of the upper contour. The cuts can now be deformed back to the configuration of fig.~\ref{halfcontours}a, and we find that the braiding transformation acts on the contours as
\beq
\left(\bea{c}\mathsf{C}'_1\\\mathsf{C}'_2\eea\right)=\left(\bea{cc}-\mathbf{t}&0\\0&1\eea\right)\left(\bea{c}\mathsf{C}_1\\\mathsf{C}_2\eea\right)\,.\label{halfbraiding}
\eeq
The Hilbert space of the four-punctured sphere, according to eq.~(\ref{hilbert4}), consists of four states~--~one of U$(1)_{\rm fl}$-charge $-1$, one of charge $+1$, and two of charge $0$. The neutral states are the ones that arise on the boundary of a three-ball with a pair of line operators inside. The state of charge $-1$ transforms under the braiding by some phase. From eq.~(\ref{hilbert4}), we would expect this phase to be the inverse square root of the determinant of the matrix in (\ref{halfbraiding}). The two U$(1)_{\rm fl}$-invariant states then transform with the matrix
\beq
\left(\bea{cc}i\mathbf{t}^{1/2}&0\\0&-i\mathbf{t}^{-1/2}\eea\right)\,.\label{shbraiding}
\eeq
Note that the braiding action (\ref{halfbraiding}) is defined only up to an overall phase, since we could make a constant U$(1)_{\rm fl}$ gauge transformation, or, equivalently, could move the cuts on fig.~\ref{halfcontours} around the sphere any number of times. Such a phase, however, would cancel out in \ref{shbraiding}, since the two states of interest are U$(1)_{\rm fl}$-invariant.
\begin{figure}
 \begin{center}
   \includegraphics[width=11cm]{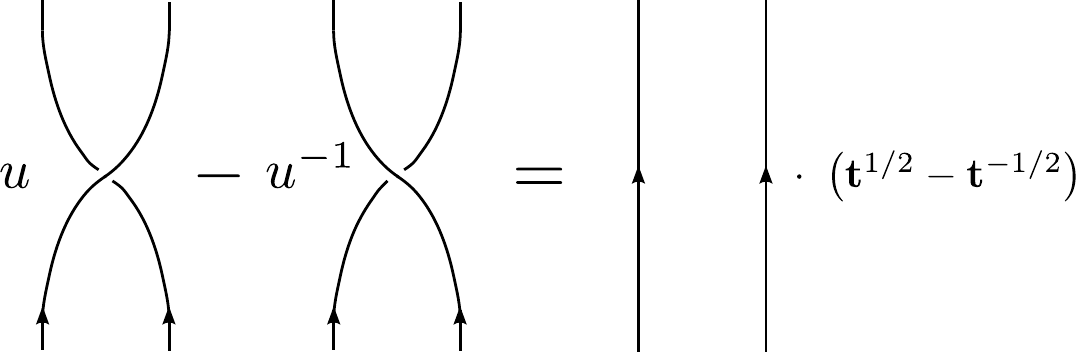}
 \end{center}
 \caption{\small A skein relation for the Alexander polynomial. In the canonical framing, $u=1$.}
  \label{halfskein}
\end{figure}
\begin{figure}
 \begin{center}
   \includegraphics[width=8cm]{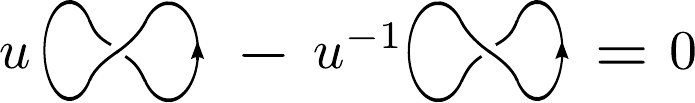}
 \end{center}
 \caption{\small The result of closing the strands in the skein relation and using the fact that the Alexander polynomial for a disjoint link is zero. The relation is consistent, if it is written in the vertical framing, and the invariant transforms by a factor of $u$ under a unit change of framing.}
  \label{framing}
\end{figure}
 
From (\ref{shbraiding}) it follows \cite{WittenCS} that the knot invariant satisfies the skein relation of fig.\ref{halfskein}, with $u=i$. (On the way, we made an arbitrary choice of the square root of the determinant of the matrix (\ref{halfbraiding}). With an opposite choice, we would get $u=-i$.) Initially, we assumed that our line operators have no framing dependence. But now we can see that this would be inconsistent with fig.~\ref{framing}, which is obtained from the skein relation by closing the braids and using the fact that the Alexander polynomial of a disjoint link is zero. We are seemingly forced to conclude that our invariant does have a framing dependence, with a framing factor $u=i$. On $S^3$, there exists a canonical choice of framing, in which the self-linking number of all components of the link is zero. If we bring all the links to this choice of framing, the polynomial would satisfy the skein relation of fig.~\ref{halfskein}, but with $u=1$. This skein relation, together with a normalization condition, which we derive later in this section, defines the single-variable Alexander polynomial (or the Conway function), as expected. But the fact that we found a non-trivial framing dependence is rather unsatisfactory. In the dual Seiberg-Witten description, the knot invariant is clearly a polynomial with real (and integral) coefficients, and there can be no factors of $i$. To get rid of the problem, we have to put an extra factor of $i$ in the braiding transformation of the highest weight state of U$(1)_{\rm fl}$-charge ${-}1$. This will multiply the matrix (\ref{shbraiding}) by $i$, and make $u=1$ in the skein relation. It would be desirable to understand the physical origin of this factor.

To be able to compute the multivariable Alexander polynomial, that is, the invariant for multicomponent links, with different components labeled by arbitrary holonomies, one needs two more skein relations \cite{Murakami}. We derive them in Appendix~\ref{multiskein}.

\subsection{$T^2$ And Line Operators}
In this section, we look more closely on the Hamiltonian quantization of the theory on a two-torus $T^2$. First we describe the Hilbert space abstractly, and then relate different states to line operators of the theory.

\subsubsection{The Torus Hilbert Space}
Let us fix a basis of cycles $a$ and $b$ on $T^2$, and denote the corresponding holonomies of the background bundle by $\mathbf{t}_a$ and $\mathbf{t}_b$. Assume first that at least one of the holonomies is non-trivial. In this case, the twisted cohomology $H^\bullet(T^2,\mathcal{L})$ is empty, and the torus Hilbert space $\mathcal{H}_{\mathbf{t}_a,\mathbf{t}_b}$ is one-dimensional. Let us choose some basis vector $|\t_a,\t_b\rangle$ for each of these Hilbert spaces. We pick a normalization such that under any SL$(2,\ZZ)$ modular transformation $\mathcal{M}$ the vectors map as
\beq
\mathcal{M}|\t_a,\t_b\rangle=|\t_{\mathcal{M}^{-1}(a)},\t_{\mathcal{M}^{-1}(b)}\rangle\,,\label{modular}
\eeq
without any extra factors. Note that the charge conjugation symmetry $\mathcal{C}$ is equivalent to the modular transformation $\mathcal{S}^2$, which flips the signs of both cycles.

A slightly more complicated case is $\t_a=\t_b=1$. The Hilbert space $\mathcal{H}_{1,1}$ is a product, with one factor being the vector space $\mathcal{H}_{\rm gh}$ of states of the ghosts, which was described before. Another factor comes from the fact that the fermionic gauge fields now have zero modes $A^+_a$, $A^+_b$, $A^-_a$ and $A^-_b$, arising from components of the one-forms $A^\pm$ along the $a$- or the $b$-cycle. With a natural choice of polarization, the modes of $A^-$ are the annihilation operators, and the modes of $A^+$ are the creation operators. The four states in the Hilbert space of the vector fields are
\beq
|{-}1\rangle, \quad |0_a\rangle\equiv A^+_a|{-}1\rangle, \quad |0_b\rangle\equiv A^+_b|{-}1\rangle, \quad |{+}1\rangle\equiv A^+_aA^+_b|{-}1\rangle\,.\label{gaugeH}
\eeq
The states $|{\pm}1\rangle$ are of charge ${\pm}1$, and are invariant under the modular group SL$(2,\ZZ)$, since they have nowhere to transform. The two states $|0_a\rangle$ and $|0_b\rangle$ are neutral, and transform under SL$(2,\ZZ)$ as a doublet.

\subsubsection{Line Operators $L_{\t\,,n}$}
Consider a solid torus with boundary $T^2$, with cycle $a$ contractible, and put a line operator of type $L_{\t_a,\,n}$ along the $b$-cycle inside. Here it is assumed that $\t_a\ne 1$. The operator is taken with the natural framing for loops in the solid torus. We can also turn on a background holonomy $\t_b$. The resulting state lives in $\mathcal{H}_{\t_a,\t_b}$, and we claim that it is
\beq
|L_{\mathbf{t}_a\,,n},\mathbf{t}_b\rangle=\mathbf{t}_b^{n-1/2}|\mathbf{t}_a,\mathbf{t}_b\rangle\,,\label{basis}
\eeq
with a suitable normalization of $|\t_a,\t_b\rangle$. (Note that $\t_b$ is not a parameter of the line operator itself, but is defined by the background bundle, and in particular by the other line operators, linked with the given one.) It is easy to see that both sides of (\ref{basis}) depend on $n$ in the same way. (In taking a half-integer power of $\mathbf{t}_b$, we ignored the sign ambiguity, since we generically do not try to fix the overall signs in this section. A more accurate treatment of signs would require keeping track of spin structures.) The non-trivial content of this equation is the statement that $|\t_a,\t_b\rangle$, defined in this way, transforms under the modular group as in (\ref{modular}), without any extra factors. For the charge conjugation symmetry $\mathcal{C}$, this is easy to see from the transformation properties of the line operators $L_{\t_a,\,n}$. For the element $\mathcal{T}$ of the modular group SL$(2,\ZZ)$, the l.h.s. changes into $\t_a^{-n+1/2}|L_{\t_a,\,n},\t_b\rangle$, where the factor of $\t_a$ is due to the change of framing. This is again consistent with (\ref{modular}). It requires a little more work to see that $|\t_a,\t_b\rangle$ transforms as in (\ref{modular}) also for the element $\mathcal{S}$ of SL$(2,\ZZ)$. Note that a pair of solid tori can be glued together to produce $S^1\times S^2$ with two parallel line operators along the $S^1$. The gluing identifies the $b$-cycles of the two tori, and maps the $a$-cycle of one torus to the $-a$ of the other. This gives a bilinear pairing between the Hilbert spaces $\mathcal{H}_{\t_a^{-1},\t_b}$ and $\mathcal{H}_{\t_a,\t_b}$. In the section \ref{punctures}, we learned that the dimension of the Hilbert space on $S^2$ with two marked points is equal to one. It follows that, under the bilinear pairing,
\beq
\left(|\t_{a}^{-1},\t_b\rangle,|\t_a,\t_b\rangle\right)=\left(|L_{\t_a^{-1},\,1-n},\t_b\rangle,|L_{\t_a,\,n},\t_b\rangle\right)=1\,.\label{product}
\eeq
Note that we can apply the elements $\mathcal{C}\mathcal{S}$ and $\mathcal{S}$ to the two vectors in this equation, and get the same gluing of the tori. Suppose that the $\mathcal{S}$-transformation of the state $|\t_a,\t_b\rangle$ gives the state $|\t_b^{-1},\t_a\rangle$ with some factor $f(\t_a,\t_b)$. It then follows that $f(\t_a,\t_b)f(\t_a^{-1},\t_b)=1$. The function $f$ should be holomorphic, and can only have zeros or singularities at $\t_a$ or $\t_b$ equal to $0$, $1$ or infinity. However, $1$ is excluded by the equation above. Then, $f$ can only be a monomial in powers of $\t_{a}$, but this possibility is excluded by the charge conjugation symmetry. We conclude that the vectors $|\t_a,\t_b\rangle$, defined as in (\ref{basis}), transform under the modular group according to (\ref{modular}). (We did not exclude the possibility of non-trivial $\t$-independent phases in (\ref{modular}), but there seem to be no possible candidates for such phases.)

As a check of the modular transformations that we have described, consider a Hopf link, formed by two unknots with some operators $L_{\t_a,\,n}$ and $L_{\t_b,\,m}$ in $S^3$. Up to powers of $\t_\bullet$, which depend on the framings, the invariant for this configuration is equal to the same scalar product (\ref{product}), that is, to one. This is the correct result for the Alexander polynomial of the Hopf link. In the discussion of the Hilbert space of empty $S^2$, we have defined a local operator $\mathcal{O}_1$. Now we can give it a geometric interpretation\footnote{This operator can be given yet another interpretation. Consider cutting out a small three-ball, and gluing in a non-compact space, which is the complement of the three-ball in $\RR^3$. The zero-modes of the ghosts cannot freely fluctuate in such geometry, so, this construction produces the desired operator. We can also give arbitrary non-zero vevs $C_0\in \CC^2$ to the fields $C$ in the asymptotic region. This would produce the operator $\lambda^+\lambda^-\delta^{(4)}(C-C_0,\bar{C}-\bar{C}_0)$.}: it can be obtained by inserting a small Hopf link of loop operators of type $L_{\t,\,n}$.

\subsubsection{Other Line Operators}
Consider again the same solid torus, and put a line operator $L_n$ along the $b$-cycle\footnote{These operators differ from the vacuum just by a factor of $\t^n$, so, we would loose nothing by considering only $n=0$. But we prefer to keep general $n$, because it will be helpful, when we come to the U$(1|1)$ theory.}. We first assume that $\t_b\ne 1$, so that the resulting state is $|L_n,\t_b\rangle=\t_b^n\, g(\t_b)|1,\t_b\rangle$, for some holomorphic function $g(\t)$. To fix it, note that the invariant for $S^1\times S^2$ with holonomy $\t_b$ around $S^1$ can be represented by
\beq
\tau(S^1\times S^2,\t_b)=\left(|L_0,\t_b\rangle,|L_0,\t_b\rangle\right)=g^2(\t_b)\,.
\eeq
On the other hand, it is equal to $(\t_b^{1/2}-\t_b^{-1/2})^{-2}$, so we find that $g(\t_b)=1/(\t^{1/2}_b-\t^{-1/2}_b)$, and therefore
\beq
|L_n,\t_b\rangle=\fr{1}{1-\t_b^{-1}}\,\t_b^{n-1/2}\,|1,\t_b\rangle\,.\label{g}
\eeq
Using this, we can find the Milnor torsion for an unknot in $S^3$. This invariant is equal to
\beq
\left(|L_{\t\,,n},1\rangle,\mathcal{S}|L_0,\t\rangle\right)=\fr{1}{\t^{1/2}-\t^{-1/2}}\,,\label{unknot}
\eeq
which is the correct result. (One can get rid of the half-integer power of $\t$ by choosing a different framing.) Another application is to find the degeneration of the operator $L_{\t\,,n}$ in the limit $\t\rightarrow 1$. From (\ref{basis}) and (\ref{g}) we find
\beq
\lim_{\t \rightarrow 1}L_{\t,\,n}=L_{n}-L_{n-1}\,,\quad \t_b\ne 1\,.\label{limit}
\eeq
(This formula is valid only in the sector $\t_b\ne 1$, that is, in presence of a non-trivial holonomy along the line operator.) This relation, when applied to invariants of links in the three-sphere, is known as the Torres formula \cite{Torres}.

Now, consider the case that $\t_b=1$, so that $L_n$ is inserted inside a solid torus with no background holonomy. The parameter $n$ then does nothing, and the resulting state corresponds just to the empty torus. We want to identify the corresponding state $|{\rm vac}\rangle$ in $\mathcal{H}_{1,1}$. In the ghost Hilbert space, it is the vector $v_0$, as defined in section \ref{ghosts}. In the gauge fields Hilbert space, it is some vector from (\ref{gaugeH}), which should have zero charge and should be invariant under the $\mathcal{T}$-transformation. The vector with these properties is $|0_a\rangle$, so we find
\beq
|L_n,1\rangle=|{\rm vac}\rangle=v_0\otimes |0_a\rangle\,.\label{g0}
\eeq

\begin{figure}
 \begin{center}
   \includegraphics[width=6.5cm]{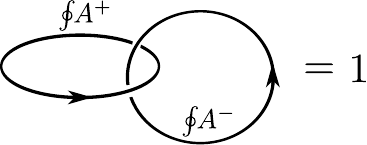}
 \end{center}
 \caption{\small The anticommutation relation for the modes of $A^+$ and $A^-$, written geometrically.}
  \label{commutation}
\end{figure}
\begin{figure}
 \begin{center}
   \includegraphics[width=8cm]{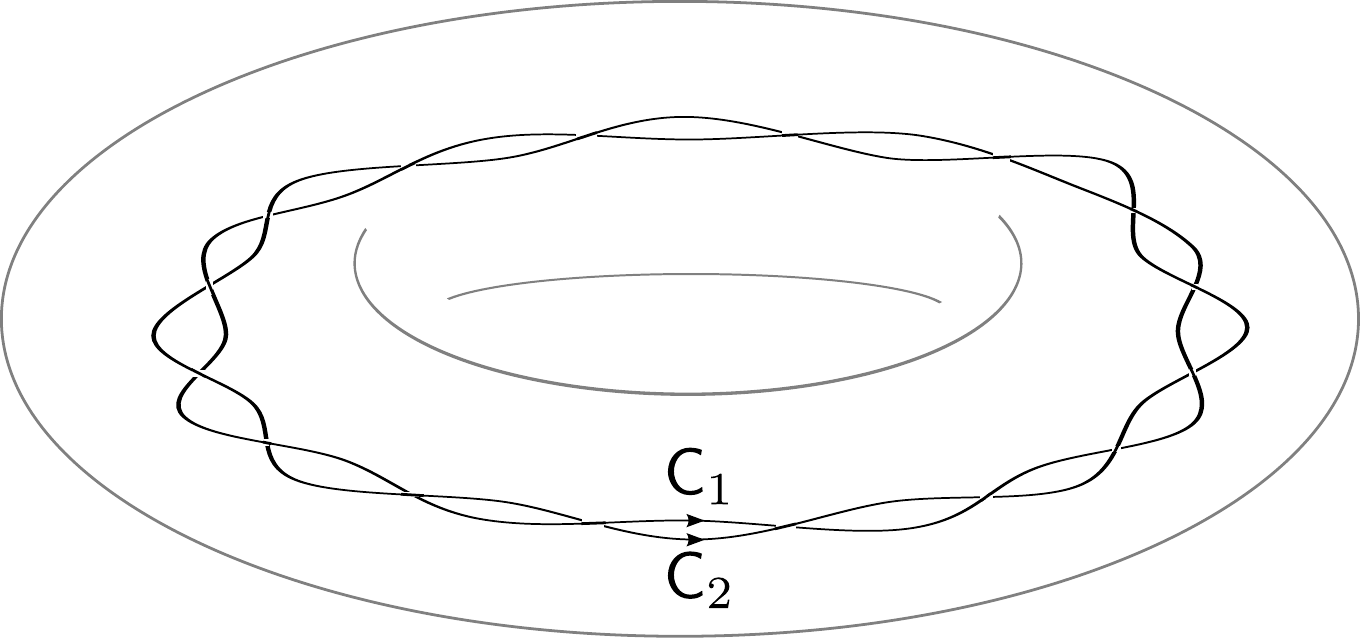}
 \end{center}
 \caption{\small By inserting the operator $\oint_{\mathsf{C}_1} A^+\oint_{\mathsf{C}_2}A^-$, with cycles $\mathsf{C}_1$ and $\mathsf{C}_2$ shown on the figure, one obtains the state $\mathcal{T}^p\mathcal{S}|{\rm vac}\rangle=(\mathcal{S}+p)|{\rm vac}\rangle$, with $p$ equal to the number of times the cycles wind around each other.}
  \label{torus}
\end{figure}

Let us also give a geometrical interpretation to some other states in  $\mathcal{H}_{1,1}$. For that, we simply need to write the modes $A^{\pm}_{a,b}$, used as the creation and annihilation operators in (\ref{gaugeH}), as integrals of $A^\pm$ over different cycles. The anticommutation relation for these operators is equivalent to a geometrical identity, shown on fig.~\ref{commutation}. To obtain the state $v_0\otimes |{-}1\rangle$, one inserts into the empty solid torus the operator $\oint_b A^-$, effectively undoing the action of $A^+_a$ in (\ref{gaugeH}). Similarly, the states $v_0\otimes |0_b\rangle$ and $v_0\otimes|{+}1\rangle$ can be obtained by inserting operators $\oint_b A^+\,\oint_b A^-$ and $\oint_a A^+$, respectively. On fig.~\ref{torus}, we show the operators needed to create the neutral states, which are obtained by applying transformations $\mathcal{T}^p$ to the $\mathcal{S}$-transform of the vacuum.

\subsubsection{OPEs of Line Operators}
\begin{figure}
 \begin{center}
   \includegraphics[width=14.2cm]{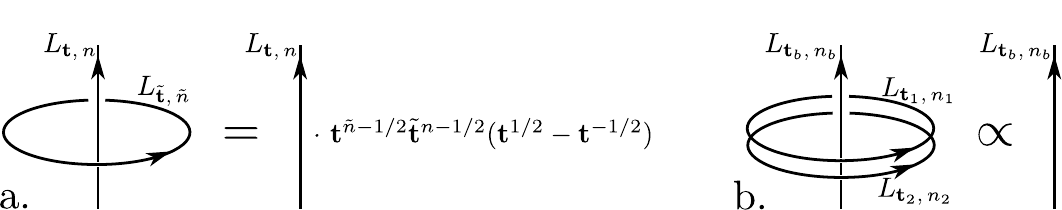}
 \end{center}
 \caption{{\large a.}~{\small A relation that follows from one-dimensionality of the Hilbert space of a sphere with two marked points.} {\large b.}~{\small The configurations on the left and on the right are proportional with some coefficient.}}
  \label{relation}
\end{figure}
We would like to find the OPEs of our line operators. For products involving the atypical operator $L_n$, the OPE is trivial: such an operator simply shifts the value of $n$ for the other operators, with which it is multiplied. More interesting are the products of the typical operators $L_{\t,\,n}$. To find their OPE, we will need  the relation of fig.~\ref{relation}a. It can be derived from the fact that the Hilbert space of the two-punctured sphere is one dimensional, and from comparison of the invariants for two linked unknots and for a single unknot in $S^3$.

To derive the expansion for the product $L_{\t_1,\,n_1}\times L_{\t_2,\,n_2}$, we place two parallel operators along the $b$-cycle inside a solid torus, and look at the resulting state on the boundary $T^2$. Assuming that $\t_1\t_2\ne 1$, the Hilbert space for the torus with this insertion is one-dimensional, and the state created by the insertion of the two operators is proportional to the state created by $L_{\t_1\t_2,\,n_1+n_2}$, with some proportionality coefficient $f$, which in general can be a holomorphic function of $\t_1$, $\t_2$, and also of the holonomy $\t_b$ of the background bundle along the $b$-cycle of the torus. To fix this coefficient, consider the configuration on fig.~\ref{relation}b. To get from the l.h.s. to the r.h.s., one can apply the relation of fig.~\ref{relation}a twice, or one can first fuse $L_{\t_1,\,n_1}$ and $L_{\t_2,\,n_2}$, and then apply the relation once. The two ways of reducing the picture should be equivalent, and this fixes the proportionality factor $f$, mentioned above, to be equal to $1-\t_b^{-1}$. This leads to the following OPE,
\beq
L_{\t_1,\,n_1}\times L_{\t_2,\,n_2}=L_{\t_1\t_2,\,n_1+n_2}-L_{\t_1\t_2,\,n_1+n_2-1}\,.\label{OPE1}
\eeq
(Here we absorbed a factor of $\t_b^{-1}$ into the shift $n_1+n_2\rightarrow n_1+n_2-1$.)

Now let us turn to the more subtle case of $\t_1\t_2=1$. Let us write the OPE as
\beq
L_{\t,\,n_1}\times L_{\t^{-1},\,n_2}=L_{P,\,n_1+n_2}\,,\label{OPE2}
\eeq
where $L_{P,\,n}$ is some new line operator, to be determined. Again, assume that the operators $L_{\t,\,n_1}$ and $L_{\t^{-1},\,n_2}$ lie along the $b$-cycle of a solid torus. In the sector $\t_b\ne 1$, the Hilbert space on $T^2$ with this insertion is  one-dimensional, and one can apply the same arguments that we used above. The result is 
\beq
L_{P,\,n_1+n_2}=L_{n_1+n_2}-2L_{n_1+n_2-1}+L_{n_1+n_2-2}\,,\quad \t_b\ne 1\,,\label{Ptne1}
\eeq
where we applied the relation (\ref{limit}) to the OPE (\ref{OPE1}). For $\t_b=1$, the product $L_{\t,\,n_1}\times L_{\t^{-1},\,n_2}$ creates some state $|L_{P,\,n_1+n_2},1\rangle$ in the Hilbert space $\mathcal{H}_{1,1}$. In the ghost sector, this state is $v_1$ (in the notations of sec. \ref{ghosts}), since the singularities in $L_{\t,\,n_1}$ and $L_{\t^{-1},\,n_2}$ do not allow the ghosts to fluctuate. We also need to find, what linear combination of the states (\ref{gaugeH}) of the fermionic gauge fields is created by $L_{P,\,n_1+n_2}$. For that, we note that gluing a solid torus with the operator $L_{P,\,n_1+n_2}$ to an empty solid torus produces $S^1\times S^2$ with two line operators $L_{\t,\,n_1}$ and $L_{\t^{-1},\,n_2}$ along $S^1$. The corresponding invariant is equal to one, so
\beq
\left(|L_{P,\,n_1+n_2},1\rangle,v_0\otimes |0_a\rangle\right)=1\,.
\eeq
On the other hand, if we glue the same tori, but with transformation $\mathcal{S}$ sliced in between, we get a three-sphere with two unlinked unknots $L_{\t,\,n_1}$ and $L_{\t^{-1},\,n_2}$ inside. The invariant for this configuration is zero, so
\beq
\left(|L_{P,\,n_1+n_2},1\rangle,v_0\otimes |0_b\rangle\right)=0\,.
\eeq
From the two equations above, we find that
\beq
|L_{P,\,n_1+n_2},1\rangle=v_1\otimes |0_b\rangle\,.\label{Pt1}
\eeq
Thus, the line operator $L_{P,\,n}$, which can be obtained from the OPE of $L_{\t,\,n_1}$ and $L_{\t^{-1},\,n_2}$, is defined by (\ref{Ptne1}) in the sector $\t_b\ne 1$, and by (\ref{Pt1}) in the sector $\t_b=1$. 

The set of line operators $L_{\t,\,n}$, $L_n$ and $L_{P,\,n}$ for different values of $n$ and $\t\ne 1$ forms a closed operator algebra. The OPEs of operators $L_{P,\,n}$ with themselves and with $L_{\t,\,n}$ follow from (\ref{OPE1}) and (\ref{OPE2}) by associativity.

\subsubsection{A Comment On Indecomposable Representations}
It is convenient to think of the operators $L_{\t,\,n}$ as of Wilson lines, coming from the typical representations of the $\mathfrak{u}(1|1)$ superalgebra, though, of course, this will be literally true only in the U$(1|1)$ theory, and not in $\psu(1|1)$. The OPE (\ref{OPE1}) of these operators agrees with the tensor product decomposition (\ref{repope1}) of the typical representations. For the second OPE (\ref{OPE2}) to agree with (\ref{repope2}), we have to assume that the line operator $L_{P,\,n}$ is actually the Wilson line for the indecomposable representation $P_n$, defined in fig.~\ref{reps}. This statement makes sense already in the $\psu(1|1)$ theory, since $P_n$ is also a representation of $\mathfrak{pl}(1|1)$. In (\ref{Ptne1}) we found that $L_{P,\,n}$ reduces in a special case to a sum of atypical line operators $L_{n}$. Comparing this statement to fig.~\ref{reps}, we see that it agrees with the decomposition that one would expect to happen for the Wilson loop in representation $P_n$. (Recall that Wilson loops in reducible indecomposable representations are naively expected to decompose into sums of Wilson loops for irreducible representations.) But we also note that this decomposition does not hold always. Indeed, if it were true also in the sector $\t_b=1$, the r.h.s. of (\ref{Ptne1}) would tell us that $L_{P,n}$ is identically zero in that sector, which is not correct, since for $\t_b=1$ the operator $L_{P,n}$  actually produces a non-zero state $v_1\otimes|0_b\rangle$. This state can be obtained by inserting the operators $\oint A^+$ and $\oint A^-$, as shown on fig.~\ref{torus}, together with the local operator $\mathcal{O}_1$, to produce the ghost wavefunction $v_1$. It is tempting to speculate that this combination of operators should arise as some point-splitting regularization of the Wilson loop in representation $P_n$, but we do not know, how to make this statement precise.

If the typical operators $L_{\t,\,n}$ are thought of as Wilson lines in the typical representations $(w,n)$, then their limit for $\t\rightarrow 1$ should correspond to Wilson lines in the (anti-)Kac modules $(0,n)_\pm$, introduced on fig.~\ref{reps}. The Torres formula (\ref{limit}) then says that the Wilson loops in these indecomposable representations actually reduce to sums of Wilson loops $L_n$ for the irreducible building blocks of the indecomposables. This statement, again, is true in the sector $\t_b\ne 1$. For $\t_b=1$, one should find some independent way to fix the state in $\mathcal{H}_{1,1}$, produced by the operator $L_{1,\,n}$. More precisely, since there are two different versions $(0,n)_+$ and $(0,n)_-$ of the limit of $(w,n)$ for $w\rightarrow 0$, one would expect that there are two versions $L_{1,n,+}$ and $L_{1,n,-}$ of the operator $\lim_{\t\rightarrow 1}L_{\t,n}$, which produce two different states in $\mathcal{H}_{1,1}$. We are not sure, what these states are.\footnote{One possible guess would be that $L_{1,n,+}$ for $\t_b=1$ is equivalent to $\lambda^{-}\delta^{(2)}(C^-,\bar{C}^+)\oint A^+$, and similarly for $L_{1,n,-}$, with plus and minus indices interchanged. The reason is that this combination is U$(1)_{\rm fl}$-invariant, and depends only on $A^+$, and not on $A^-$, as the Wilson line in representation $(0,n)_+$ should.}.

The general situation with Wilson loops in reducible indecomposable representations is the following. It is consistent to assume that they do split into sums of Wilson loops $L_n$, if the background monodromy $\t_b$ along the knot is non-trivial. When $\t_b=1$, one has to find some independent way to determine, what states in $\mathcal{H}_{1,1}$ they produce. For $P_n$, we used the OPE of two typical operators, and for the (anti-)Kac modules $(0,n)_\pm$, one could possibly use the relation to the degeneration limit of the typical operators. But for general indecomposable representations, there seems to be no natural way to determine the state in $\mathcal{H}_{1,1}$, and therefore it does not make much sense to consider such Wilson loops as separate operators at all.

\subsection{U$(1|1)$ Chern-Simons}\label{Ham11}
Since the U$(1|1)$ theory is the $\ZZ_k$-orbifold of the $\psu(1|1)$ Chern-Simons,  it is completely straightforward to write out its Hamiltonian quantization, once it is known for $\psu(1|1)$. For that, one simply needs to restrict to states with U$(1)_{\rm fl}$-charge divisible by $k$, and to sum over winding sectors. 

For illustration, we consider explicitly the torus Hilbert space. The windings around the two cycles will be labeled by integers $w$ and $w'$, which we take to lie in the range $0\le w,w'\le k-1$. The corresponding holonomies are $\t_w=\exp(2\pi iw/k)$ and $\t_{w'}=\exp(2\pi i{w'}/k)$. Let $\mathcal{H}_{0,0}$ be the $\ZZ_k$-invariant subspace of the $\psu(1|1)$ zero-winding Hilbert space $\mathcal{H}_{1,1}$, and let $\mathcal{H}_{w,w'}\equiv \mathcal{H}_{\t_w,\t_{w'}}$ be the one-dimensional Hilbert spaces in the sectors with windings $w$ and $w'$. The Hilbert space of the U$(1|1)$ theory on $T^2$ is the direct sum $\mathcal{H}_{T^2}=\oplus_{w,w'} \mathcal{H}_{w,w'}$.

To find the states that are created by loop operators $L_{w,\,n}$, $L_n$ and $L_{P,\,n}$, we take corresponding states in the $\psu(1|1)$ theory, set the longitudinal holonomy $\t_b$ to be equal to $\exp(2\pi i w'/k)$, and sum over the winding sectors $w'=0,\dots k-1$. Setting $|w,w'\rangle\equiv|\t_w,\t_{w'}\rangle$, from the equations (\ref{basis}), (\ref{g}), (\ref{g0}), (\ref{Ptne1}) and (\ref{Pt1}) we find
\begin{align}
|L_{w,n}\rangle&=\sum_{w'=0}^{k-1} \exp(2\pi i(n-1/2)w'/k)|w,w'\rangle,\quad w\ne 0\,;\nnr
|L_n\rangle&=v_0\otimes|0_a\rangle+\fr{1}{2i}\sum_{w'=1}^{k-1}\fr{\exp(2\pi inw'/k)}{\sin(\pi w'/k)}|0,w'\rangle\,;\nnr
|L_{P,n}\rangle&=v_1\otimes|0_b\rangle+2i\sum_{w'=1}^{k-1}\sin(\pi w'/k)\exp(2\pi i(n-1)w'/k)|0,w'\rangle\,.
\end{align}
The parameter $n$ is periodic with period $k$, and we take it to belong to the interval $0\le n\le k-1$. If we project out the subspace $\mathcal{H}_{0,0}$, the states $|L_{n}\rangle$ and $|L_{w,n}\rangle$ with $n=0,\dots k-1$, $w=1\dots k-1$, corresponding to a restricted set of irreducible representations, would form a basis in the remaining Hilbert space. This is what one would have in the ordinary, bosonic Chern-Simons theory. In the full Hilbert space $\mathcal{H}_{T^2}$, the states created by the line operators that we have discussed do not form a basis. More precisely, it is not even clear, what one would mean by such a basis, due to the rather weird nature of $\mathcal{H}_{0,0}$.

The bilinear product of states in U$(1|1)$ theory is $1/k$ times the product in the $\psu(1|1)$ theory, where the factor $1/k$ comes from eq.~(\ref{orbifolded}). In particular, we have 
\beq
(|w,w'\rangle,|\tilde{w},\tilde{w}'\rangle)=\fr{1}{k}\delta_{w+\tilde{w}\,{\rm mod}\,k,\,0}\,\delta_{w'-\tilde{w}'\,{\rm mod}\,k,\,0}\,,
\eeq
and therefore
\beq
(|L_{w,n}\rangle,|L_{\tilde{w},\tilde{n}}\rangle)=\delta_{w+\tilde{w}\,{\rm mod}\,k,\,0}\,\delta_{n+\tilde{n}-1\,{\rm mod}\,k,\,0}\,.
\eeq

Let us look at the modular properties of the states, created by the line operators. Under the transformation $\mathcal{T}$, the state $|w,w'\rangle$ transforms into $|w,w'-w\rangle$. The operator $L_{w,\,n}$ thus picks a phase $\exp(2\pi i w(n-1/2)/k)$. The combination $w(n-1/2)$ is the quadratic Casimir for the typical representation $(w,n)$, and the framing factor that we got is what one would expect from the conformal field theory. The operator $L_n$ is invariant under $\mathcal{T}$.
The operator $L_{P,\,n}$ does not transform with a simple phase, but rather is shifted as
\beq
\mathcal{T}|L_{P,n}\rangle = |L_{P,\,n}\rangle+v_1\otimes |0_a\rangle\,.
\eeq
Geometrically, the reason is that the operator, which defines the state $|0_b\rangle$, is given by integration of $A^+$ and $A^-$ over the contours of fig.~\ref{torus}. Under the $\mathcal{T}$-transformation, the winding number of the two contours changes. We note that in the sector $\mathcal{H}_{0,0}$ the operator $\mathcal{T}$ is not diagonalizable. This is the signature of the logarithmic behavior of the CFT, which presumably corresponds to our Chern-Simons theory.

Under the modular transformation $\mathcal{S}$, the state $|L_{w,\,n}\rangle$ changes into $\sum_{R'}S_{w,n}^{R'}|L_{R'}\rangle$ with
\begin{align}
S_{w,n}^{w'n'}&=\fr{1}{k}\exp(-2\pi i((n-1/2)w'+(n'-1/2)w)/k)\,,\label{typ1}\\
S_{w,n}^{n'}&=\fr{2i\sin(\pi w/k)}{k}\exp(-2\pi i n'w/k)\,.\label{typ2}
\end{align}
The other line operators transform as $\mathcal{S}|L_n\rangle=v_0\otimes|0_b\rangle+\sum_{R'}S^{R'}_n|L_{R'}\rangle$ and 
$\mathcal{S}|L_{P,n}\rangle=-v_1\otimes|0_a\rangle+\sum_{R'}S^{R'}_{P,n}|L_{R'}\rangle$, with
\begin{align}
S_{n}^{w',n'}&=-\fr{1}{2ik\sin(\pi w'/k)}\exp(-2\pi i n w'/k)\label{atyps1}\,,\\
S_{P,n}^{w',n'}&=-\fr{2i\sin(\pi w'/k)}{k}\exp(-2\pi i(n-1)w'/k)\label{atyps2}
\end{align}
Modular transformations very similar to (\ref{typ1})-(\ref{atyps2}) were previously derived in the U$(1|1)$ WZW model in \cite{RS2}. There are, however, some differences. The transformations most similar to ours, but with $\mathcal{H}_{0,0}$ part omitted, are called ``naive'' in that paper. A slightly different version of transformations is derived using a particular regularization, whose role is essentially to avoid dealing with $\mathcal{H}_{0,0}$. (The Chern-Simons interpretation of this regularization is explained on fig.~11-12 of that paper.) We will not attempt to rederive the modular transformations with the regularization of \cite{RS2}, since in our approach a regularization is not needed.

\section{Some Generalizations}\label{gen}
In this section, we make some brief comments on supergroup Chern-Simons theories other than $\psu(1|1)$ or U$(1|1)$. Many of the facts that are presented here have appeared previously in \cite{MW}. The reason we decided to make this summary is that in \cite{MW} the focus was not on the three-dimensional, but on the analytically-continued version of the theory. Here we would also like to emphasize the importance of coupling to a background flat bundle. Our understanding of the supergroup Chern-Simons theories is very limited, and this section will contain more questions than answers.

\subsection{Definition And Brane Constructions}\label{compact}
To define a supergroup Chern-Simons theory, one needs to choose a complex Lie superalgebra\footnote{For a brief review of Lie superalgebras, with a view towards Chern-Simons theory, the reader can consult section 3.1 of \cite{MW}.} $\mathfrak{g}$, which possesses a non-degenerate invariant bilinear form. The bosonic and the fermionic parts of $\mathfrak{g}$ will be denoted by $\mathfrak{g}_\bos$ and $\mathfrak{g}_\ferm$, respectively. One also needs to choose a real form $\mathfrak{g}_\bos^\RR$ for $\mathfrak{g}_\bos$, and a global form $G_\bos$ for the corresponding ordinary real Lie group\footnote{One could also imagine defining a complex supergroup Chern-Simons theory, in which the bosonic gauge fields would be valued in the complex Lie algebra $\g_\bos$, and the fermions~--~in two copies of $\g_\ferm$. More generally, it should be possible to define quivers of supergroup Chern-Simons theories, as mentioned in sec.~2.2.6 of \cite{MW}.}. A real form for the whole superalgebra $\mathfrak{g}$ is not needed. The action of the theory is the usual Chern-Simons action, except that the gauge field is a sum of an ordinary $\g_\bos^\RR$-valued gauge field and a Grassmann $\mathfrak{g}_\ferm$-valued one-form. The action is multiplied by a level $k$, whose quantization condition is determined by the global form $G_\bos$, as in the usual Chern-Simons theory. More precisely, the fermionic part of the action can have a global anomaly, in which case the quantization condition for $k$ should be shifted by $1/2$, to cancel the anomaly. To state exactly what we mean by $k$, we 
have to specify the regularization scheme. In flat space, one can make the path-integral absolutely convergent by adding a Yang-Mills term, at the expense of breaking the supersymmetry from $\cN=4$ to $\cN=3$. The Chern-Simons level then receives no one-loop renormalization. By $k$ we mean this ``quantum-corrected'' level\footnote{Note that different notations were used in \cite{MW}. What we call $k$ here is equal to what was called $\calK$ in that paper.}. An equivalent definition of $k$ is by a brane construction, which is presented below. 
On a curved space, the correct treatment of the theory at one-loop is not entirely clear (see {\it e.g.} Appendix~E of \cite{MW}.)

By analogy with the ordinary Chern-Simons theory, one can define an ``uncorrected'' level $k'$ by
\beq
k=k'+|h_{\mathfrak{g}}|\,{\rm sign}(k')\,,\label{qcor}
\eeq
where $h_\mathfrak{g}$ is the dual Coxeter number of the superalgebra. One expects that this $k'$ is the level of the current algebra, which one would find in the Hamiltonian quantization of the theory, but that remains to be shown. We note that, while $k$ can be a half-integer, with definition (\ref{qcor}) $k'$ is always an integer.

Completely analogously to the $\psu(1|1)$ case, the fermionic part of the gauge symmetry can be globally gauge-fixed. This introduces $\g_\ferm$-valued bosonic superghost $C$ and antighost $\bar{C}$, as well as a fermionic $\g_\ferm$-valued Lagrangian multiplier $\lambda$. Observables of the topological theory are then in the cohomology of a BRST charge $\Qb$. This partial gauge-fixing procedure for supergroup Chern-Simons was first described in \cite{KapustinSaulina}.

\begin{figure}
 \begin{center}
   \includegraphics[width=16cm]{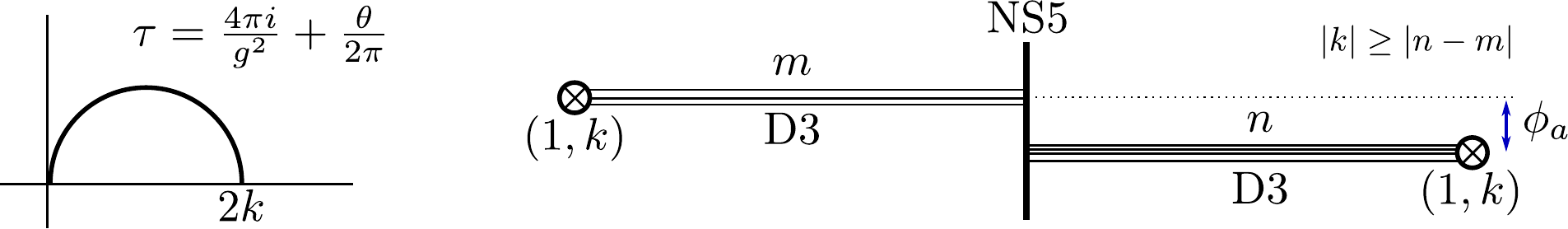}
 \end{center}
 \caption{\small Brane construction for an $\cN=4$ Gaiotto-Witten theory. The complexified type~IIB string coupling should belong to a semicircle of radius $k$, as shown on the left. The relative displacement $\phi_a$ of the two $(1,k)$-branes is the SU$(2)_X$-triplet of masses. The D3-branes are shown slightly displaced along the direction of the NS5-brane just for clarity of the picture.}
  \label{branes1}
\end{figure}
\begin{table}[t]
 \beq
 \begin{array}{l|c|c|c|c|c|c|c|c|c|c}\toprule
               &0&1&2&3&4&5&6&7&8&9                               \\ \hline
           {\rm D3}  &\checkmark&\checkmark&\checkmark&(\checkmark)&-&-&-&-&-&-         \\ \hline
           {\rm NS5} &\checkmark&\checkmark&\checkmark&-&\checkmark&\checkmark&\checkmark&-&-&- \\ \hline
           (1,k)&\checkmark&\checkmark&\checkmark&-&-&-&-&\checkmark&\checkmark&\checkmark \\ \bottomrule
  \end{array}\nonumber
 \eeq
 \caption{\label{directions} \small Details on the brane configuration of fig.~\ref{branes1}. The D3-branes span a finite interval in the third direction. The R-symmetry groups SU$(2)_X$ and SU$(2)_Y$ act on the directions 456 and 789, respectively.}
 \end{table}
As was found in \cite{KapustinSaulina}, supergroup Chern-Simons theories can be obtained by topological twisting from the $\cN=4$ Chern-Simons-matter theories of \cite{Janus}. For unitary and orthosymplectic gauge groups, the latter can be engineered in type~IIB string theory by brane constructions \cite{srule1}, \cite{ABJM}, \cite{ABJ}. For the U$(m|n)$ theory, the brane configuration is shown on fig.~\ref{branes1}. Table~\ref{directions} shows, in which directions the branes are stretched. For eight supersymmetries to be preserved, the complexified type~IIB coupling should lie on a semicircle of radius $k$, as shown on the left of fig.~\ref{branes1}. The coupling constant can thus be of order $g^2\sim 1/k$, so, the theory has a well-defined perturbative expansion in $1/k$. The level $k$ for U$(m|n)$ should satisfy the generalized s-rule condition $|k|\ge|n-m|$, and otherwise the theory breaks supersymmetry \cite{HW}, \cite{srule1}, \cite{srule2}, \cite{srule3}. One can also turn on an SU$(2)_X$-triplet of masses $\phi_a$, which correspond in the brane picture to the relative displacement of the $(1,k)$-branes in directions 456, as shown on fig.~\ref{branes1}. For this deformation to preserve supersymmetry, the generalized s-rule requires $|k|\ge{\rm max}(m,n)$.

\begin{figure}
 \begin{center}
   \includegraphics[width=13cm]{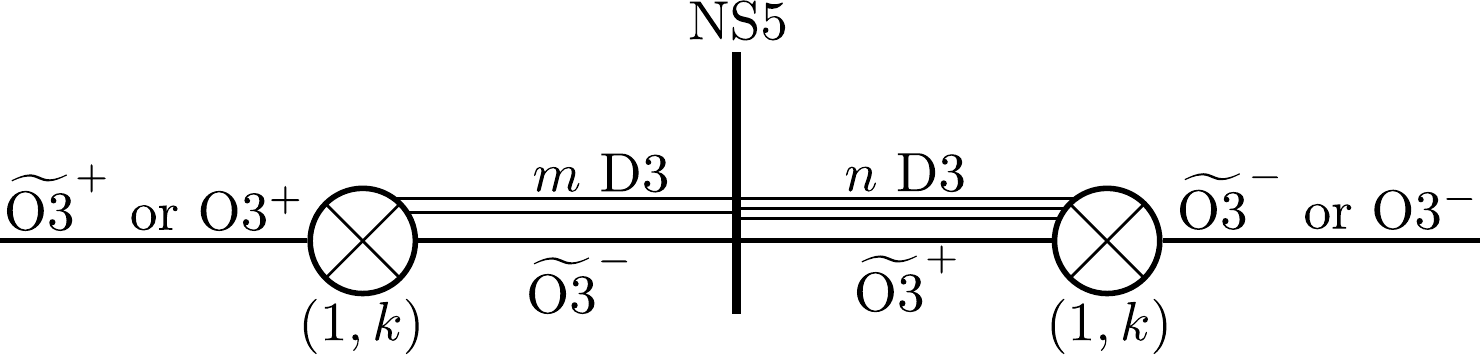}
 \end{center}
 \caption{\small The brane construction for the $\cN=4$ Gaiotto-Witten theory, which upon twisting would give OSp$(2m+1|2n)$ Chern-Simons. The leftmost and rightmost orientifold planes are $\tilde{{\rm O}3}^\pm$, if $k$ is even, and $O3^\pm$, if $k$ is odd.}
  \label{osp1}
\end{figure}

Let us also discuss brane construction for the orthosymplectic theories. For that, we add an orientifold three-plane to the configuration of fig.~\ref{branes}. 
(For a review of orientifold planes, see \cite{Baryons}, \cite{HananyOrientifolds}, or section 5.1 of \cite{MW}.) Recall that the orientifold three-planes have two  $\ZZ_2$-charges, one of which is usually denoted by plus or minus, and the other by a tilde. Upon crossing a $(p,q)$-fivebrane, the type of the orientifold changes: if $p$ mod $2\ne0$, then plus is exchanged with minus, and if $q$ mod $2\ne0$, then the tilde is added or removed. A possible configuration is shown on fig.~\ref{osp1}. In the interval between the two $(1,k)$-fivebranes, the gauge group is O$(2m+1)$ on the left and Sp$(2n)$ on the right. The leftmost and rightmost orientifold planes on the figure have a tilde, if $k$ is even, and do not have it, if $k$ is odd. If the $\widetilde{{\rm O}3}^-$-plane would appear on the far right, the theory would have an extra three-dimensional hypermultiplet, coming from the fundamental strings that join the D3-branes and the $\widetilde{{\rm O}3}^-$-plane. That would give a theory different from what we want. Therefore, we have to take $k$ to be an odd integer. In the OSp$(2m+1|2n)$ Chern-Simons, we normalize the action to be
\beq
\fr{k_{\osp}}{4\pi}\int\Str \left(A{\rm d}A+\fr{2}{3}A^3\right)\,,\label{ospact}
\eeq
where $\Str$ is the supertrace in the fundamental representation of the supergroup. Here $k_\osp=k/2$, where the factor of $1/2$ comes from the orientifolding. Let us call a bosonic Chern-Simons term canonically-normalized, if it transforms by arbitrary multiples of $2\pi$ under large gauge transformations, assuming that the gauge group is connected and simply-connected. With the normalization (\ref{ospact}), the level $k_\osp$ multiplies the canonically-normalized Chern-Simons term for the Sp$(2n)$ subgroup, and twice the canonically-normalized action\footnote{More precisely, this is true for $m>1$. For $m=1$, it is four times the canonically-normalized action.} for Spin$(2m+1)$. From what we have said about the brane configuration, we see that $k$ is odd, and thus $k_{\osp}\in 1/2+\ZZ$. Therefore, the Sp$(2n)$ part of the bosonic action is anomalous under large gauge transformations. But that precisely compensates for the anomaly for $2m+1$ hypermultiplets in the fundamental of Sp$(2n)$, so, the theory is well-defined. For any supergroup Chern-Simons theory, one expects the analog of the s-rule to be  $|k_\g|\ge|h_{\g}|$. This is equivalent to the requirement that $k_\g'$, as defined in (\ref{qcor}), does exist. For OSp$(2m+1|2n)$, this condition reads as $|k|\ge |2(n-m)+1|$.

\begin{figure}
 \begin{center}
   \includegraphics[width=13cm]{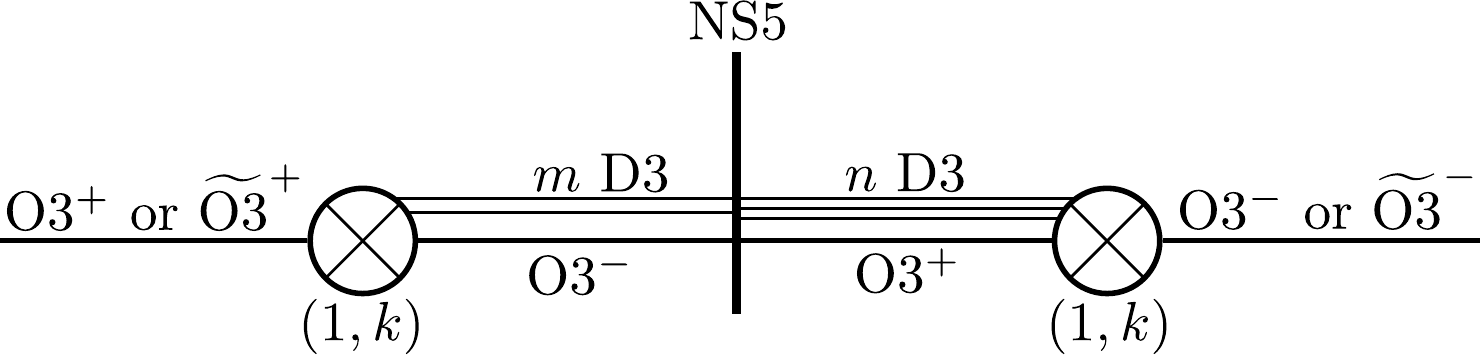}
 \end{center}
 \caption{\small The brane construction for the $\cN=4$ Gaiotto-Witten theory, which upon twisting would give OSp$(2m|2n)$ Chern-Simons. The leftmost and rightmost orientifold planes are $O3^\pm$, if $k$ is even, and $\tilde{{\rm O}3}^\pm$, if $k$ is odd.}
  \label{osp2}
\end{figure}

For the even orthosymplectic group OSp$(2m|2n)$, the brane configuration is shown on fig.~\ref{osp2}. To avoid having an $\widetilde{{\rm O}3}^-$-plane and an extra hypermultiplet, this time we have to take $k$ to be even, and therefore $k_{\osp}=k/2$ is an arbitrary integer, consistently with the fact that the fermionic determinant has no global anomaly. The generalized s-rule is $|k|\ge 2|n-m+1|$.

\subsection{Some Properties}
Importantly, for Lie superalgebras there exist automorphisms, which commute with the bosonic subalgebra. For the so-called type I superalgebras, the group of these automorphisms is U$(1)$. Type~I superalgebras are $\mathfrak{gl}(m|n)$, together with the subquotients $\mathfrak{sl}$ and $\mathfrak{psl}$, and the orthosymplectic superalgebras $\osp(2|2n)$. The fermionic part $\g_\ferm$ for type I decomposes under the action of $\g_\bos$ into a direct sum of two representations. The U$(1)$-automorphism acts on them with charges $\pm1$. For superalgebras of type II, which are all the other superalgebras, the relevant group of automorphisms is only $\ZZ_2$. It acts trivially on $\g_\bos$, and flips the sign of elements of $\g_\ferm$. In Chern-Simons theory, one can use these automorphisms to couple the theory to a background flat connection. For type I, this can be a complex flat line bundle, just as we found for $\psu(1|1)$ and U$(1|1)$. The partition function of the theory depends on the background complex flat connection holomorphically. In flat space, the imaginary part of the background flat connection can be identified with the SU$(2)_X$-triplet of masses, mentioned above. For a theory with a type~II superalgebra, the background bundle can only be a $\ZZ_2$-bundle. Equivalently, one can assign antiperiodic boundary conditions around various cycles of the three-manifold for the $\g_\ferm$-valued fields.

Line observables of the supergroup Chern-Simons theory include Wilson lines in various representations of the supergroup, as well as vortex operators, which are expected \cite{Elitzur}, \cite{MW} to be equivalent to the Wilson lines, at least modulo $\Qb$. One can also construct twist line operators by turning on a singular holonomy for the background flat gauge field, as we did in simple examples in the present paper. For special values of the holonomy, those operators can be equivalent to ordinary vortex operators.

Consider the theory on $\RR^3$, or other space with three non-compact directions, and assume that the background flat bundle was turned off. It is then possible to give vevs to the scalar superghost fields $C$ and $\bar{C}$ and to partially Higgs the theory. For example, the U$(m|n)$ gauge supergroup can in this way be reduced down to U$(|n-m|)$. (In the brane picture, this corresponds to recombining a number of D3-branes and taking them away from the NS5-brane in the directions 789.) Since the superghosts appear only in $\Qb$-exact terms, this procedure does not change the expectation values of observables in the $\Qb$-cohomology. By this Higgsing argument one can see that the expectation values of Wilson loops vanish for almost all representations, except for the maximally-atypical ones. The classes of maximally-atypical representations are in a natural correspondence with representations of U$(|n-m|)$, and the Wilson loops in those representations reduce to Wilson loops of the ordinary, bosonic U$(|n-m|)$ Chern-Simons theory upon Higgsing. Thus, on $\RR^3$ the U$(m|n)$ supergroup theory does not produce new knot invariants. (A similar story holds for other supergroups\footnote{Almost all supergroup Chern-Simons theories can be reduced in this way to bosonic Chern-Simons. One exception is the series OSp$(2m+1|2n)$, which can be Higgsed only to OSp$(1|2n)$. However, one finds \cite{MW} that the analytically-continued version of OSp$(1|2n)$ Chern-Simons is dual to the ordinary Chern-Simons with gauge group O$(2n+1)$.}.) It is however interesting to turn on a background flat bundle, which in flat space means just a constant SU$(2)_X$-triplet of mass terms. Looking at the brane picture, one would expect that for large $\phi_a$ the U$(m|n)$ theory would reduce to ${\rm U}(m)\times{\rm U}(n)$ Chern-Simons. If this were true, then, in particular, we would have a knot invariant, which interpolates between the U$(|n-m|)$ and the ${\rm U}(m)\times{\rm U}(n)$ invariants. This is certainly very puzzling. Unfortunately, we cannot test this in the simple examples considered in this paper, since the atypical representations of U$(1|1)$ do not produce non-trivial knot invariants.

On a compact closed three-manifold, the theory has both bosonic and fermionic zero modes. To get a well-defined invariant, one needs to turn on a background flat bundle. The partition function is then a holomorphic function thereof. Alternatively, one can insert loops with vortex operators. As discussed in \cite{MW}, to remove all the zero modes by a single vortex operator, it has to be labeled by a typical weight of the superalgebra.

\subsection{Dualities}\label{branes}
A reader familiar with the brane construction of the analytically-continued Chern-Simons theory \cite{5knots}, \cite{MW} would notice its similarity to the configuration of fig.~\ref{branes1}. If we moved the $(1,k)$-branes along the third direction away to infinity, we would recover precisely the configuration studied in \cite{5knots} and \cite{MW}. In the language of the analytically-continued theory, the role of the $(1,k)$-branes is to choose the real integration contour for the path-integral. Indeed, the fluctuations of the D3-branes in the directions 456 are described in the 4d $\cN=4$ Yang-Mills theory by three components of the adjoint-valued scalar field. Upon twisting, those become the imaginary part of the gauge field of the analytically-continued Chern-Simons theory. At the positions of the $(1,k)$-branes these fields are set to zero, which means that we are working with the real integration contour.

\begin{figure}
 \begin{center}
   \includegraphics[width=11cm]{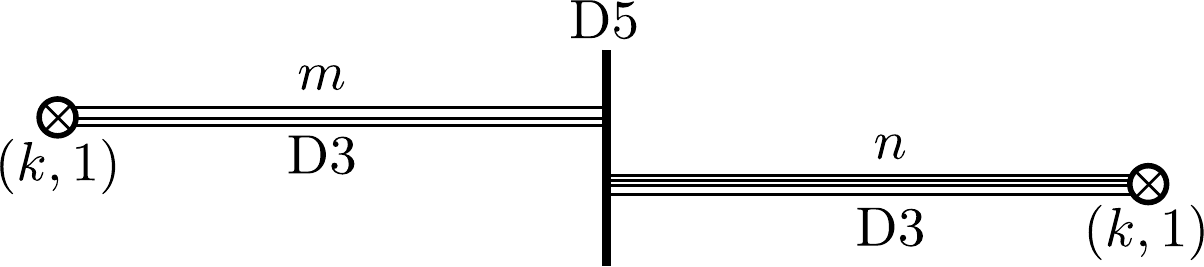}
 \end{center}
 \caption{\small The S-dual of the brane configuration, which describes the U$(m|n)$ Chern-Simons theory.}
  \label{k1brane}
\end{figure}
Having a brane construction, one can apply various string theory dualities. In the analytically-continued Chern-Simons, it has been shown that the S-dual theory gives a new way to compute the Chern-Simons invariants \cite{5knots}, \cite{GaiW}. One might ask, whether we can obtain anything useful by considering the S-dual of our configuration of fig.~\ref{branes1}, which is shown on fig.~\ref{k1brane}. Unfortunately, this does not seem to be the case, beyond the duality for the $\psu(1|1)$ and U$(1|1)$ theory, which has been considered in previous sections.

\begin{figure}
 \begin{center}
   \includegraphics[width=4.3cm]{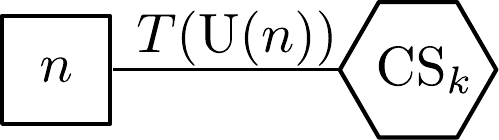}
 \end{center}
 \caption{\small The interaction of $n$ D3-branes with a $(k,1)$-brane is described by coupling the D3-brane gauge fields to the $T({\rm U}(n))$ theory via the U$(n)$ symmetry of the Higgs branch of $T({\rm U}(n))$, and gauging the Coulomb branch of $T({\rm U}(n))$ with a U$(n)$ Chern-Simons gauge field at level $k$.} \label{k1separately}
\end{figure}

The problem is that the S-dual configuration of fig.~\ref{k1brane} contains D3-branes ending on $(k,1)$-fivebranes. The low energy field theory for such a ``tail'' has been described in \cite{GWfour}, and is shown on fig.~\ref{k1separately}. The U$(n)$ gauge theory of $n$ D3-branes is coupled to the Higgs branch of the three-dimensional theory $T({\rm U}(n))$, the Coulomb branch of which is gauged by a level $k$ Chern-Simons gauge field. The $T({\rm U}(n))$ theory with non-abelian symmetries of the Coulomb branch gauged does not have a Lagrangian description, and therefore the configuration of fig.~\ref{k1brane} does not seem to be particularly useful for the purpose of studying supergroup topological invariants. 

More precisely, there exists one case, where gauging the Coulomb branch of $T({\rm U}(n))$ is easy \cite{GWfour}~---~namely, $n=1$. Using the description of this case in \cite{GWfour}, one can readily see that the configuration of fig.~\ref{k1brane} for $m=n=1$ gives the mirror of U$(1|1)$ Chern-Simons, which was considered in section~\ref{mirroru11}.
\begin{figure}
 \begin{center}
   \includegraphics[width=13cm]{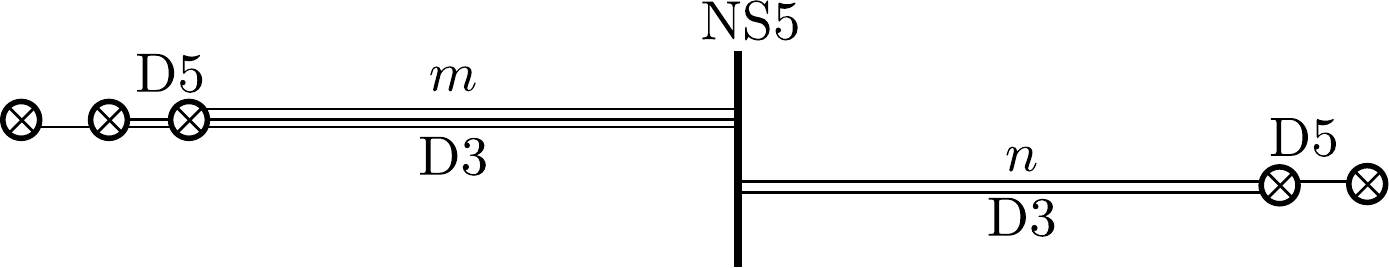}
 \end{center}
 \caption{\small A brane configuration, which produces a free U$(n)\times$U$(m)$ bifundamental hypermultiplet. There are $m$ and $n$ D5-branes on the left and on the right, arranged so as to impose the Dirichlet boundary condition in the 4d $\cN=4$ Yang-Mills theory.} \label{dirichlet}
\end{figure}

\begin{figure}
 \begin{center}
   \includegraphics[width=12.5cm]{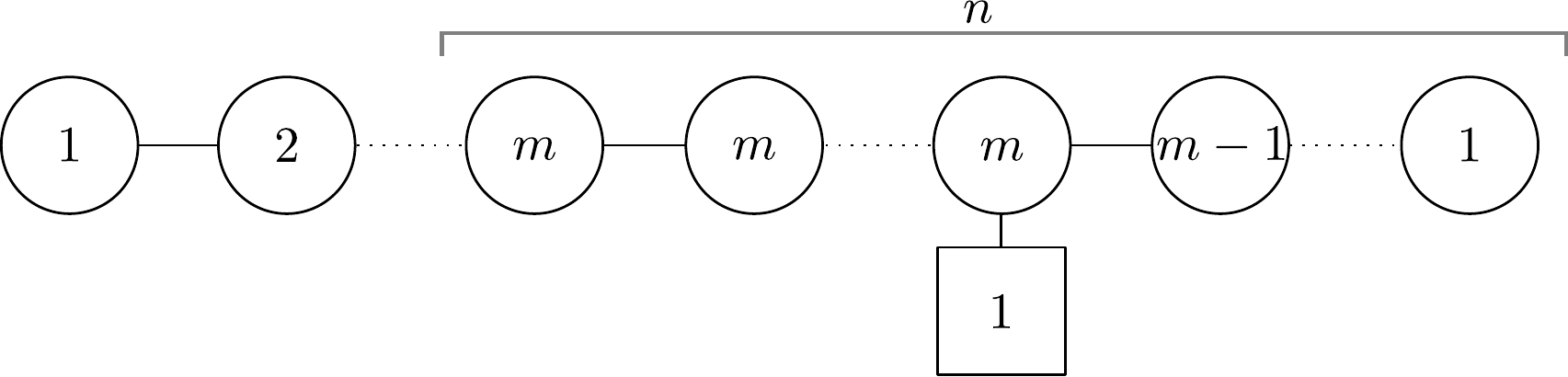}
 \end{center}
 \caption{\small A quiver gauge theory, which is obtained by S-duality and a sequence of Hanany-Witten moves from the brane configuration of fig.~\ref{dirichlet}.. We follow the notations of \cite{GWfour}: the circles denote unitary gauge groups, the square is the fundamental hypermultiplet, and connecting lines are bifundamental hypermultiplets.}
  \label{quiver}
\end{figure}

One can alternatively view the mirror transformation of the U$(m|n)$ theory as follows. We represent the bifundamental hypermultiplet of the U$(m|n)$ theory as the IR limit of the Coulomb branch of some UV theory, and then couple it to bosonic Chern-Simons gauge fields. The relevant UV theory can be found by replacing the $(1,k)$-fivebranes on fig.~\ref{branes1} by a bunch of D5-branes, so as to impose the Dirichlet boundary condition (see fig.~\ref{dirichlet}), and then applying the S-duality and making some Hanany-Witten moves. (For $n=m=1$, this procedure would give the $\psu(1|1)$ theory and its mirror.) The resulting UV theory is given by the quiver of fig.~\ref{quiver}. It is an ``ugly'' quiver, in the terminology of \cite{GWfour}. As demonstrated in section 2.4 of that paper, it has $nm$ monopole operators, which in the IR give rise to $nm$ free hypermultiplets, as expected.

Again, this description is not useful for non-abelian supergroup Chern-Simons theories, since the non-abelian symmetry of the Coulomb branch of the quiver emerges only in the IR. We can nevertheless play a game similar to what we did for the single hypermultiplet. We can couple the quiver theory to $n+m-1$ flat GL$(1)$ gauge fields, using the dual photon translation symmetries and FI terms of the UV theory. On the one hand, it is clear from the IR theory that the resulting invariant is a product of $nm$ abelian torsions. On the other hand, it can be computed by solving non-abelian Seiberg-Witten equations\footnote{Those equations are completely analogous to the abelian ones, and are written out in Appendix\ref{magdetails}.} for the quiver of fig.~\ref{quiver}. One expects that the solutions to those equations, in the limit of large FI terms, can be obtained by embedding $nm$ solutions of the abelian equations, so as to reproduce a product of abelian torsions. Since, anyway, this invariant does not produce anything new, we will not consider it in more detail.
 
 There is one last case, where the mirror symmetry can be useful for supergroup Chern-Simons. This is when the level $k$ is equal to one. The reason is that a $(1,1)$-fivebrane can be related by S-duality, say, to a D5-brane, while preserving the NS5-brane in the configuration of fig.~\ref{branes1}. The generalized s-rule requires in this case that $|n-m|\le 1$. By applying a further S-duality, the theory can be mapped to an $\cN=4$ Yang-Mills with no matter or with a single fundamental hypermultiplet. In this way, {\it e.g.}, the U$(n|n)$ Chern-Simons theory at level one would be related to the non-abelian U$(n)$ Seiberg-Witten equations. The problem, however, is that the s-rule in this case does not allow us to turn on a background flat bundle, except for the case of the U$(1|1)$ theory. Therefore, even if the mirror theory does compute some non-trivial invariant, it will not be computable in the U$(m|n)$ supergroup Chern-Simons. It is possible that in the orthosymplectic OSp$(2m+1|2n)$ case the situation is better, and one can turn on a background $\ZZ_2$-bundle and get a non-trivial duality of invariants, but we will not explore this here.

\section*{Acknowledgments}
I thank E.~Witten for many helpful discussions and for collaboration on a related project \cite{MW}. I also thank N.~Dedushenko and J.~Gomis for useful discussions. I am supported by the Charlotte Elizabeth Procter Fellowship.

\appendix
\section{Details On The $\cN=4$ QCD}\label{magdetails}
Here we describe the fields, the BRST transformations and the Lagrangian for the topologically twisted $\cN=4$ SQCD with one fundamental flavor. The bosonic fields of the theory are the gauge field $A$, the triplet of scalars, which we write as a complex scalar $\sigma$ and a real field $Y_1$, and the hypermultiplet scalar fields, which upon twisting become a spinor $Z^\alpha$. The fermions of the vector multiplet transform in the $(2,2,2)$ representation of the Lorentz and R-symmetry groups, and upon twisting produce fermionic scalars $\eta$ and $\tilde{\psi}$ of ghost numbers $-1$ and $+1$, a one-form $\psi$ of ghost number $+1$, and a two-form $\chi$ of ghost number $-1$. The fermions of the hypermultiplet after twisting remain spinors, and will be denoted by $\zeta_u$ (of ghost number $+1$) and $\zeta_v$ (of ghost number $-1$). 

The BRST transformations of the fields can be obtained by dimensional reduction\footnote{Our notations slightly differ from \cite{MW} in that here the adjoint-valued fields are Hermitian. The covariant differential is $d_A=d+iA$.} from the formulas of \cite{MW},
\beqn
&&\delta A=\psi\,,\quad\delta\sigma=0\,,\quad\delta\bar\sigma=\eta\,,\quad\delta Y_1=\tilde\psi\,,\quad \delta Z=\zeta_u\nnr
&&\delta\eta=i[\sigma,\bar{\sigma}]\,,\quad\delta\psi=-d_A\sigma\,,\quad\delta\tilde\psi=i[\sigma,Y_1]\,,\quad \delta\chi=H\,,\quad \delta\zeta_u=i\sigma Z\,,\quad\delta\zeta_v=f\,.\nonumber
\eeqn
Here $\mu=iZ^\alpha\otimes \bar{Z}_\beta\sigma_\alpha^\beta$ is the moment map, and $H$ and $f$ are auxiliary fields. The equations of motion set $H=F+\star \left(d_A Y_1+\fr{1}{2}\mu-\fr{1}{2}e^2\phi\right)$ and $f=\slashed{D}Z+iY_1Z$, and the Seiberg-Witten equations are
\beqn
&&F+\star \left(d_A Y_1+\fr{1}{2}\mu-\fr{1}{2}e^2\phi\right)=0\,,\label{SW11}\\ &&\slashed{D}Z+iY_1Z=0\,.\label{SW12}
\eeqn
The FI one-form $\phi$ is valued in the center of $\u(n)$. Here are a couple of useful identities,
\beqn
&&\int\rmd^3x\sqrt{\gamma}\left(D_i\bar{Z}_\alpha D^iZ^\alpha+\bar{Z}_\alpha\left(Y_1^2+\fr{1}{4}R\right)Z^\alpha\right)\nnr
&=&\int{\rm d}^3x\sqrt{\gamma}\,|f|^2-\int\rmd^3x\sqrt{\gamma}\,\tr\left(Y_1D_i\mu^{i}\right)+\int\rmd^3x\,\tr\left(F\wedge\mu\right)\,,\\\label{lichnerowicz}
&&\int{\rm d}^3x\sqrt{\gamma}\,\tr\left(\fr{1}{2}F^2_{ij}+(D_iY)^2+\fr{1}{4}(\mu_i-e^2\phi_i)^2-e^2\phi^iD_iY_1\right)\nnr
&=&\int\tr\,(H\wedge\star H)+\int\tr\,(F\wedge e^2\phi-F\wedge\mu)+\int{\rm d}^3x\sqrt{\gamma}\,\tr\,(Y_1D_i\mu^i)\,.
\eeqn
where $R$ is the scalar curvature. These identities allow to rewrite the SQCD action in the form (\ref{imprecise0})-(\ref{imprecise}). (Our normalization of the coupling constant is such that the gauge field kinetic term is $\int\tr\,F_{ij}^2/4\pi e^2$.)

The action of the twisted theory in general contains the term $Y_1D^i\phi_i$, which breaks the SU$(2)_Y$-symmetry. This, in fact, is the same term that we saw in section \ref{simplest} in the electric theory. If ${\rm d}\star\phi=0$, the symmetry is restored. For an irreducible solution of the Seiberg-Witten equations, one then has $Y_1=\sigma=\bar{\sigma}=0$, and the equations (\ref{SW11})-(\ref{SW12}) can be simplified to (\ref{SW}). For a more general $\phi$, the field $Y_1$ is non-zero and can be found by applying $d_A$ to the equation (\ref{SW11}).

We focused on the QCD with one fundamental flavor, but this twisting procedure generalizes in an obvious way to an arbitrary quiver theory with vector multiplets and hypermultiplets.

\section{Boundary Conditions Near A Line Operator}\label{bc}
In general, in giving a definition of a disorder operator, one needs to specify the boundary conditions for the fields near the singularity, to ensure that the Hamiltonian in presence of the operator remains self-adjoint. (A closely related condition is that in Euclidean signature the kinetic operator of the fields should remain Fredholm.) For that, the boundary conditions should satisfy two requirements. First, to verify the Hermiticity of the Hamiltonian, one integrates by parts, and the boundary term should vanish. Second, the boundary conditions should set to zero half of the modes near the boundary. Here we would like to sketch these boundary conditions for our disorder operators $L_{\mathbf{t},\,n}$, since we use them explicitly in section~\ref{punctures}. (Note that sometimes in similar problems there exist families of possible boundary conditions, and this leads to important physical consequences \cite{leak1}, \cite{leak2}. In our case, nothing like this happens.)

We consider an operator $L_{\mathbf{t},\,n}$, stretched along a straight line in $\RR^3$. The coordinate along the operator will be denoted by $t$ and will be treated as time, and the polar coordinates in the transverse plane will be denoted by $r$ and $\theta$. For the background gauge field, we choose the gauge in which $B$ is zero, but fields with positive U$(1)_{\rm fl}$-charge are multiplied by $\mathbf{t}$ in going around the operator. We assume $\mathbf{t}$ to be unimodular and write it as $\mathbf{t}=\exp(2\pi ia)$, with $a\in(0,1)$. 

For the scalar field $C^+$, we want to impose a boundary condition with which the two-dimensional Laplacian $\Delta$ would be self-adjoint. The field can be expanded in modes of different angular momentum $\ell$, valued in $a+\ZZ$. Near $r=0$, the modes with angular momentum $\ell$ behave like $r^{\pm|\ell|}$. We impose the boundary condition $C|_{r\rightarrow 0}=0$. It actually implies that $C$ vanishes at least as $r^{{\rm min}(a,1-a)}$. This boundary condition has the required properties.

The $\Qb$ transformations act as
\beq
\delta A_0=-\partial_t C\,,\quad \delta A=-{\rm d}C\,,\quad \delta \bar{C}=\lambda\,,
\eeq
where we separated the fermionic gauge field into its time component $A_0$ and components $A$ in the transverse plane. The boundary condition for the fermions, which is compatible with vanishing of $C$ and with $\Qb$-invariance, is to require that $\lambda$ and $A_0$ vanish at $r=0$, and that $A$ is less singular than $1/r$, in an orthonormal frame. Then, in fact, the fields $\lambda$, $A_0$ and $rA$ vanish at least as $r^{{\rm min}(a,1-a)}$, and are square-integrable. The fermionic Hamiltonian is the operator $d+d^*$ in two dimensions, acting on the field $\cA=A_0+A+\star\lambda$, where $\star$ is the 2d Hodge operator. It is easy to see (on the physical level of rigor) that with our boundary condition the Hamiltonian is self-adjoint. If $z=r\exp(i\theta)$ is the complex coordinate, then the operator reduces to 
\beq
\left(\begin{array}{cc}0&-\bar\partial\\ \partial & 0\end{array}\right)\,,
\eeq
acting on the doublet $\left((A_0+i\lambda)/2, A_z\right)$, plus a similar operator for the other pair of fields $\left((A_0-i\lambda)/2, A_{\bar{z}}\right)$. In verifying the Hermiticity of this operator, the boundary term in the integration by parts vanishes. The boundary condition sets to zero a minimal possible number of modes, so one expects that the operator is not only Hermitian, but is self-adjoint.

\section{Skein Relations For The Multivariable Alexander Polynomial}\label{multiskein}
Here we derive two skein relations for the multivariable Alexander polynomial, which are known \cite{Murakami} to define it completely, together with the skein relation of fig.~\ref{halfskein}, the normalization (\ref{unknot}), the formula of fig.~\ref{relation}a, and the fact that the invariant is zero for a disjoint link.

\begin{figure}
 \begin{center}
   \includegraphics[width=9cm]{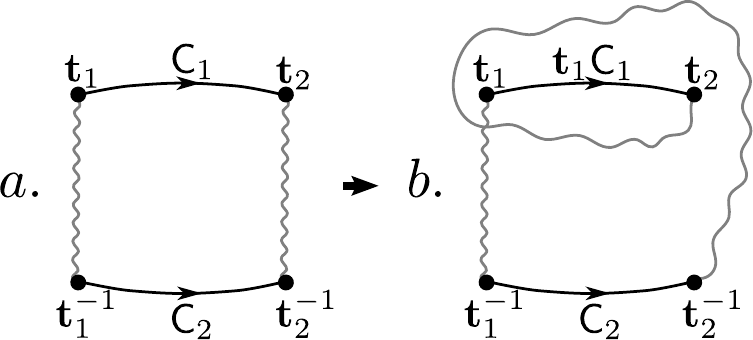}
 \end{center}
 \caption{\small {\large a.}~Two cuts and two basis contours for a four-punctured sphere. {\large b.}~The result of the braiding transformation. The contour $\mathsf{C}_1$ got multiplied by $\mathbf{t}_1$ in crossing the left cut. It will also get a factor of $\mathbf{t}_2$, when the right cut is moved back to its place.}
  \label{contours4}
\end{figure}
\begin{figure}
 \begin{center}
   \includegraphics[width=12cm]{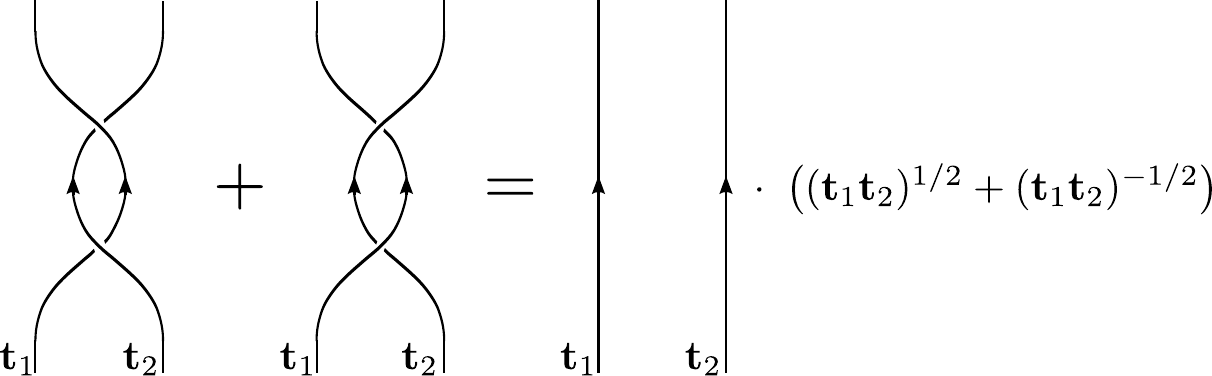}
 \end{center}
 \caption{\small A skein relation for the multivariable Alexander polynomial.}
  \label{skein}
\end{figure}

Consider the case of two strands, labeled by holonomies $\t_1$ and $\t_2$. The sphere with four punctures $\t_1$, $\t_1^{-1}$, $\t_2$, $\t_2^{-1}$ and two basis contours is shown on fig.~\ref{contours4}a. Upon performing a braiding transformation, which brings the marked point $\t_2$ around the point $\t_1$, we arrive at the picture on fig.~\ref{contours4}b. The contour $\mathsf{C}_1$ gets a factor of $\mathbf{t}_1$ in crossing the left cut. To compare to fig.~\ref{contours4}a, we also need to move the right cut back to its place. That will multiply the contour $\mathsf{C}_1$ by a factor of $\mathbf{t}_2$. Overall, the transformation acts on the contours as
\beq
\left(\begin{array}{c}\mathsf{C}'_1\\\mathsf{C}'_2\end{array}\right)=\left(\begin{array}{cc}\t_1\t_2&0\\0&1\end{array}\right)\left(\begin{array}{c}\mathsf{C}_1\\\mathsf{C}_2\end{array}\right)\,.
\eeq
Therefore, the state of  U$(1)_{\rm fl}$-charge $-1$ transforms by a factor $(\t_1\t_2)^{-1/2}$, and the two U$(1)_{\rm fl}$-neutral states are transformed by a matrix with eigenvalues $(\t_1\t_2)^{1/2}$ and $(\t_1\t_2)^{-1/2}$. (In taking a square root, we made a choice of sign such that the resulting skein relation for $\t_1=\t_2$ is consistent with fig.~\ref{halfskein}.) The skein relation that we find is shown on fig.~\ref{skein}.

\begin{figure}
 \begin{center}
   \includegraphics[width=12.5cm]{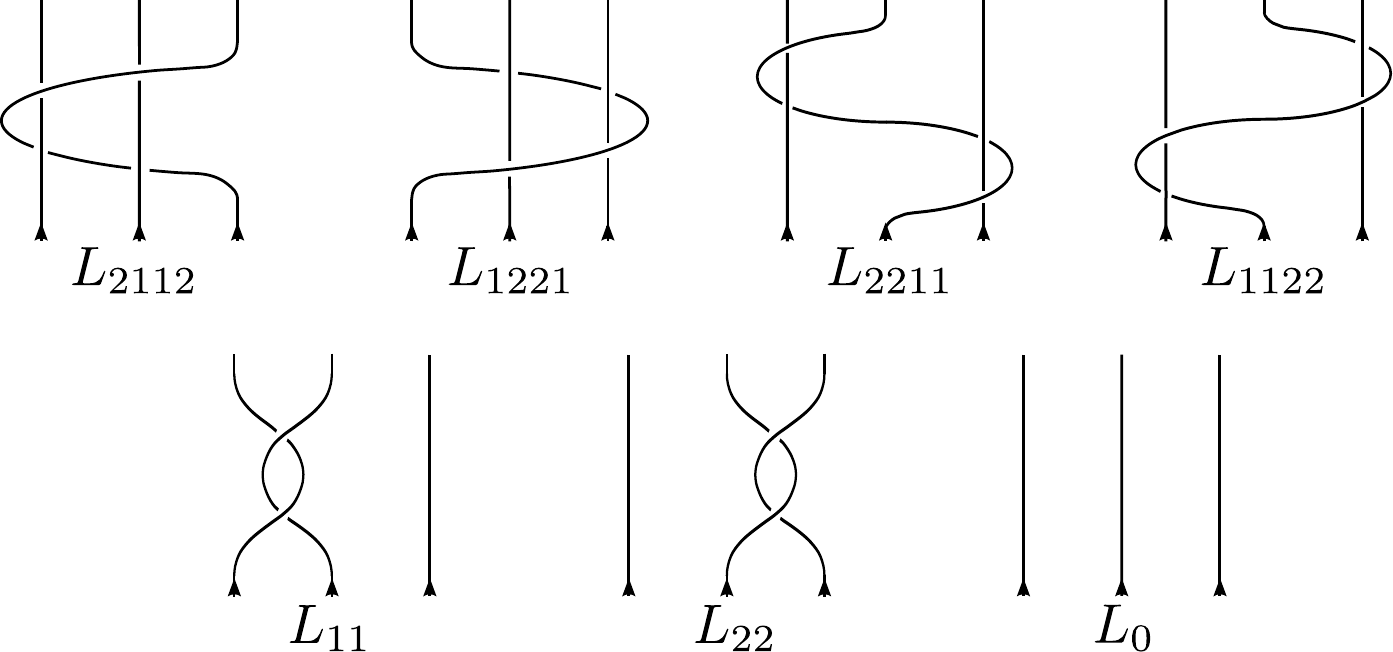}
 \end{center}
 \caption{\small Braids for the 3-strand skein relation.}
  \label{largeskein}
\end{figure}

To completely characterize the multivariable Alexander polynomial, one more skein relation is needed \cite{Murakami}. It relates seven three-strand configurations, shown on fig.~\ref{largeskein}. The existence of this skein relation follows from the fact that the dimension of the U$(1)_{\rm fl}$-invariant subspace of the Hilbert space of the six-punctured sphere, according to (\ref{dimension}), is ${4\choose 2}=6$.

\begin{figure}
 \begin{center}
   \includegraphics[width=14cm]{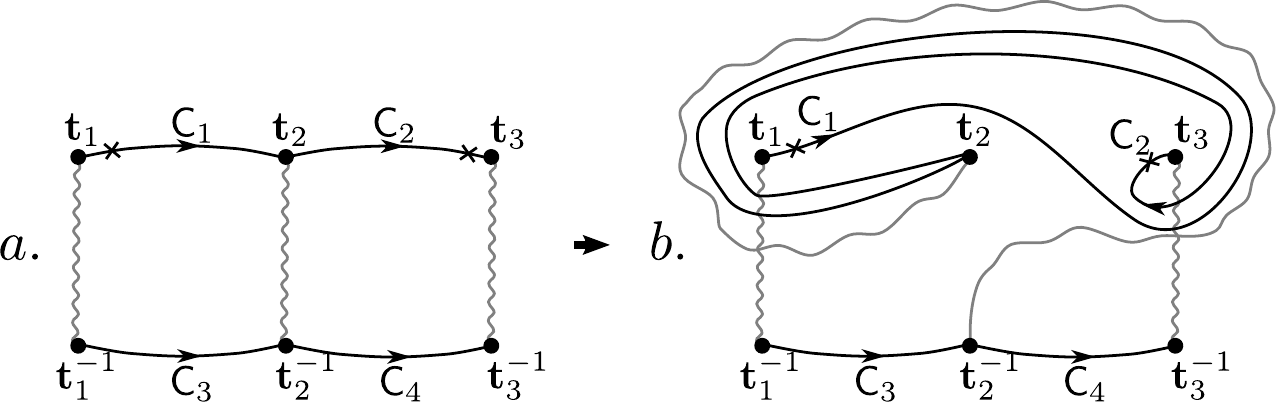}
 \end{center}
 \caption{\small {\large a.}~A particular choice of cuts and basis contours for the six-punctured sphere. {\large b.}~The result of the braiding transformation, corresponding to the link $L_{2211}$.}
  \label{largecontours}
\end{figure}

\begin{figure}
 \begin{center}
   \includegraphics[width=6cm]{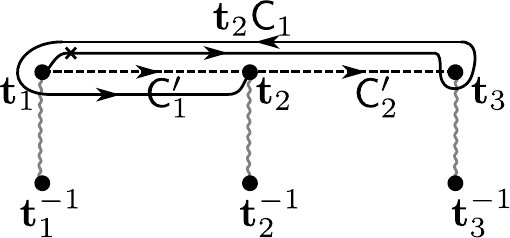}
 \end{center}
 \caption{\small Contour $\t_2\mathsf{C}_1$, which comes from $\mathsf{C}_1$ of fig.~\ref{largecontours} after moving the middle cut back to its place. We show the new basis contours $\mathsf{C}_1'$ and $\mathsf{C}_2'$ by dashed lines.}
  \label{t2c1}
\end{figure}

We need to find the action of the braiding transformations of fig.~\ref{largeskein} on the four contours that generate the twisted homology of the six-punctured sphere. For example, let us consider the link $L_{2211}$. The basis contours and the result of the braiding transformation are shown on fig.~\ref{largecontours}. On the contours $\mathsf{C}_{1}$ and $\mathsf{C}_2$ we put cross-marks at some points, which are not moved in the transformation. At these points the one-form, which is being integrated over the contour, is taken on the first sheet, and on the rest of the contour it is defined by analytic continuation. The first step in comparing figures \ref{largecontours}b and \ref{largecontours}a is to bring the middle cut back to its place. On the way, it will cross the contours $\mathsf{C}_1$ and $\mathsf{C}_2$, and that will multiply them by $\t_2$. On fig.~\ref{t2c1}, we show the contour $\t_2\mathsf{C}_2$. We need to expand it in the new basis $\mathsf{C}_1'$ and $\mathsf{C}_2'$, which is shown by dashed lines. We start comparing the contours from the cross-mark, and add a factor of $\t^{-1}$ each time we cross a cut counterclockwise around a puncture $\t$. We find
\beq
\t_2\mathsf{C}_1=\mathsf{C}_1'+\mathsf{C}_2'+\t_3^{-1}(-\mathsf{C}_2'-\mathsf{C}_1'+\t_1^{-1}\mathsf{C}_1')\,.
\eeq
Repeating the same steps for $\mathsf{C}_2$, and for each link from fig.~\ref{largeskein}, we find the braiding matrices
\begin{align}
&B_{2112}=\t_3\left(\begin{array}{cc}1&0\\\t_2(\t_1-1)&\t_1\t_2\end{array}\right)\,,\quad\quad\quad B_{1221}=\t_1\left(\begin{array}{cc}\t_2\t_3&\t_3-1\\0&1\end{array}\right)\,,\nonumber\\
&B_{2211}=\t_2\left(\begin{array}{cc}\t_1&\t_1(1-\t_3)\\1-\t_1&1-\t_1+\t_1\t_3\end{array}\right)\,,\quad B_{1122}=\t_2\left(\begin{array}{cc}1-\t_3+\t_1\t_3&\t_2^{-1}(1-\t_3)\\\t_2\t_3(1-\t_1)&\t_3\end{array}\right)\,,\nonumber\\
&B_{11}=\left(\begin{array}{cc}\t_1\t_2&0\\\t_2(1-\t_1)&1\end{array}\right)\,,\quad B_{22}=\left(\begin{array}{cc}1&1-\t_3\\0&\t_2\t_3\end{array}\right)\,,\quad B_{0}=\left(\begin{array}{cc}1&0\\0&1\end{array}\right)\,.\label{braidingmatr}
\end{align}
Here we defined the matrices by $(\mathsf{C}_1',\mathsf{C}_2')^T=B(\mathsf{C}_1,\mathsf{C}_2)^T$. The contours $\mathsf{C}_3$ and $\mathsf{C}_4$ are transformed trivially.

Let $a_{1,2,3,4}^+$ be the four creation operators, obtained by integrating the fermionic gauge field $A^+$ over the corresponding contours. The Hilbert space of the six-punctured sphere contains one state of charge $-2$, from which we build the other states by applying $a^+_\bullet$. The six neutral states, which we are interested in, are
\begin{align}
&a^+_1a^+_2|{-}2\rangle,\nonumber\\
&a^+_1a^+_3|{-}2\rangle,\quad a^+_2a^+_3|{-}2\rangle,\nonumber\\
&a^+_1a^+_4|{-}2\rangle,\quad a^+_2a^+_4|{-}2\rangle,\nonumber\\
&a^+_3a^+_4|{-}2\rangle\,.\label{6hilbert}
\end{align}
The highest weight state $|{-}2\rangle$ transforms under braiding by a factor $\det^{-1/2}B$, and therefore so does the state $a_3^+a^+_4|{-}2\rangle$. The state $a^+_1a^+_2|{-}2\rangle$ transforms by a factor $\det^{1/2}B$. The states in the second  and the third lines of (\ref{6hilbert}) transform in doublets by the matrix $B\det^{-1/2}B$. In total, for each braiding transformation, the $6\times 6$ braiding matrix has $1+1+4=6$ independent matrix elements. We can collect them in a $7\times 6$ matrix, in which the rows correspond to the diagrams of fig.~\ref{largeskein}. The null-vector of this matrix will give us the skein relation. Let us set $g_{\pm}(\t)=\t^{1/2}\pm\t^{-1/2}$. Using the explicit expressions for the braiding matrices (\ref{braidingmatr}), one finds the skein relation to be
\begin{align}
&g_+(\t_1)g_-(\t_2)L_{2112}-g_-(\t_2)g_+(\t_3)L_{1221}+g_-(\t_1\t_3^{-1})(L_{2211}+L_{1122})\nonumber\\
&+g_-(\t_2\t_3\t_1^{-1})g_+(\t_3)L_{11}-g_-(\t_1\t_2\t_3^{-1})g_+(\t_1)L_{22}+g_-(\t_1^2\t_3^{-2})L_0=0\,.
\end{align}
This, indeed, is the correct skein relation for the multivariable Alexander polynomial. Together with other relations and normalization conditions that we have found, it fixes the knot invariant completely \cite{Murakami}. We should note, however, that we did not explain, how to properly choose the square root of the determinant of the braiding matrix in the transformation of the highest weight state. Thus, our derivation does not allow to unambiguously fix relative signs of different diagrams in the skein relation.

\enddocument